\documentclass[prd,showpacs,nofootinbib,preprint]{revtex4}
\usepackage{graphicx}
\usepackage{epsfig}
\usepackage{amsmath}
\usepackage{amsfonts}
\usepackage{amssymb}
\usepackage{url}
\usepackage{hyperref}
\usepackage{subfigure}
\usepackage{hhline}
\usepackage{color}
\usepackage{bm}
\usepackage{cancel}
\newcommand{\beqa}{\begin{eqnarray}}
\newcommand{\eeqa}{\end{eqnarray}}
\newcommand{\beq}{\begin{equation}}
\newcommand{\eeq}{\end{equation}}

\newcommand{\nn}{\nonumber}
\newcommand{\bmt}{\begin{pmatrix}}
\newcommand{\emt}{\end{pmatrix}}
\usepackage[toc,page]{appendix}
\usepackage{comment}
\usepackage{multirow}
\newcommand{\be}{\begin{equation}}
\newcommand{\ee}{\end{equation}}
\newcommand{\bea}{\begin{eqnarray}}
\newcommand{\eea}{\end{eqnarray}}

\def\Bbar{\overline{B}}

\def\ubar{\overline{u}}

\def\Qbar{\overline{Q}}

\def\nubar{{\overline{\nu}}}

\def\L{\mathcal{L}}

\def\A{\mathcal{A}}

\def\th{{\theta}}
\begin{document}
\title{ Efficacy of scalar leptoquark on $B_s \to (K^{(*)}, D_s^{(*)}) \tau \bar \nu_\tau$ decay modes}

\author{Suchismita Sahoo$^{a}$}
\email{suchismita8792@gmail.com}

\author{Anupama Bhol$^{b}$}
\email{anupama.phy@gmail.com}


  \affiliation{$^a$\, Department of Physics, Central University of Karnataka, Kalaburagi-585367, India\\ $^b$ Govt. Women's College Baripada-757001, India}

\begin{abstract}
We  scrutinize  the impact of various relevant scalar leptoquarks on the physical observables associated with the rare semileptonic decay processes of $B_s$ meson involving  $b \to (u, c) l\bar \nu_l$ quark level transitions. We constrain the new parameter space consistent with the experimental limit on   Br($B_{u,c} \to \tau \nu_l$), Br($B \to \pi \tau \bar \nu_\tau$), $R_{D^{(*)}}, R_{J/\psi}$ and  $R_\pi^l$ observables. Using the allowed parameter space, we  compute the branching ratios, forward-backward asymmetries,  lepton and hardon polarization asymmetries of $B_s \to (K^{(*)}, D_s^{(*)}) \tau \bar \nu_\tau$ decay modes.  Forbye, we look at the possibility of existence of lepton non-universality in these processes.

\end{abstract}
\maketitle

\section{Introduction}
Even though the Standard Model (SM) is currently the best theory of particle physics, it does not explain the complete picture of subatomic world. There are still some fundamental questions which are not answered within the domain of SM, so one has to search for new physics (NP) beyond it. In this aspect, the investigation of  weak decays of $B$ mesons provide an excellent window. Recently various challenging anomalies in the sector of the violation of  lepton flavor universality (LFU) in semileptonic B meson decays, especially in the measurements of $R_{D^{(*)}}, R_{J/\psi}$ have been observed. The measurements in the observables like
$R_{D} = \frac{\mathcal {\rm Br}( B \to D\tau \bar{\nu}_\tau)}{\mathcal {\rm Br}( B \to D\, l \bar{\nu}_l)}\,$, $ R_{D^{\ast}}= \frac{\mathcal {\rm Br}( B \to {D}^{\ast}\tau \bar{\nu}_\tau)}{\mathcal {\rm Br}(\bar B \to {D}^{\ast}\,l \bar{\nu}_l)}\,$ by BaBar \cite {Lees:2012xj, Lees:2013uzd}, Belle \cite {Huschle:2015rga,Abdesselam:2016cgx, Abdesselam:2016xqt, Hirose:2017dxl, Hirose:2016wfn} and LHCb \cite{Aaij:2015yra, Aaij:2017deq, Aaij:2017uff} experiments  deviate from their standard model predictions at  $3.8\sigma$ level \cite{HFLAV}. Another observable, $R_{J/\Psi} = \frac{\mathcal {\rm Br}( B_c \to {J/\Psi}\tau \bar{\nu}_\tau)}{\mathcal {\rm Br}( B_c \to {J/\Psi}\, l \bar{\nu}_l)}\,$ measured by LHCb shows a discrepancy at $1.7\sigma$  \cite{Aaij:2017tyk} from the SM value. The combined data of $R_{D^{(*)}}$  by HFLAV Collaboration \cite{HFLAV} are
\bea
&&R_D^{\rm Expt}=0.340\pm 0.027 \pm 0.013\,,\\
&&R_{D^*}^{\rm Expt}=0.295\pm 0.011\pm 0.008\,,
\eea 
respectively, whereas the value of $R_{J/\Psi}$ by LHCb is  \cite{Aaij:2017tyk}
\bea
R_{J/\psi}^{\rm Expt}= 0.71 \pm 0.17 \pm 0.18\,,
 \eea
  with the first error as statistical and the second one as systematic. The SM predictions of all these observables $R_D$, $R_{D^*}$ \cite{Na:2015kha, Fajfer:2012vx, Fajfer:2012jt}, and $R_{J/\Psi}$~\cite{Wen-Fei:2013uea, Ivanov:2005fd, Dutta:2017xmj} are $0.299 \pm 0.003$, $ 0.258 \pm 0.005$, and  $0.289\pm 0.01$ respectively. Since the uncertainties from the Cabibbo-Kobayashi-Maskawa (CKM) matrix and the hadronic transition form factors are cancelled out to a large extent in  all these ratios of branching fractions of these decay modes, any deviation in these observables from their  SM values would definitely point out towards the signals of new physics. Besides these ratios, another discrepancy in $b \to u l \bar \nu_l$ processes  is also noticed in the measured ratio \cite{Fajfer:2012jt}
\bea
R_\pi^l = \frac{\tau_{B^0}}{\tau_{B^-}}\frac{{\rm Br}(B^- \to \tau^-  \bar \nu_\tau)}{{\rm Br}(B^0 \to \pi^+ l^- \bar \nu_l)}, ~~~l=e, \mu\,,
\eea
where $\tau_{B^0}~(\tau_{B^-})$ is the life time of $B^0~(B^-)$ meson.  
Taking the experimental results of the branching ratios of $B_u^- \to \tau^- \bar \nu_\tau$ and  $B^0 \to \pi^+ l^- \bar \nu_l$ decay processes  
\bea
&&{\rm Br}(B_u^- \to \tau^- \bar \nu_\tau)^{\rm Expt}=(1.09\pm 0.24)\times 10^{-4}, \\
&&{\rm Br}( B^0 \to \pi^+ l^- \bar \nu_l)^{\rm Expt} =(1.45\pm 0.05)\times 10^{-4}, 
\label{Bpil-Exp}
\eea
with $\tau_{B^-}/\tau_{B^0}=1.076\pm 0.004$ from \cite{Patrignani:2016xqp}, we obtain  
\bea
R_\pi^l |^{\rm Expt} =0.699\pm 0.156,
\eea
which has also  nearly $1 \sigma$ discrepancy from its SM value $R_\pi^l |^{\rm SM}=0.583 \pm 0.055$.

Over the last few years, most of the research works have focused on the $B\to D^{(*)}\tau \bar \nu_\tau$ problems rather than the issues  in the $B_s \to {D_s}^{(*)}l\bar \nu_l$ channels. Though both of these decays are driven by the  $b\to c$ transtion, the difference is in the spectator quark. The decay channels $B_s \to {D_s}^{(*)}l\bar \nu_l$ have the spectator strange quark whereas $B\to D^{(*)}l\bar \nu_l$ have either up or down quark. These decays have $SU(3)$ flavor symmetry with the dependence on CKM matrix element $V_{cb}$ and hence they should show similar properties in the limit of $SU(3)$ flavor symmetry. Further, studies of $B_{(s)}\to D_{(s)}^{(*)}\tau\bar \nu_\tau $ may help to determine the value of $|V_{cb}|$ with higher precision in both exclusive and inclusive measurements. Several authors have investigated the semileptonic $B_s \to {D_s}^{(*)}\tau\bar \nu_\tau$~\cite{Bhol:2014jta}  decays within SM and the branching fractions have been computed using the constituent quark meson (CQM) model, QCD sum rules approach, the light cone sum rules
(LCSR) approach \cite{Li:2009wq}, the covariant light-front quark model (CLFQM) \cite{ Li:2010bb}, lattice QCD method \cite{Atoui:2013mqa, Atoui:2013zza, Bailey:2012rr, Monahan:2016qxu, Na:2012kp, Monahan:2018lzv} and in the perturabative QCD
factorization approach \cite{Chen:2011ut, Fan:2013kqa,  Monahan:2017uby}. Very recently the problem has been studied in Ref. \cite{Dutta:2018jxz} using effective theory formalism in presence of NP in a model independent way.
Moreover the recent results of ${R}_{D^{(*)}}$ also motivates to analyse $b\to u \tau \bar\nu_\tau$ counterparts and look for the lepton non-universality (LNU) observables. 
  With strange quark as spectator quark another possible decay channel of $B_s$ meson is $B_s \to K^{(*)}\tau\bar \nu_\tau$ decays with $b\to u \tau \bar \nu_\tau$ charged current interaction and which again, have been studied by a few authors~\cite{Wang:2012ab, Meissner:2013pba, Faustov:2013ima, Horgan:2013hoa, Bouchard:2014ypa} within SM. Recently $B_s \to K^{(*)}\tau \bar \nu_\tau$ decays have been worked out in the Ref.~\cite{Sahoo:2017bdx, Rajeev:2018txm} within SM and beyond. 
 The transition $b\to u \tau \bar \nu_\tau$ process is also potentially sensitive due to the presence of a charged Higgs boson in new physics models like the two-Higgs-doublet model (2HDM)~\cite{Lee:1973iz, Branco:2011iw} and in the minimal supersymmetric model (MSSM)~\cite{Wess:1974tw, Golfand:1971iw} and again, a systematic study of process both experimentally and theoretically can be helpful for the measurement of the smallest element of CKM matrix, $|V_{ub}|$. 

In this concern, we would like to study these  particular rare semileptonic decays of $B_s$ mesons with a lepton and a neutrino in the final state i.e., $B_s \to M \tau\bar \nu_\tau$ where $M$ is either ${D_s}^{(*)}$ mesons or $K^{(*)}$ mesons in the scalar leptoquark (SLQ) model.  Leptoquark (LQ) is a unique  hypothetical  color triplet scalar (spin=0) or vector (spin=1) bosonic particle, which acts as a bridge between  the quark and  lepton sectors, thus carries  both baryon $(B)$ and lepton $(L)$ quantum numbers. The $B$ and $L$ numbers conserving LQs avoid rapid proton decays and  are light enough to be seen in the present experiments.  The existence of LQ is proposed  in many new theoretical frameworks, such as the grand unified theories \cite{Georgi:1974sy, Fritzsch:1974nn, Langacker:1980js, Georgi:1974my}, Pati-Salam model \cite{Pati:1974yy, Pati:1973uk, Pati:1973rp, Shanker:1981mj, Shanker:1982nd}, quark and lepton composite model \cite{Kaplan:1991dc} and the technicolor model \cite{Schrempp:1984nj, Gripaios:2009dq}. The phenomenology of SLQs  in connection to only flavor anomalies, and   both flavor and dark matter sectors has been studied extensively in the literature \cite{Alok:2017sui, Becirevic:2017jtw, Hiller:2017bzc, DAmico:2017mtc, Becirevic:2016yqi, Bauer:2015knc, Li:2016vvp, Calibbi:2015kma, Freytsis:2015qca, Dumont:2016xpj, Dorsner:2016wpm, Varzielas:2015iva, Dorsner:2011ai, Davidson:1993qk,  Saha:2010vw,  Mohanta:2013lsa, Sahoo:2015fla, Sahoo:2015pzk, Sahoo:2015qha, Sahoo:2015wya, Kosnik:2012dj, Singirala:2018mio, Chauhan:2017ndd, Becirevic:2018afm, Angelescu:2018tyl, Sahoo:2017lzi, Sahoo:2016nvx, Sahoo:2016edx}. In this work, we consider three relevant SLQs, singlet $S_1(\bar 3, 1,1/3)$, doublet $R_2(3,2,7/6)$ and triplet $S_3(\bar 3,3,1/3)$, which provide additional vector, scalar and tensor type couplings contributions to the SM. Constraining the new parameter space from Br($B_{u,c} \to \tau \nu_\tau$), Br($B \to \pi \tau \bar \nu_\tau$), $R_{D^{(*)}}$, $R_{J/\psi}$ and $R_\pi^l$ parameters, we compute the branching ratios, forward-backward asymmetries, polarization asymmetries and lepton non-universality parameters of $B_s \to (D^{(*)}_s, K^{(*)}) \tau \bar \nu_\tau$  processes.

The  paper is organised as follows. We present the theoretical framework and the  most general effective Hamiltonian   associated with $b \to (u,c)\tau \bar \nu_\tau$ processes in section II.  In section III, we provide the detailed discussion on the new scalar leptoquarks and the constrained on new couplings from the available experimetal results on feasible observables. The numerical analysis of all the physical observables of $B_s \to (D^{(*)}_s, K^{(*)}) \tau \bar \nu_\tau$ decay modes in the presence of leptoquarks are given in section IV and section V summarize our estimated results. 

\section{Theoretical Framework for the analysis of $b \to (u,c ) l \bar \nu_l$ decay processes}
Considering  the neutrinos to be left-handed, the most general effective Lagrangian describing the $b\to q \tau \bar \nu_l$,  $(q=u,~c)$ transition is given by \cite{Bhattacharya:2011qm, Cirigliano:2009wk}
\begin{eqnarray}
\mathcal L_{\rm eff} &=&
-\frac{4\,G_F}{\sqrt{2}}\,V_{q b}\,\Bigg\{(1 + V_L)\,\bar{l}_L\,\gamma_{\mu}\,\nu_L\,\bar{q}_L\,\gamma^{\mu}\,b_L +
V_R\,\bar{l}_L\,\gamma_{\mu}\,\nu_L\,\bar{q}_R\,\gamma^{\mu}\,b_R 
\nn \\
&&+
S_L\,\bar{l}_R\,\nu_L\,\bar{q}_R\,b_L +
S_R\,\bar{l}_R\,\nu_L\,\bar{q}_L\,b_R + 
T_L\,\bar{l}_R\,\sigma_{\mu\nu}\,\nu_L\,\bar{q}_R\,\sigma^{\mu\nu}\,b_L  \Bigg\} + {\rm h.c.}\,,\label{ham}
\end{eqnarray}
where $l=e,\mu,\tau$ are the flavor of neutrinos. 
Though the new Wilson coefficients $V_{L, R}, S_{L, R}, T_L$ are zero  in the SM, they can have nonvanishing values in the presence of new physics.

Using the generalized effective Lagrangian \ref{ham}\,,  the branching ratios of $ B \to P l \bar \nu_l$ processes, where $P=D, D_s, \pi, K$ are the pseudoscalar mesons are given by \cite{Sakaki:2013bfa, Tanaka:2012nw}
\bea
\label{pilnu}
\frac{d{\rm Br}(B_s \to P l\bar \nu_l)}{dq^2} &=&\tau_{B_s} {G_F^2 |V_{qb}|^2 \over 192\pi^3 M_{B_s}^3} q^2 \sqrt{\lambda_P(M_{B_s}^2,\,M_{P}^2,\,q^2)} \left( 1 - {m_l^2 \over q^2} \right)^2 \times \nn \\ && \biggl\{ \biggr. |1 + V_L + V_R|^2 \left[ \left( 1 + {m_l^2 \over2q^2} \right) H_{0}^{2} + {3 \over 2}{m_l^2 \over q^2} \, H_{t}^{2} \right] \nn \\                                                      &&+ {3 \over 2} |S_L + S_R|^2 \, H_S^{2} + 8|T_L|^2 \left( 1+ {2m_l^2 \over q^2} \right) \, H_T^{2} \nn \\ &&+ 3{\rm Re}[ ( 1 + V_L + V_R ) (S_L^* + S_R^* ) ] {m_l \over \sqrt{q^2}} \, H_S H_{t} \nn \\
 &&- 12{\rm Re}[ ( 1 + V_L + V_R ) T_L^* ] {m_l \over \sqrt{q^2}} \, H_T H_{0} \biggl.\biggr\} \,,
\eea
where the helicity amplitudes in terms of form factors $(F_{0,+})$ are expressed as
\bea
&&H_0 = \sqrt{ \frac{\lambda_P(M_{B_s}^2,\,M_{P}^2,\,q^2)}{q^2}}\,F_{+}(q^2), ~~~~H_t = \frac{M_{B_s}^2 - M_P^2}{\sqrt{q^2}}\,F_0(q^2)\,, \nn \\
&&H_S=\frac{M_{B_s}^2 - M_P^2}{m_b - m_{q}}\,F_0(q^2)\,,~~~~~~~~~~~~~~H_T=-\frac{\sqrt{\lambda_P(M_{B_s}^2,\,M_{P}^2,\,q^2)}}{M_{B_s}+M_P}F_T(q^2),
\eea
with 
\bea
\lambda_{P}(a,b,c) = ((a - b)^2 - c)((a + b)^2 - c).
\eea
 The branching ratios of $B_s \to V l \nubar_l$ with respect to $q^2$, where $V=K^*, D^*, D_s^*$ are the vector bosons are given by \cite{Sakaki:2013bfa, Tanaka:2012nw}   
\bea
\label{vlnu1}
 {d{\rm Br}( B_s \to V l\bar \nu_l) \over dq^2} &=& \tau_{B_s} {G_F^2 |V_{qb}|^2 \over 192\pi^3 M_{B_s}^3} q^2 \sqrt{\lambda_V(M_{B_s}^2,\,M_{V}^2,\,q^2)} \left( 1 - {m_l^2 \over q^2} \right)^2 \times \nn \\ && \biggl\{ \biggr. 
      ( |1 + V_L|^2 + |V_R|^2 ) \left[ \left( 1 + {m_l^2 \over2q^2} \right) \left( H_{V,+}^2 + H_{V,-}^2 + H_{V,0}^2 \right) + {3 \over 2}{m_l^2 \over q^2} \, H_{V,t}^2 \right] \nn \\ &&  - 2{\rm Re}[(1 + V_L) V_R^*] \left[ \left( 1 + {m_l^2 \over 2q^2} \right) \left( H_{V,0}^2 + 2 H_{V,+} H_{V,-} \right) + {3 \over 2}{m_l^2 \over q^2} \, H_{V,t}^2 \right] \nn \\ &&  + {3 \over 2} |S_L - S_R|^2 \, H_S^2 + 8|T_L|^2 \left( 1+ {2m_l^2 \over q^2} \right) \left( H_{T,+}^2 + H_{T,-}^2 + H_{T,0}^2  \right) \nn \\ &&  + 3{\rm Re}[ ( 1 + V_L - V_R ) (S_L^* - S_R^* ) ] {m_l \over \sqrt{q^2}} \, H_S H_{V,t} \nn \\ &&  - 12{\rm Re}[ (1 + V_L) T_L^* ] {m_l \over \sqrt{q^2}} \left( H_{T,0} H_{V,0} + H_{T,+} H_{V,+} - H_{T,-} H_{V,-} \right) \nn \\ &&  + 12{\rm Re}[V_R T_L^* ] {m_l \over \sqrt{q^2}} \left( H_{T,0} H_{V,0} + H_{T,+} H_{V,-} - H_{T,-} H_{V,+} \right) \biggl.\biggr\} \,,
\eea
where the hadronic amplitudes $H_{V, \pm}$, $H_{V, 0}$, $H_{V, t}$  and $H_{S}$ in terms of the form factors $V, A_{0,1,2}, T_{1,2,3}$ can be found in  the Refs.  \cite{Sakaki:2013bfa, Tanaka:2012nw}.
 Besides the branching ratios, we also explore more physical observables in these decay modes in order to probe the structure of NP. The zero crossing of lepton forward-backward asymmetry is one of the interesting observable, which is defined as \cite{Sakaki:2013bfa, Biancofiore:2013ki}
\bea
\A_{\rm FB} = { \int_0^1 {d\Gamma \over d\cos\th}d\cos\th-\int^0_{-1}{d\Gamma \over d\cos\th}d\cos\th \over \int_{-1}^1 {d\Gamma \over d\cos\th}d\cos\th }\,.
\eea
Here $\theta$ is the angle between the three-momenta of $\tau$ and $B_s$ in the lepton rest frame. As like the $R_{D^{(*)}}$ LNU parameters, we also define the lepton universality violating parameters $R_{P(V)}^{\tau l}$ as 
\bea
&&R_P^{\tau l}=\frac{{\rm Br}(B_s \to P \tau \bar \nu_\tau)}{{\rm Br}(B_s \to P l \bar \nu_l)}, \\
&&R_V^{\tau l}=\frac{{\rm Br}(B_s \to V \tau \bar \nu_\tau)}{{\rm Br}(B_s \to V l \bar \nu_l)},~~~~l=e, \mu.
\eea 
Other amazing observables are the polarization asymmetry parameters. The  $\tau$ polarization asymmetry of $B_s \to P \tau \bar \nu_l$ decay modes are  given as \cite{Sakaki:2013bfa},
\bea
P_\tau = { \Gamma(\Bbar \to D^{(*)} \tau \nubar)|_{\lambda_\tau=1/2} - \Gamma(\Bbar \to D^{(*)} \tau \nubar)|_{\lambda_\tau=-1/2} \over \Gamma(\Bbar \to D^{(*)} \tau \nubar)|_{\lambda_\tau=1/2} + \Gamma(\Bbar \to D^{(*)} \tau \nubar)|_{\lambda_\tau=-1/2} } \,,
\eea
and the  $V(=K^*, D_s^*)$ longitudinal polarization parameters are defined as \cite{Biancofiore:2013ki},
\bea
P_V = { \Gamma(B_s \to V \tau \bar \nu_l)|_{\lambda_V=0} \over \Gamma(B_s \to V \tau \bar \nu_l)|_{\lambda_V=0} + \Gamma(B_s \to V \tau \bar \nu_l)|_{\lambda_V=1} + \Gamma(B_s \to V \tau \bar \nu_l)|_{\lambda_V=-1} } \,.
\eea
 The detailed expressions for the $q^2$ distributions for various $\tau$ and $V$ polarization states  can be found in the Ref.  \cite{Sakaki:2013bfa}.

\section{Model with scalar leptoquarks }
In the presence of scalar LQ, the interaction Lagrangian responsible for the $b \to q \ell\nubar$ processes are given by \cite{Sakaki:2013bfa},
\bea
\L_{\rm LQ} &=& \left( y_{2L}^{ij}\,\ubar_{iR} L_{jL} + y_{2R}^{ij}\,\Qbar_{iL} i\sigma_2 \ell_{jR} \right)R_2 \,,+\left( y_{1L}^{ij}\,\Qbar_{iL}^c i\sigma_2 L_{jL} + y_{1R}^{ij}\,\ubar_{iR}^c \ell_{jR} \right)S_1 \nn \\ && + y_{3L}^{ij}\,\Qbar_{iL}^c i\sigma_2{\bm\sigma} L_{jL}{\bm S}_3\,,
 \eea
where  $i$ and $j$ are the generation indices,  $Q_{iL}~(u_{iR}, ~d_{iR})$ and $L_{jL}~(\ell_{jR})$ are the left (right) handed quark and lepton $SU(2)_L$ doublets (singlets) respectively.  Here  $Q_{iL}^c$ and $u_{iR}^c$ are the charge-conjugated fermion fields. After performing the Fierz transformations, we obtain the additional Wilson coefficients contributions to the $b \to q_m \tau \bar \nu_l$ processes as \cite{Sakaki:2013bfa}
\begin{subequations}
   \label{eq:LQ_coefficients}
   \bea
      \label{eq:CV1}
     V_L &=& { 1 \over 2\sqrt2 G_F V_{mb} } \sum_{k=1}^3 V_{k3} \left[ {y_{1L}^{kl}y_{1L}^{m3*} \over 2M_{S_1^{1/3}}^2} - {y_{3L}^{kl}y_{3L}^{m3*} \over 2M_{S_3^{1/3}}^2} \right] \,, \\
     V_R&=&0\,,\\
      \label{eq:CS2}
      S_L&=&0\,,\\
      S_R &=& { 1 \over 2\sqrt2 G_F V_{mb} } \sum_{k=1}^3 V_{k3} \left[ -{y_{1L}^{kl}y_{1R}^{m3*} \over 2M_{S_1^{1/3}}^2} - {y_{2L}^{ml}y_{2R}^{k3*} \over 2M_{R_2^{2/3}}^2} \right] \,, \\
      \label{eq:CT}
      T_L &=& { 1 \over 2\sqrt2 G_F V_{mb} } \sum_{k=1}^3 V_{k3} \left[ {y_{1L}^{kl}y_{1R}^{m3*} \over 8M_{S_1^{1/3}}^2} - {y_{2L}^{ml}y_{2R}^{k3*} \over 8M_{R_2^{2/3}}^2} \right] \,,
   \eea
\end{subequations}
where $V_{k3}$ denotes the elements of CKM matrix.   Here $y_{xL}^{ij}$ and $y_{xR}^{ij}$ ($x=1,2,3$) are the leptoquark couplings in the mass basis of the down type quarks and charged leptons.  The upper index in the LQ mass  denotes the electric charge of LQ.

\subsection{Constraint on leptoquark couplings}

With the idea on new Wilson coefficients in mind, we  now move on to constrain the new parameters from the  available experimentally feasible flavor observables like ${\rm Br}(B_{u,c} \to \tau \bar \nu_\tau)$, ${\rm Br}(B \to\pi \tau \bar \nu_\tau)$,  $R_\pi^l$, $R_{D^{(*)}}$ and $R_{J/\psi}$. We assume that only  the third generation lepton receives  the additional new physics contributions arising due to the scalar leptoquarks exchange and the couplings with light leptons are considered to be SM like.  The SM branching ratios of $B_q \to \pi l \nu_l$ processes obtained by using the masses of all the particles, lifetime of $B_q$ meson, CKM matrix elements from  \cite{Patrignani:2016xqp} and the $B \to \pi$ form factors  from \cite{Khodjamirian:2011ub, Bourrely:2008za, Boyd:1994tt, Boyd:1995cf}, are given by
\bea \label{Bpil-SM}
&&{\rm Br}( B^0 \to \pi^+ \mu^- \bar \nu_\mu)^{\rm SM}= (1.35\pm 0.10)\times 10^{-4},\\
&&{\rm Br}( B^0 \to \pi^+ \tau^- \bar \nu_\tau)^{\rm SM}=(9.40 \pm 0.75)\times 10^{-5}.
\eea
Although,  the branching ratio of the muonic channel agrees reasonably well  with the experimental value \ref{Bpil-Exp}\,,  the tau-channel is within its  current experimental limit  \cite{Patrignani:2016xqp} 
\bea
{\rm Br}( B^0 \to \pi^+ \tau^- \bar \nu_\tau)^{\rm Expt} < 2.5\times 10^{-4}.
\eea
Includig the new physics contribution, the branching ratios of $B_q \to l\bar \nu_l$ processes  are given by \cite{Biancofiore:2013ki}
\bea \label{leptonic}
{\rm Br}(B_q \to l \bar \nu_l) &=&
\frac{G_F^2\,|V_{qb}|^2}{8\,\pi}\,\tau_{B_q} f_{B_q}^2\,m_l^2\,M_{B_q}\,\Big(1 - \frac{m_l^2}{M_{B_q}^2}\Big)^2\,  \nn \\ &\times&
\Big |\left( 1+V_L-V_R\right) - \frac{M_{B_q}^2}{m_l\,(m_b + m_q)}\,\left(S_L-S_R\right) \Big |^2\,.
\eea
Using  the decay constants $f_{B_u}=190.5\pm 4.2$ MeV, $f_{B_c}=489\pm 4\pm 3$ MeV from \cite{Aoki:2013ldr, Chiu:2007km} and rest input parameters from \cite{Patrignani:2016xqp}, the branching ratios of $B_{u,c}^+ \to \tau^+ \nu_\tau$ processes in the SM are found to be
\bea \label{butau-SM}
&&{\rm Br}(B_u^+ \to \tau^+ \nu_\tau)^{\rm SM}=(8.48\pm 0.5)\times 10^{-5},\\
&&{\rm Br}(B_c^+ \to \tau^+ \nu_\tau)^{\rm SM}=(3.6\pm 0.14)\times 10^{-2}\,.
\eea
Considering the current world average of the $B_c$ lifetime, the  upper limit on the branching ratio of $B_c^+ \to \tau^+ \nu_\tau$ process  is  \cite{Akeroyd:2017mhr}
\bea
{\rm Br}(B_c^+ \to \tau^+ \nu_\tau)\lesssim 10\%.
\eea   
To compute the allowed regions of  new parameters associated with $b \to u \tau\bar \nu_\tau$ processes, we compare the theoretically predicted  values of ${\rm Br}(B_u^+ \to \tau^+ \nu_\tau)$, $R_\pi^l$  with their corresponding $3\sigma$  range of observed experimental results and for $b \to c \tau\bar \nu_l$ transitions, we use the experimental limits on  $R_{D^{(*)}}, ~R_{J/\psi}$ parameters and the branching ratio of $B_c \to \tau \nu_\tau$ channel.  The upper limit on the branching ratio of $B^0 \to \pi^+ \tau^- \bar \nu_\tau$ process is also used to constrain the new couplings of  $b \to u \tau\bar \nu_l$. By reason of zero contribution of tensor coupling to the branching ratios of $B_{u,c}^+ \to \tau^+ \nu_\tau$ processes and the lack of  precise determination of the form factors associated with tensorial operators
for $B_c \to J/\psi \tau \bar \nu_\tau$ decay mode,  the tensor operator part contribution from  Br($B^0 \to \pi^+ \tau^- \bar \nu_\tau$) and $R_{D^{(*)}}$ observables are only included in this analysis.   We consider two cases  of couplings, (a) couplings as real and (b) couplings as imaginary. 
\subsubsection{Real couplings}
\begin{figure}[htb]
\includegraphics[scale=0.45]{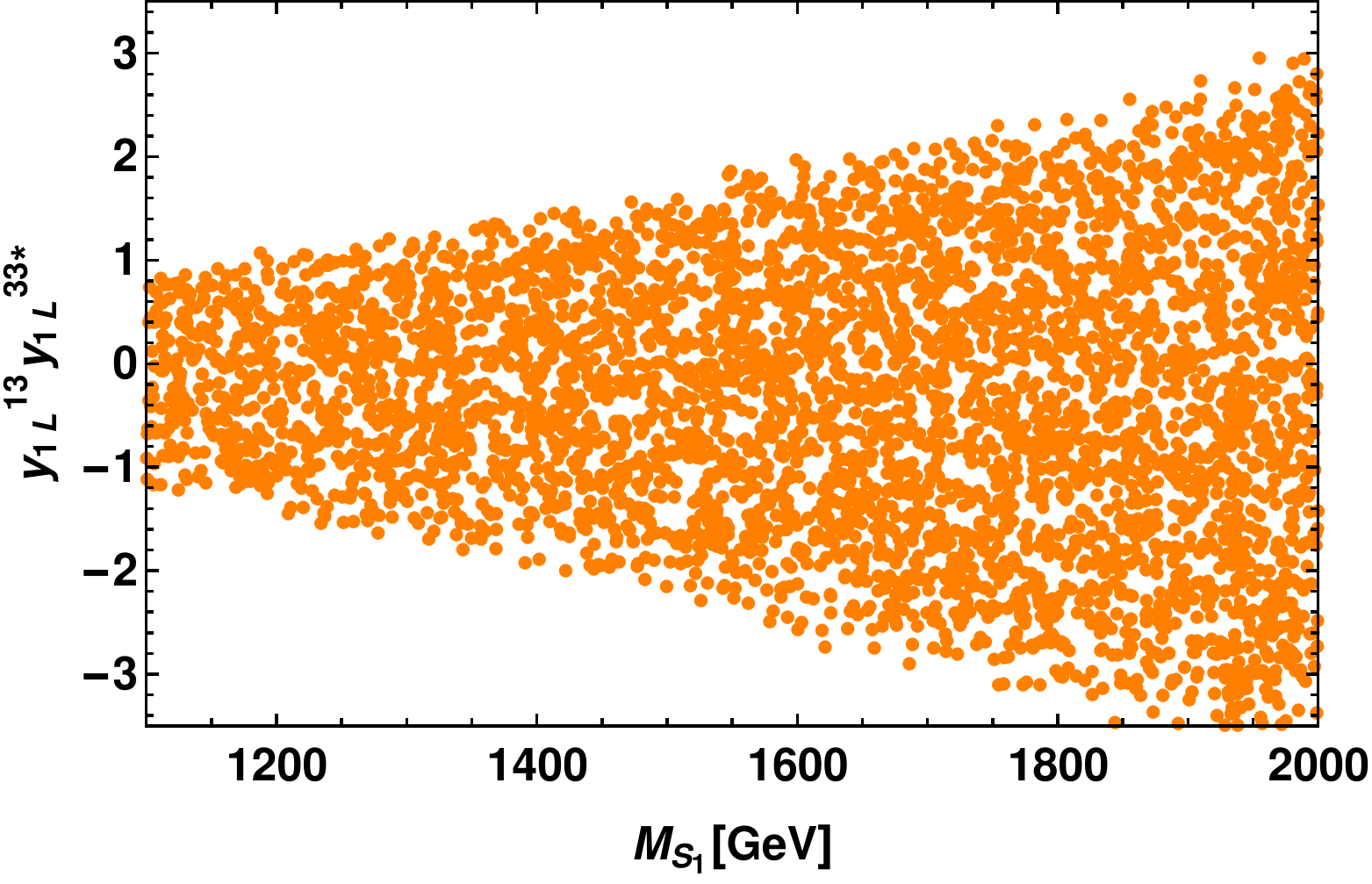}
\quad
\includegraphics[scale=0.45]{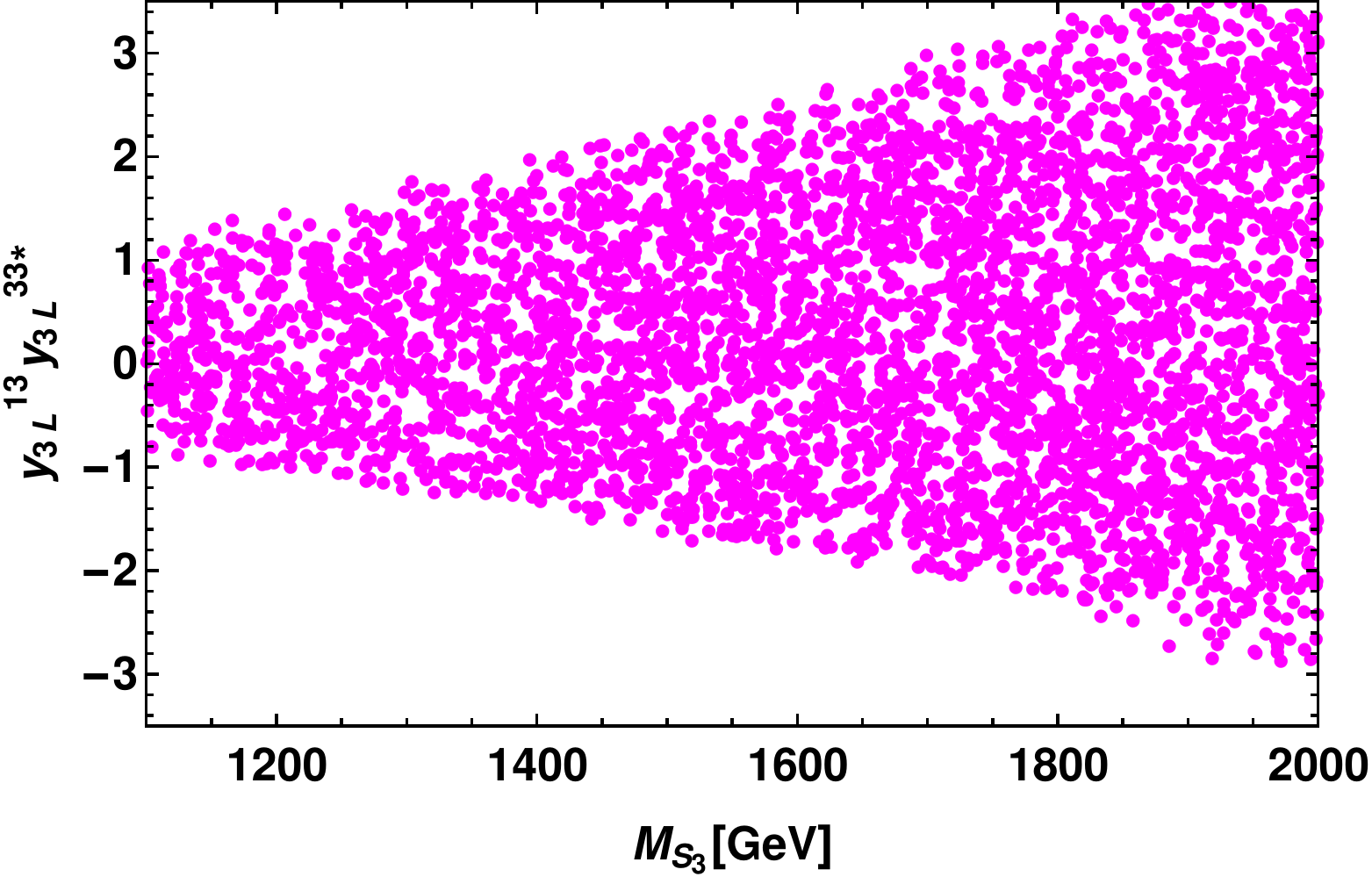}
\quad
\includegraphics[scale=0.45]{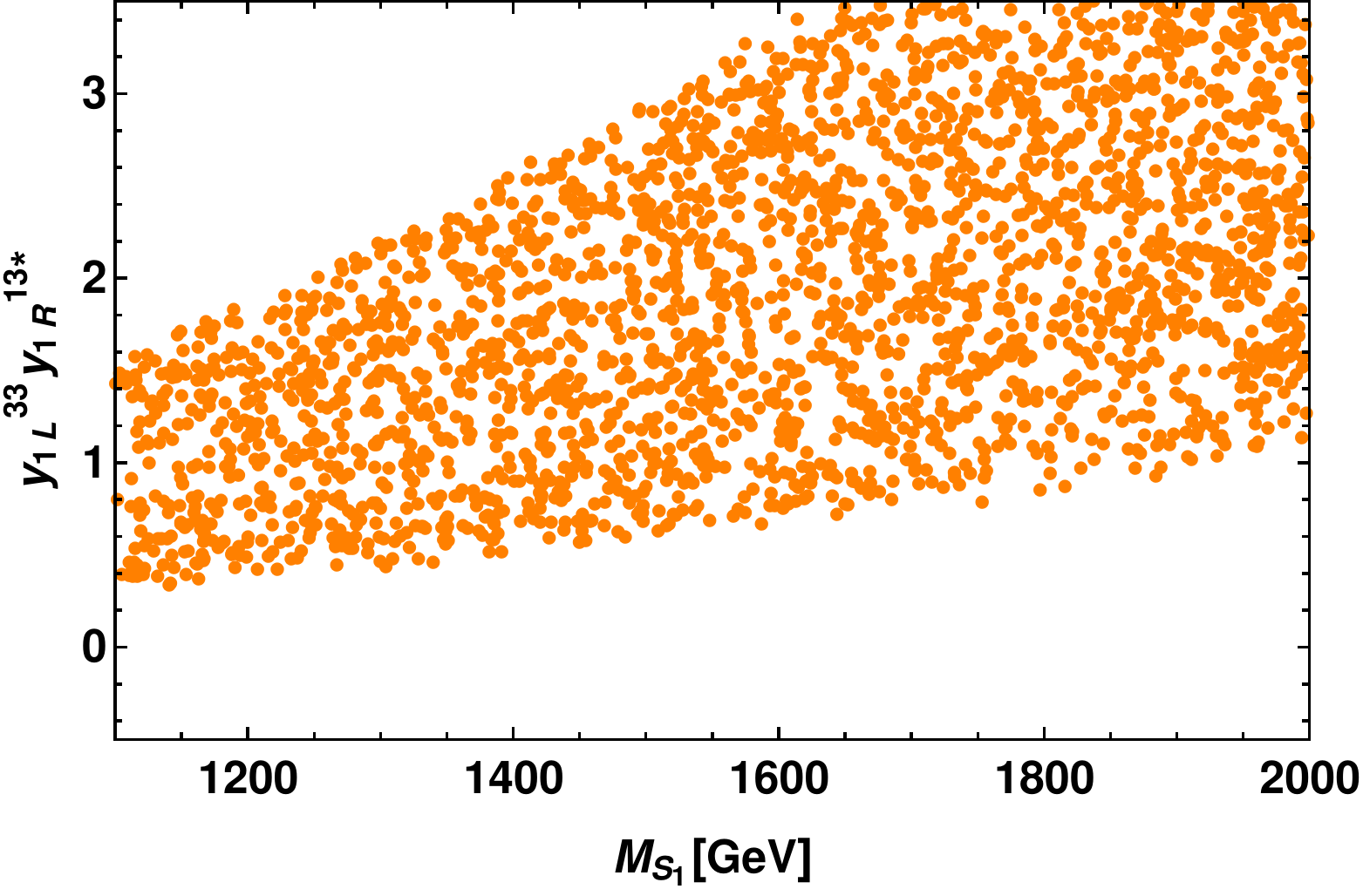}
\quad
\includegraphics[scale=0.45]{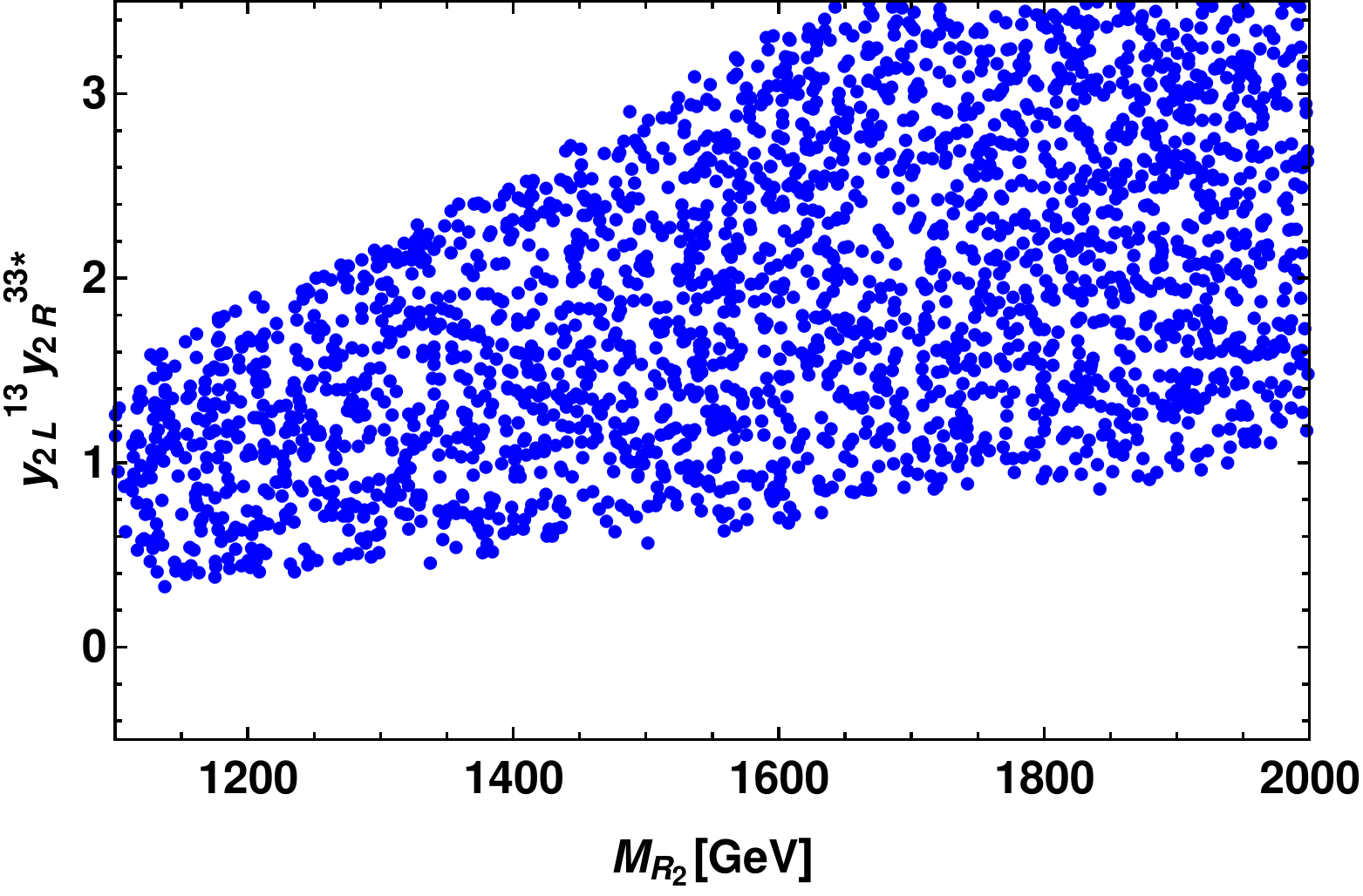}
\caption{Constraints on scalar leptoquark masses and  couplings (real) associated with $b \to u \tau \bar \nu_\tau$ decay processes.} \label{Fig:real-bulnu}
\end{figure}

\begin{figure}[htb]
\includegraphics[scale=0.45]{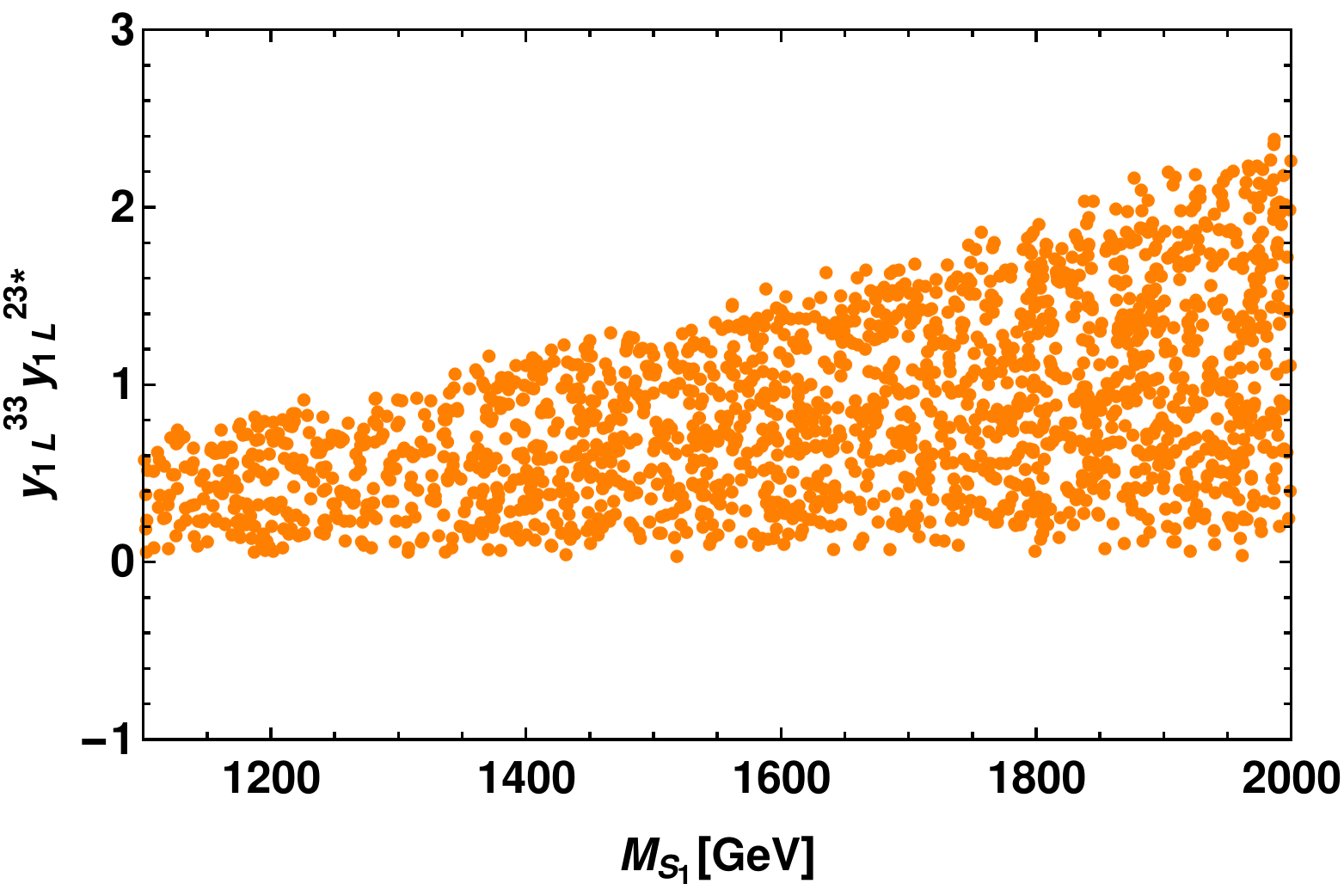}
\quad
\includegraphics[scale=0.45]{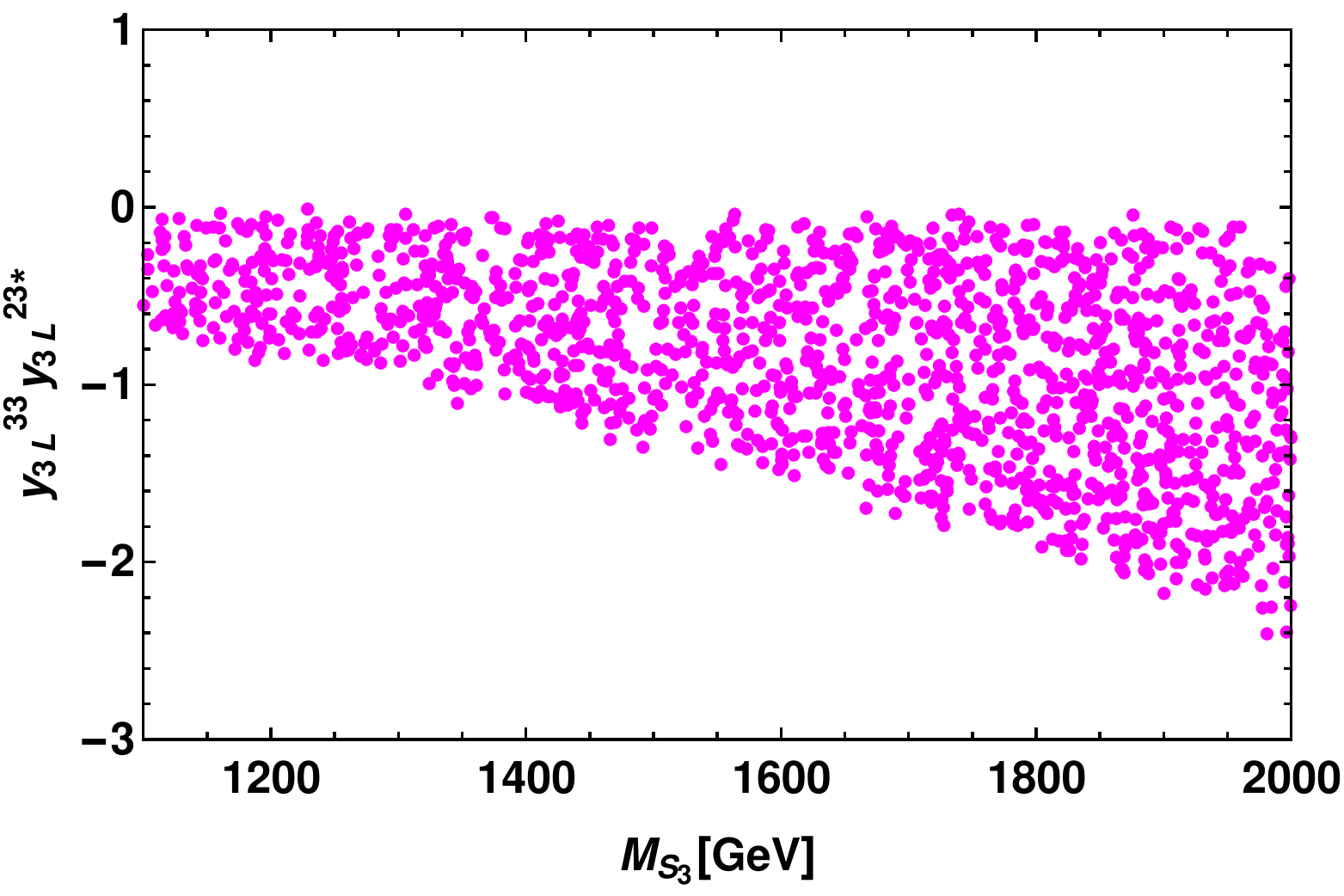}
\quad
\includegraphics[scale=0.45]{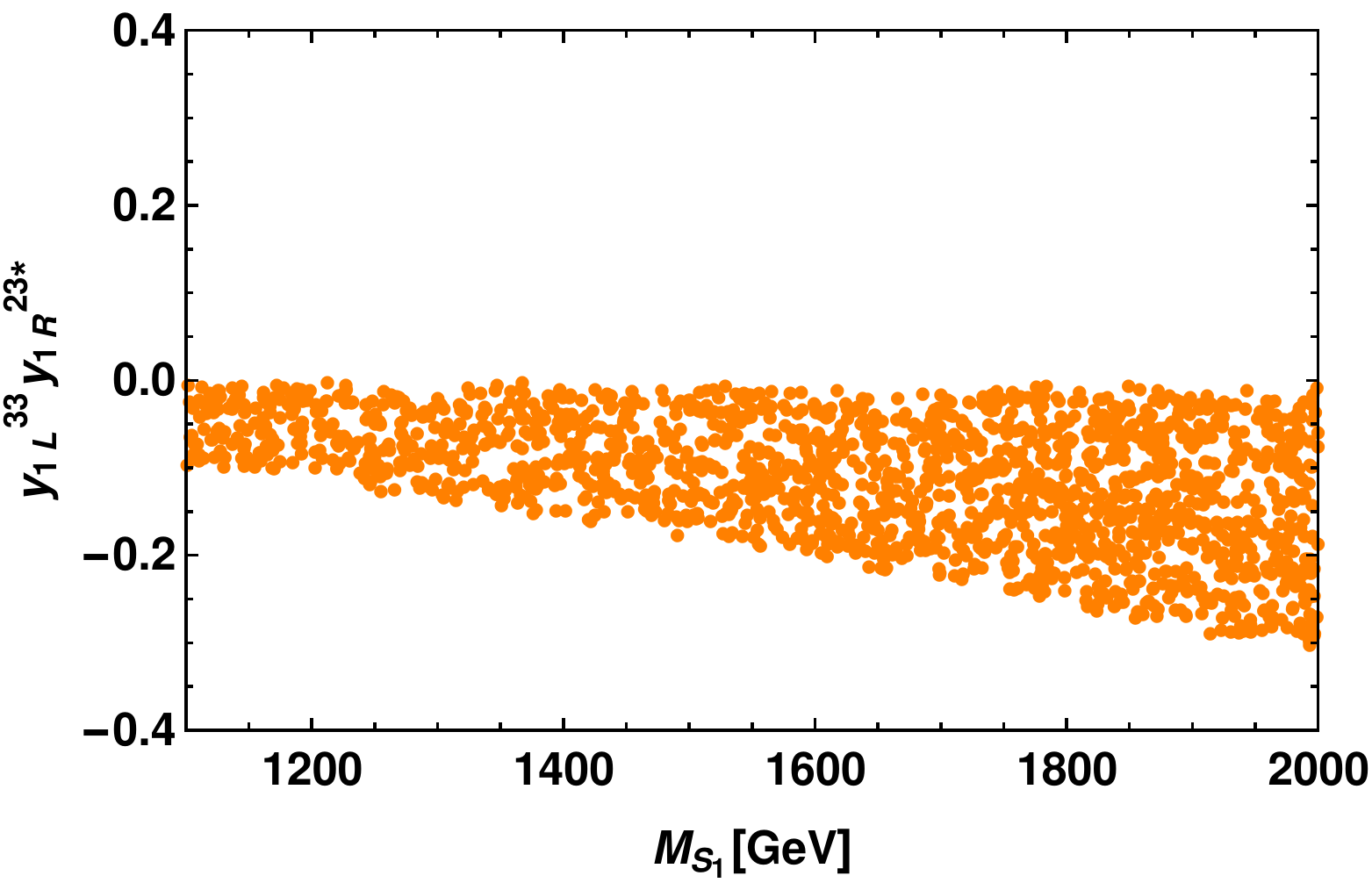}
\quad
\includegraphics[scale=0.45]{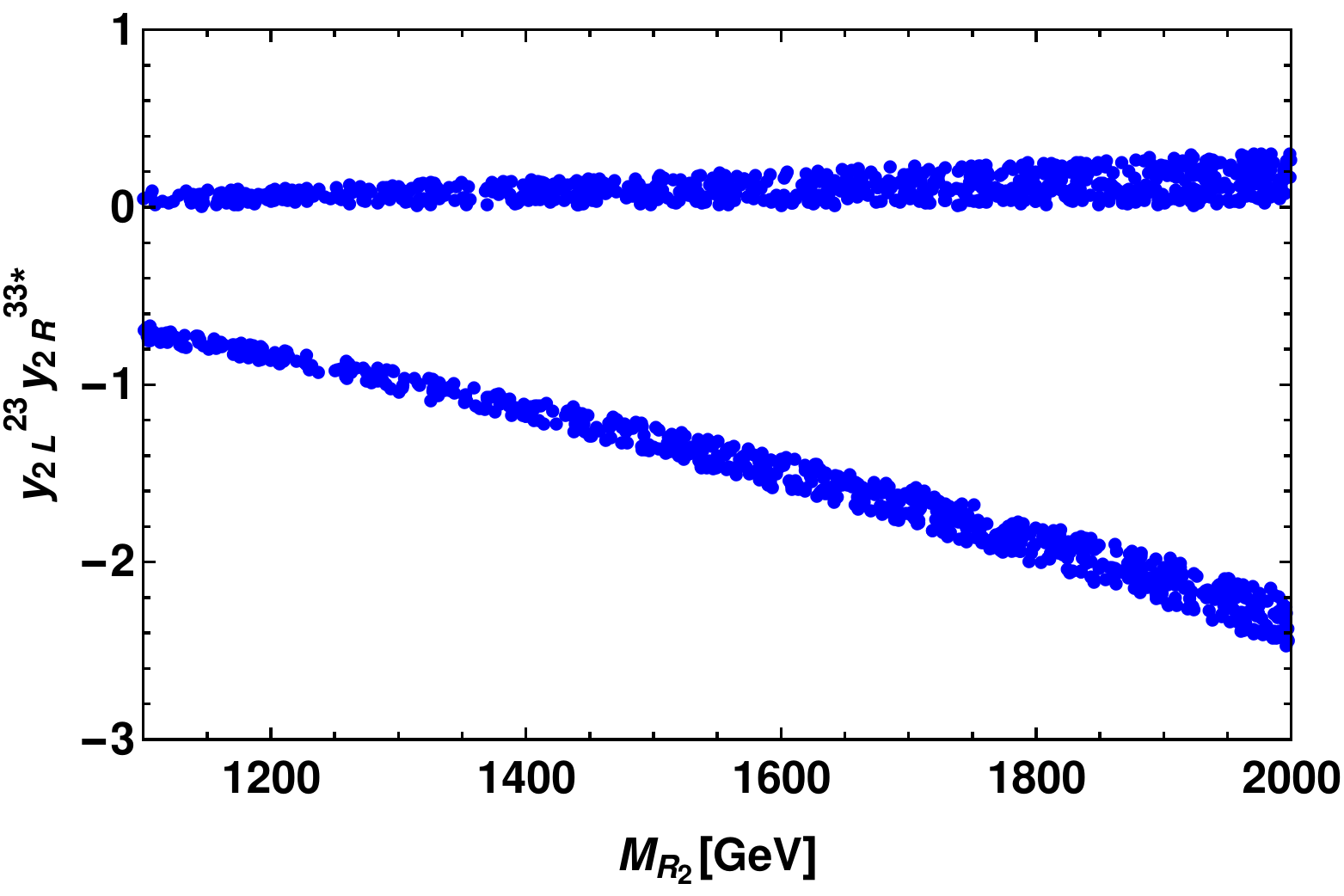}
\caption{Constraints on scalar leptoquark masses and  couplings (real) associated with $b \to c \tau \bar \nu_\tau$ decay processes.} \label{Fig:real-bclnu}
\end{figure}

Considering  the  leptoquarks couplings as real, the constrained plots of the $S_1$ (top-left panel) and  $S_3$ (top-right panel)  SLQ masses and their corresponding new couplings  related with $b \to u \tau\bar \nu_\tau$ process are  presented in Fig. \ref{Fig:real-bulnu}\,.  In the bottom-left (bottom-right) panel of Fig. \ref{Fig:real-bulnu}\,, we show the allowed space of scalar type couplings and $S_1~(R_2)$ SLQ masses. 
 For the case of $b \to c \tau \bar \nu_\tau$, the constraints on  $y_{1L}^{33} y_{1L}^{23^*}~(y_{3L}^{33} y_{3L}^{23^*})$  product of  couplings  of the $S_1 (S_3)$ leptoquarks  and their corresponding masses are depicted in the top-left (top-right) panel of Fig. \ref{Fig:real-bclnu}\,.  The bottom-left panel of Fig. \ref{Fig:real-bclnu} represents the allowed region of $y_{1L}^{33} y_{1R}^{23^*}$ couplings  and masses of $S_1$ leptoquarks and the corresponding plot for $R_2$ leptoquark is shown in the bottom-right panel.
  In   Table \ref{Tab:real}\,, we give the   allowed ranges of leptoquark masses and  real couplings obtained by imposing the extrema conditions. 
\begin{table}[htb]
\centering
\caption{Minimum and maximum values  of the leptoquark couplings (real) and masses.} \label{Tab:real}
\begin{tabular}{|c|c|c|c|c|}
\hline
Decay processes~&~Leptoquarks~&~Couplings~&~Real part~&~Mass of leptoquark\\
~&~~&~&~(Min, Max)~&~(Min, Max)\\
\hline
\hline
\multirow{4}{*}{} $b \to u \tau\bar \nu_\tau$~&~$S_1 $~&~$y_{1L}^{33}y_{1L}^{13^*}$~&~$(-3.5,3.0)$~&~$(1100,2000)$\\
&~~&~$y_{1L}^{33}y_{1R}^{13^*}$~&~$(0.3,3.5)$~&~$(1100,2000)$\\ \cline{2-5}
&~$S_3 $~&~$y_{3L}^{33}y_{3L}^{13^*}$~&~$(-3.0,3.5)$~&~$(1100,2000)$\\ 
\cline{2-5}
&~$R_2 $~&~$y_{2L}^{13}y_{2R}^{33^*}$~&~$(0.3,3.5)$~&~$(1100,2000)$\\

\hline
\multirow{4}{*}{} $b \to c \tau\bar \nu_\tau$~&~$S_1 $~&~$y_{1L}^{33}y_{1L}^{23^*}$~&~$(0.0,2.2)$~&~$(1100,2000)$~\\
&~~&~$y_{1L}^{33}y_{1R}^{23^*}$~&~$(-0.32,0.0)$~&~$(1100,2000)$~\\ \cline{2-5}
&~$S_3 $~&~$y_{3L}^{33}y_{3L}^{23^*}$~&~$(-2.2,0.0)$~&~$(1100,2000)$\\ \cline{2-5}
&~$R_2 $~&~$y_{2L}^{23}y_{2R}^{33^*}$~&~$(0.0,0.32)$~&~$(1100,2000)$\\
\hline
\end{tabular}
\end{table}

\subsubsection{Complex couplings}

\begin{figure}[htb]
\includegraphics[scale=0.45]{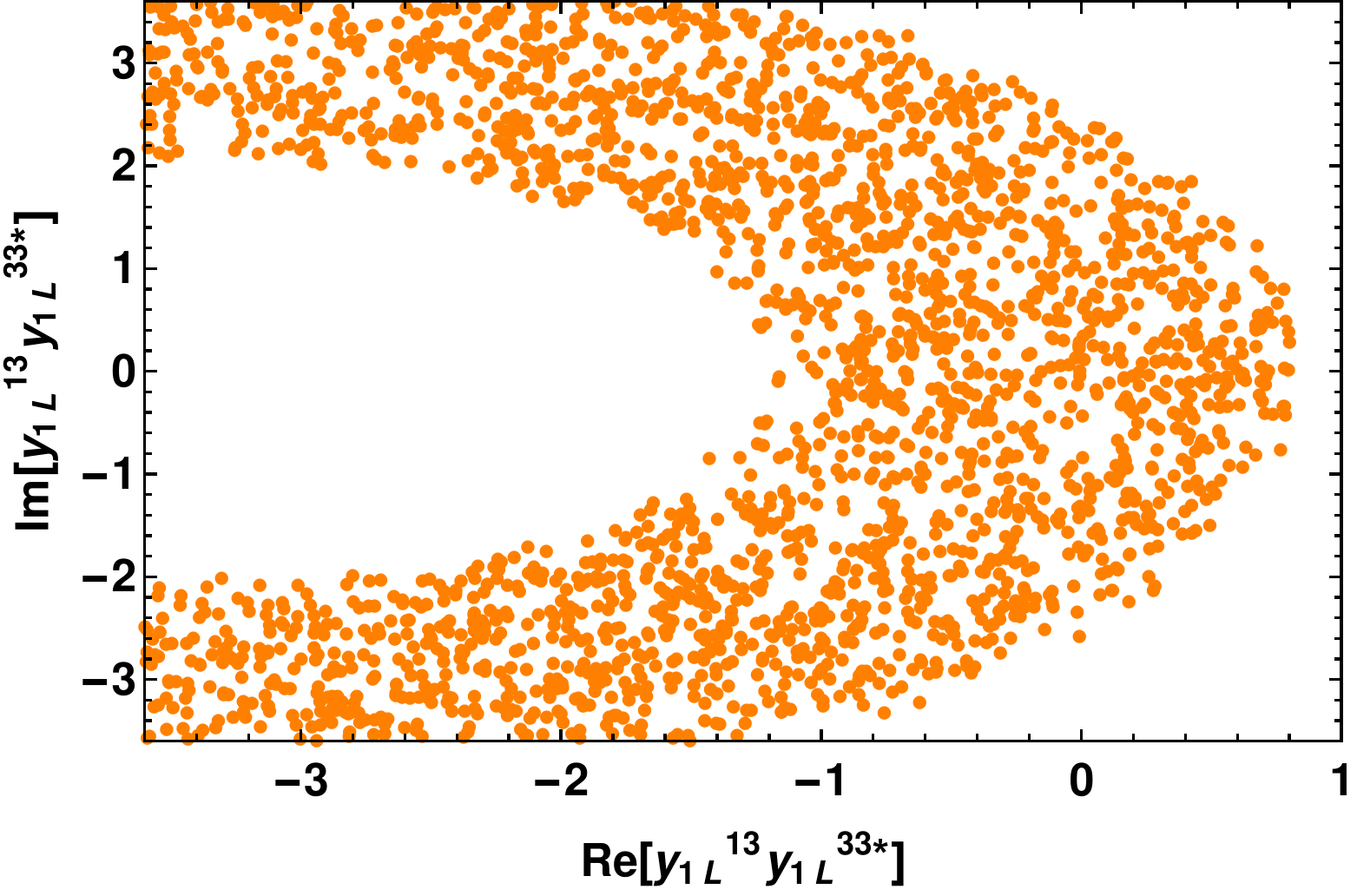}
\quad
\includegraphics[scale=0.45]{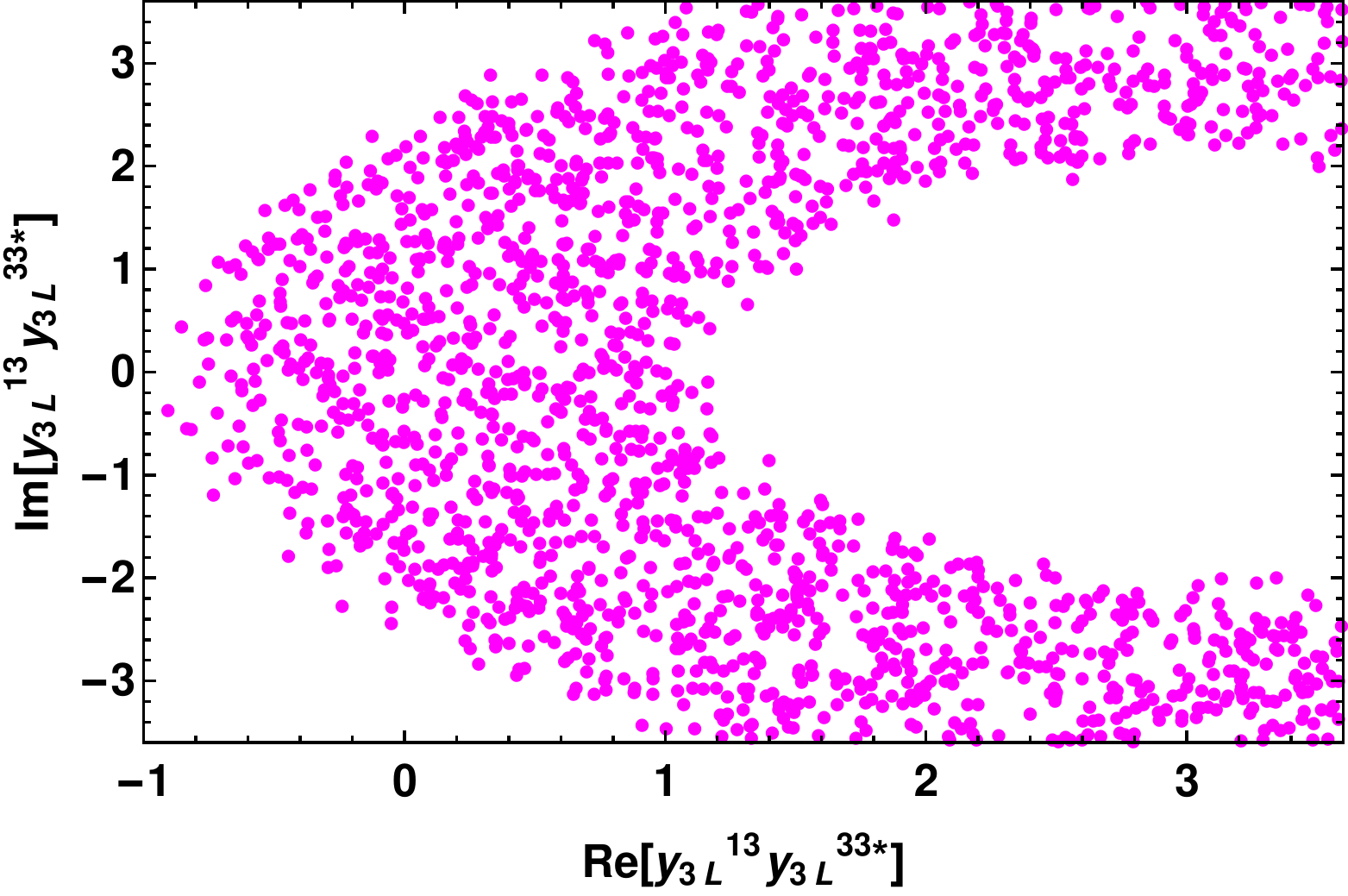}
\quad
\includegraphics[scale=0.45]{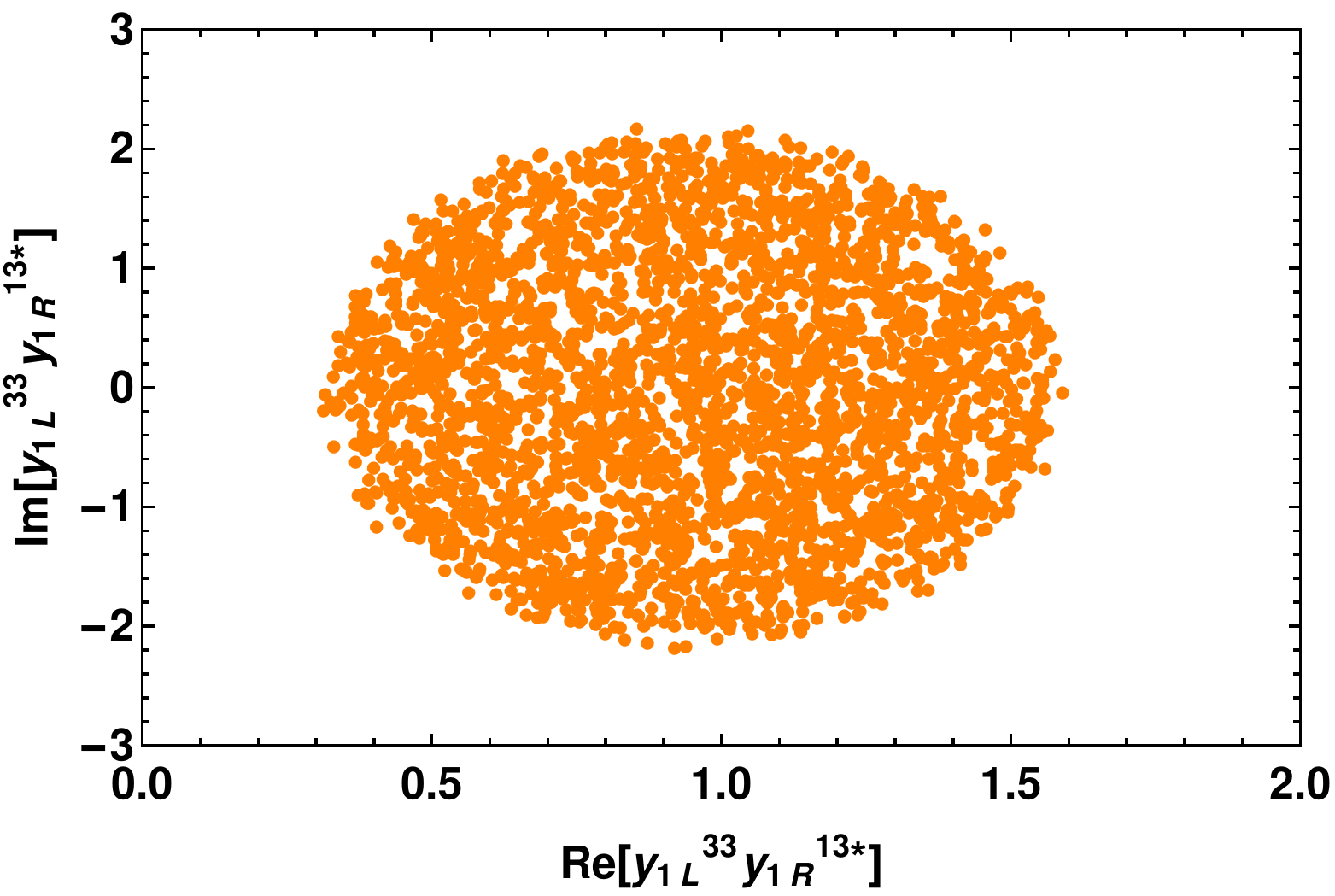}
\quad
\includegraphics[scale=0.45]{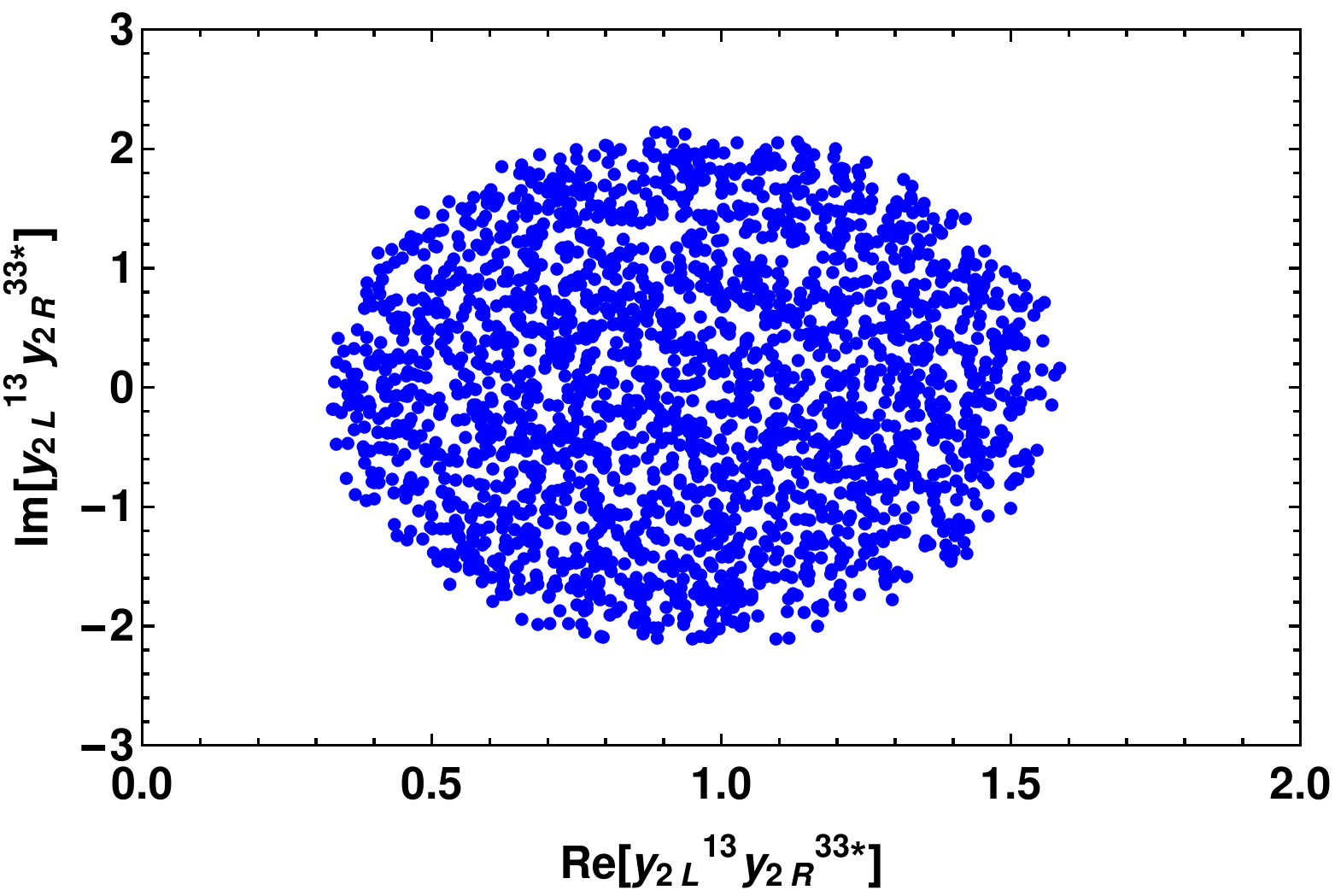}
\caption{Constraints on  real and imaginary part of the leptoquark couplings associated with $b \to u \tau \bar \nu_\tau$ decay processes.} \label{Fig:complex-bulnu}
\end{figure}

\begin{figure}[htb]
\includegraphics[scale=0.45]{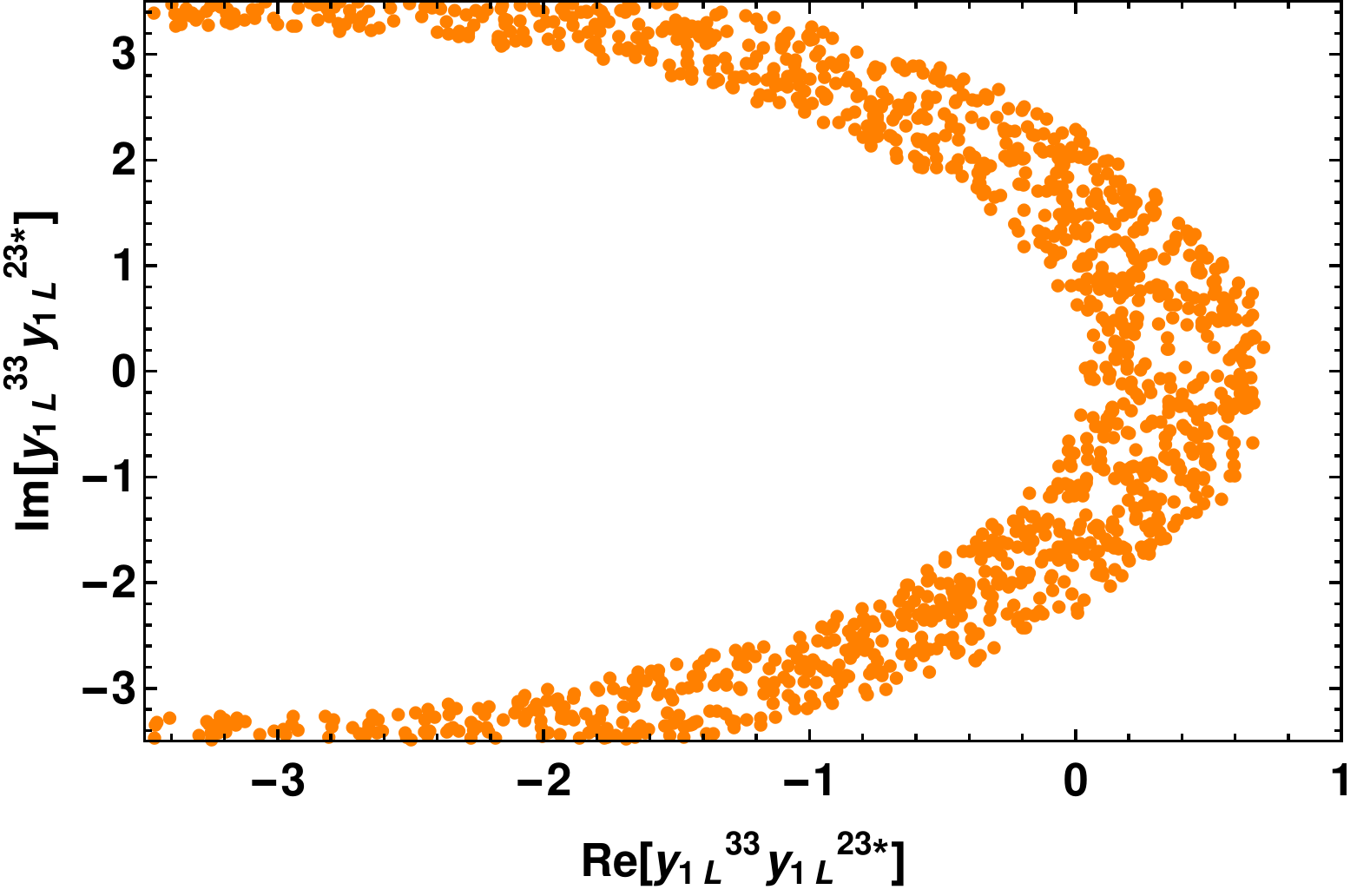}
\quad
\includegraphics[scale=0.45]{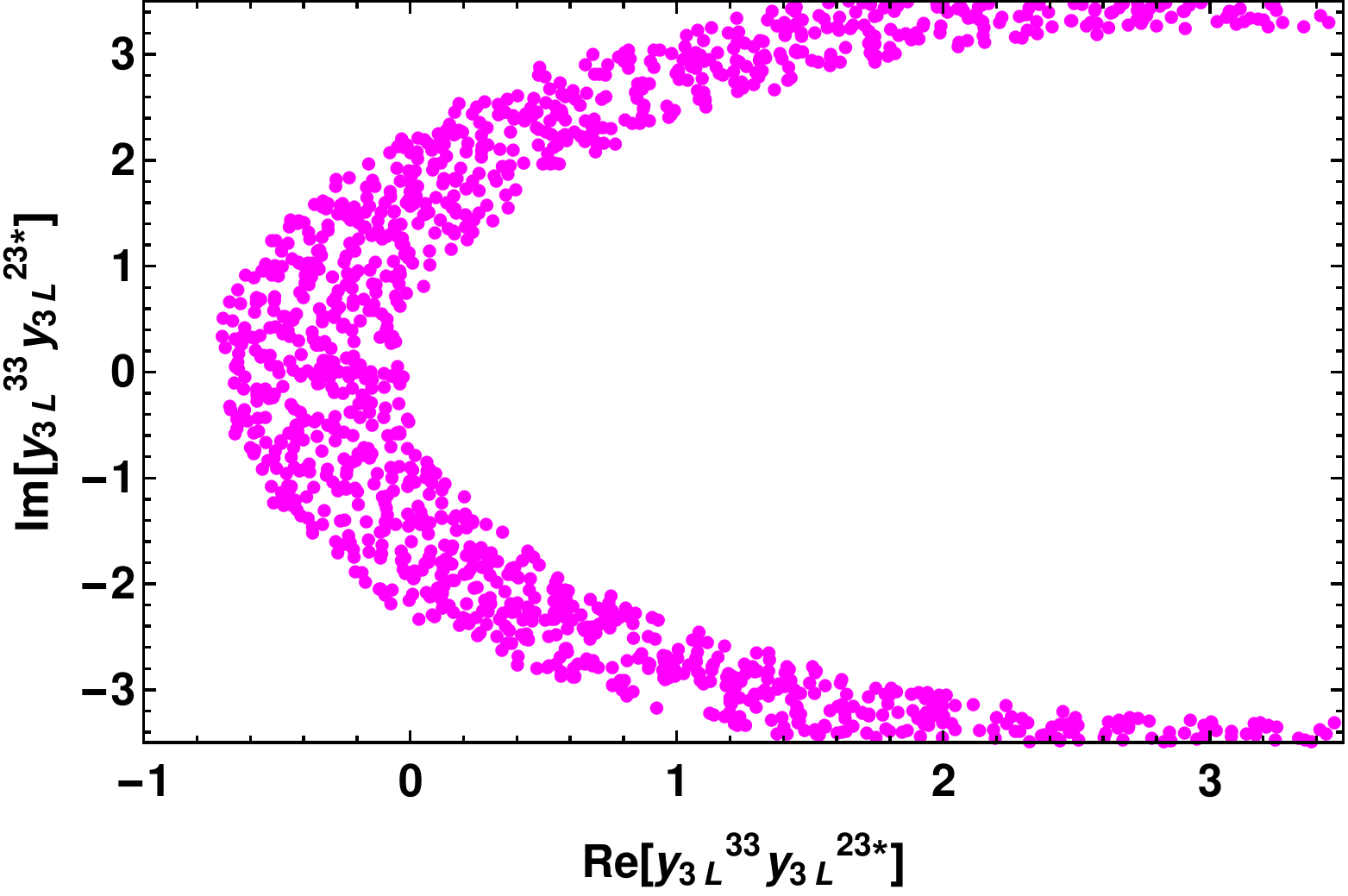}
\quad
\includegraphics[scale=0.45]{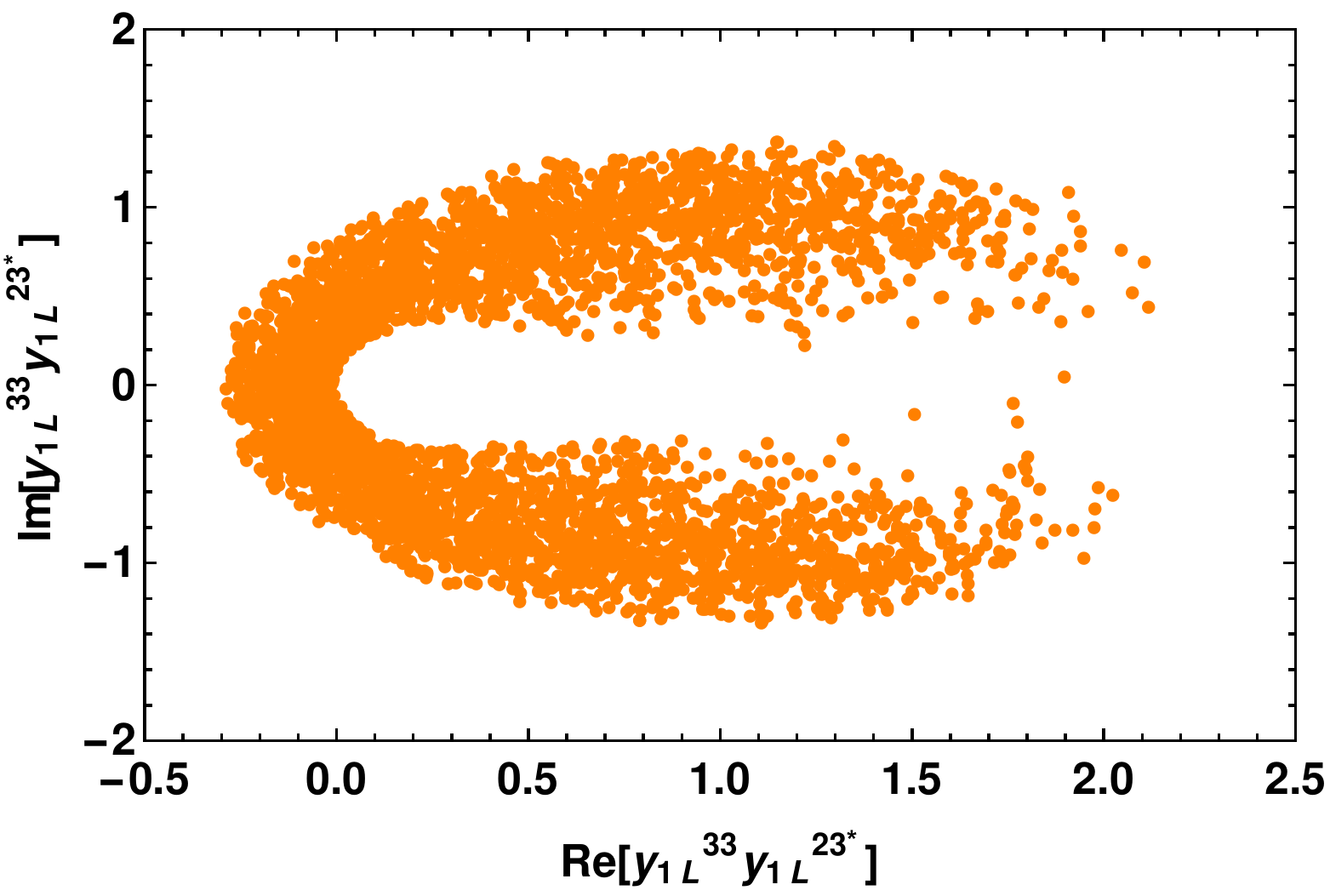}
\quad
\includegraphics[scale=0.45]{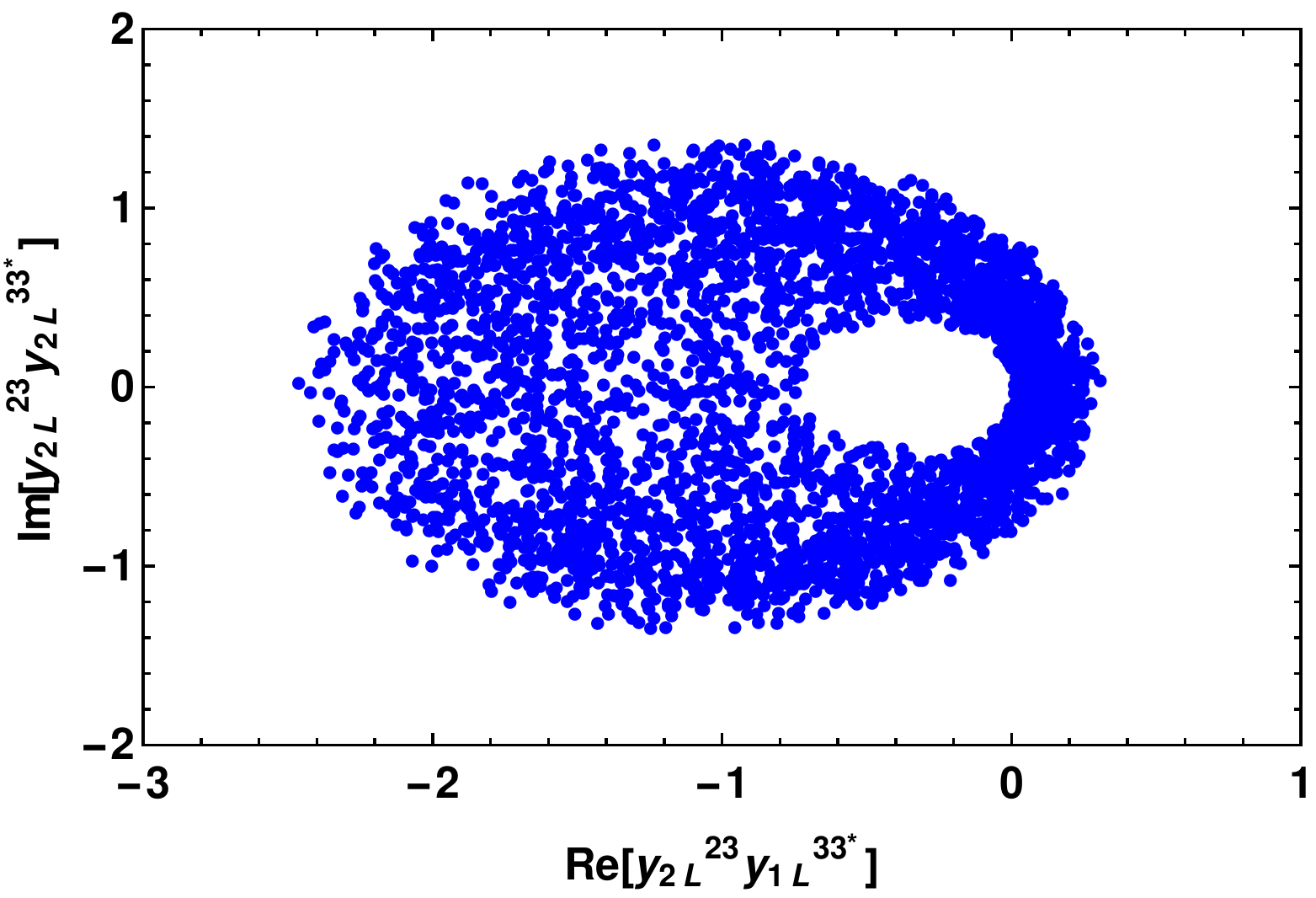}
\caption{Constraints on  real and imaginary part of the leptoquark couplings associated with $b \to c \tau \bar \nu_\tau$ decay processes.} \label{Fig:complex-bclnu}
\end{figure}

Considering the leptoquarks couplings as complex, the constraint on the real and imaginary parts of $y_{1L}^{33}y_{1L}^{13^*}~(y_{1L}^{33}y_{1R}^{13^*})$ couplings of $S_1$ leptoquark is depicted in the top-left (bottom-left) panel of Fig. \ref{Fig:complex-bulnu}\,. The allowed space of $y_{3L}^{33}y_{3L}^{13^*}$ and $y_{2L}^{13}y_{2L}^{33^*}$ couplings obtained from the Br($B_u \to \tau \nu_l$), Br($B \to \pi \tau \nu_l$) and $R_\pi^l$ experimental data are shown in the top-right and bottom-right panels of Fig. \ref{Fig:complex-bulnu}\,. 
Imposing the experimental limit on Br($B_c \to \tau \nu_\tau$), $R_{D^{(*)}}$ and $R_{J/\psi}$ observables, the constraints on real and imaginary parts of the vectorial type couplings of $S_1~(S_3)$ leptoquark are presented in the top-left (top-right) panel of Fig. \ref{Fig:complex-bclnu}\,. The corresponding constrained plots for the scalar  couplings of $S_1$ and $R_2$ leptoquarks are given in the bottom-left and bottom-right panel of this figure. 
 The obtained allowed new parameter space is presented in Table \ref{Tab:complex}\,.
\begin{table}[htb]
\centering
\caption{Minimum and maximum values of the real and imaginary parts of the leptoquarks couplings.} \label{Tab:complex}
\begin{tabular}{|c|c|c|c|c|}
\hline
Decay processes~&~Leptoquarks~&~Couplings~&~Real part~&~Imaginary part\\
~&~~&~&~(Min, Max)~&~(Min, Max)\\
\hline
\hline
\multirow{4}{*}{} $b \to u \tau\bar \nu_\tau$~&~$S_1 $~&~$y_{1L}^{33}y_{1L}^{13^*}$~&~$(-1.4,0.6)$~&~$(-3.5,3.5)$\\
&~~&~$y_{1L}^{33}y_{1R}^{13^*}$~&~$(0.3,1.6)$~&~$(-2.0,2.0)$\\ \cline{2-5}
&~$S_3 $~&~$y_{3L}^{33}y_{3L}^{13^*}$~&~$(-0.6,1.4)$~&~$(-3.5,3.5)$\\ \cline{2-5}
&~$R_2 $~&~$y_{2L}^{13}y_{2R}^{33^*}$~&~$(0.3,1.6)$~&~$(-2.0,2.0)$\\

\hline
\multirow{4}{*}{} $b \to c \tau\bar \nu_\tau$~&~$S_1 $~&~$y_{1L}^{33}y_{1L}^{23^*}$~&~$(-0.2,0.6)$~&~$(-3.5,3.5)$\\
&~~&~$y_{1L}^{33}y_{1R}^{23^*}$~&~$(-0.35,0.1)$~&~$(-1.4,1.4)$\\ \cline{2-5}
&~$S_3 $~&~$y_{3L}^{33}y_{3L}^{23^*}$~&~$(-0.6,0.2)$~&~$(-3.5,3.5)$\\ \cline{2-5}
&~$R_2 $~&~$y_{2L}^{23}y_{2R}^{33^*}$~&~$(-2.4,-0.6)$~&~$(-1.4,1.4)$\\
\hline
\end{tabular}
\end{table}

\section{Numerical analysis}
In this section, we perform the numerical analysis of the branching ratios and physical observables of $B_s \to (K^{(*)}, D_s^{(*)}) \tau \bar \nu_\tau$ processes. For numerical estimation, we use the particle masses, life time of $B_s$ meson and the CKM matrix elements from \cite{Patrignani:2016xqp}\,.   The  predictions for the various observables require sufficient knowledge of the associated hadronic form factors.  For  $\bar B_s \to K^+ l^- \bar \nu_l$ decay processes, we consider  the perturbative QCD (PQCD) calculation \cite{Wang:2012ab, Meissner:2013pba}  based on the $k_T$ factorization \cite{Keum:2000ph, Keum:2000wi, Lu:2000em, Lu:2000hj} at next-to-leading order (NLO) in $\alpha_s$ \cite{Li:2012nk}, which gives
\bea
F_1^{B_s\to K}(q^2) &=&  {F_1^{B_s\to K}(0)} \left(\frac{1}{(1-q^2/M_{B_s}^2)} + \frac{a_1 q^2/M_{B_s}^2}{(1-q^2/M_{B_s}^2)(1-b_1 q^2/M_{B_s}^2)}\right),\nn \\
F_0^{B_s\to K}(q^2) &=& \frac{F_0^{B_s\to K}(0)} {(1-a_0 q^2/M_{B_s}^2+ b_0 q^4/M_{B_s}^4)}~.
\eea
The values of the parameters $a_{0,1}$, $b_{0,1}$ and $F_{0,1}^{B_s \to K}$ can be found in the Ref. \cite{Wang:2012ab}. 
The $q^2$ dependence of the form factors $V, A_{0,1,2},T_{1,2,3}$ associated with $B_s \to K^* \tau \bar \nu_\tau$ decay modes can be parametrized as \cite{Ball:2004rg}
 \begin{eqnarray}
 f_1^{B_s\to K^*}\left(q^2\right)&=& \frac{r_1}{1-q^2/m^2_R} + \frac{r_2}{1-q^2/m^2_{fit}}\;,~~~~~~~f_1=V, A_0, T_1,\nn\\
  f_2^{B_s\to K^*}\left(q^2\right) &=&\frac{r_2}{1-q^2/m^2_{fit}}\;,\hspace{3.5cm} f_2=A_1, T_2,\nn\\
 f_3^{B_s\to K^*}\left(q^2\right)& =& \frac{r_1}{1-q^2/m^2_{fit}} + \frac{r_2}{\left(1-q^2/m^2_{fit}\right)^2},~~~ f_3=A_2, \tilde{T_3}\;,
 \end{eqnarray}
where $T_3$ is related to $\tilde{T}_3$ and the values of the parameters involved in the calculation of form factors are taken from Ref. \cite{Ball:2004rg}. 
The $B_s \to D_s^{(*)} \tau \bar \nu_\tau$ form factors, computed by using the PQCD approach \cite{Fan:2013kqa} are used in this analysis, which can be parametrized as 
\bea
F^{B_s\to D_s^{(*)}}(q^2) &=& \frac{F^{B_s\to D_s^{(*)}}(0)} {(1-a q^2/M_{B_s}^2+ b q^4/M_{B_s}^4)}\,,
\eea
where $F^{B_s\to D_s^{(*)}}$ stand for the form factors $F_{0,+}, V, A_{0,1,2}$. The fitting values of the  $a$ and $b$ parameters can be found in \cite{Fan:2013kqa}. Since there is no PQCD results on the  form factors associated with tensor operator, we show our results for $B_s \to D_s^{(*)} l \bar \nu_l$ process with vanishing tensor form factors. 

Now the stage is ready with detailed expressions of physical observables,   required input parameters and  the constrained new Wilson coefficients for complete numerical anaysis.
Using these input values, the predicted  branching ratios $B_s \to (K^{(*)}, D_s^{(*)}) \mu \bar \nu_l$ processes in the SM are given by
\bea
&&{\rm Br}(B_s \to K \mu \bar \nu_\mu)^{\rm SM}= (1.044\pm 0.084)\times 10^{-4},\\
&&{\rm Br}(B_s \to K^* \mu \bar \nu_\mu)^{\rm SM}=(3.43\pm 0.275)\times 10^{-4} ,\\
&&{\rm Br}(B_s \to D_s \mu \bar \nu_\mu)^{\rm SM}= (2.17\pm 0.174)\times 10^{-2},\\
&&{\rm Br}(B_s \to D_s^* \mu \bar \nu_\mu)^{\rm SM}= (4.82\pm 0.386)\times 10^{-2}.
\eea
 In the following subsection, we discuss the impact of individual scalar leptoquarks on the branchings ratios and the optimized physical observables of semileptonic $B_s$ decay modes.   
\subsection{$S_1$ scalar leptoquark}
\begin{figure}[htb]
\centering
\includegraphics[scale=0.5]{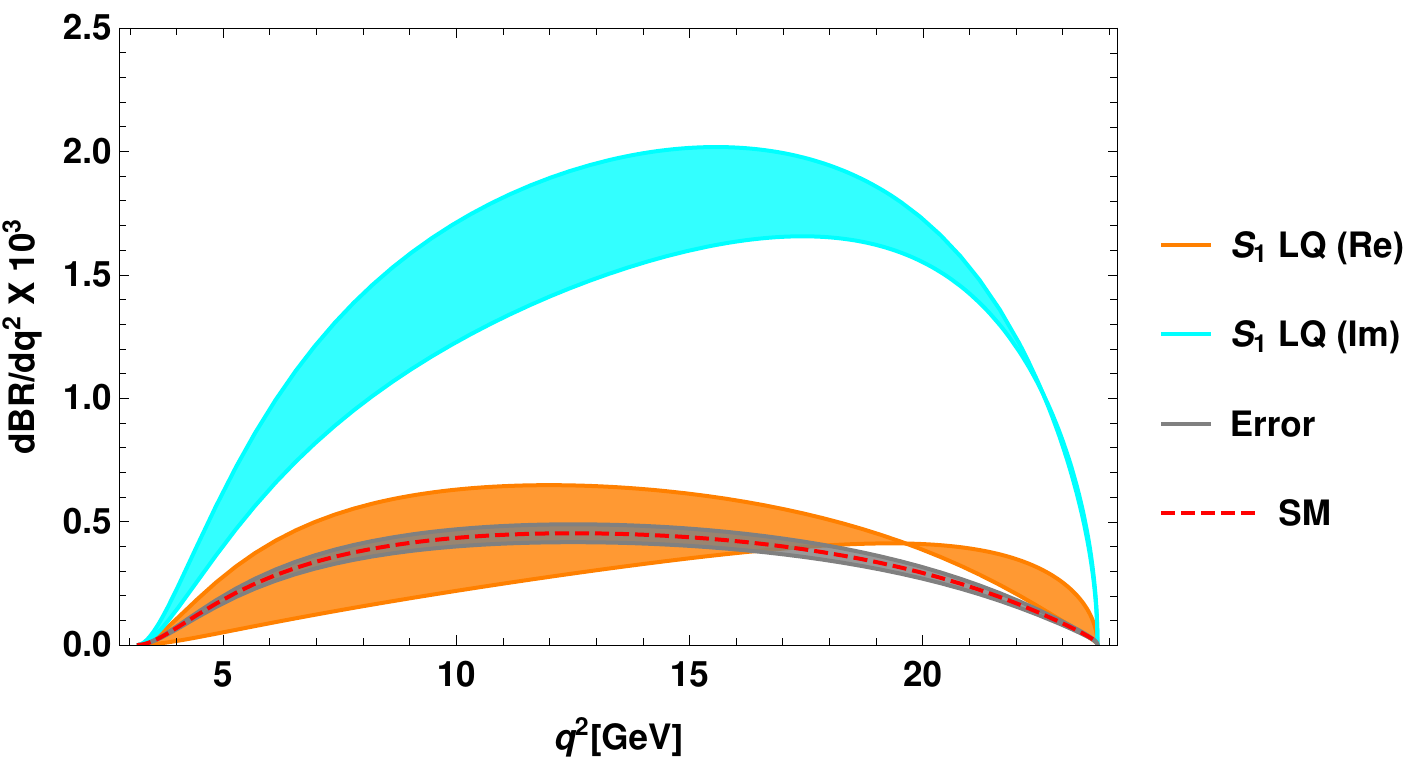}
\quad
\includegraphics[scale=0.5]{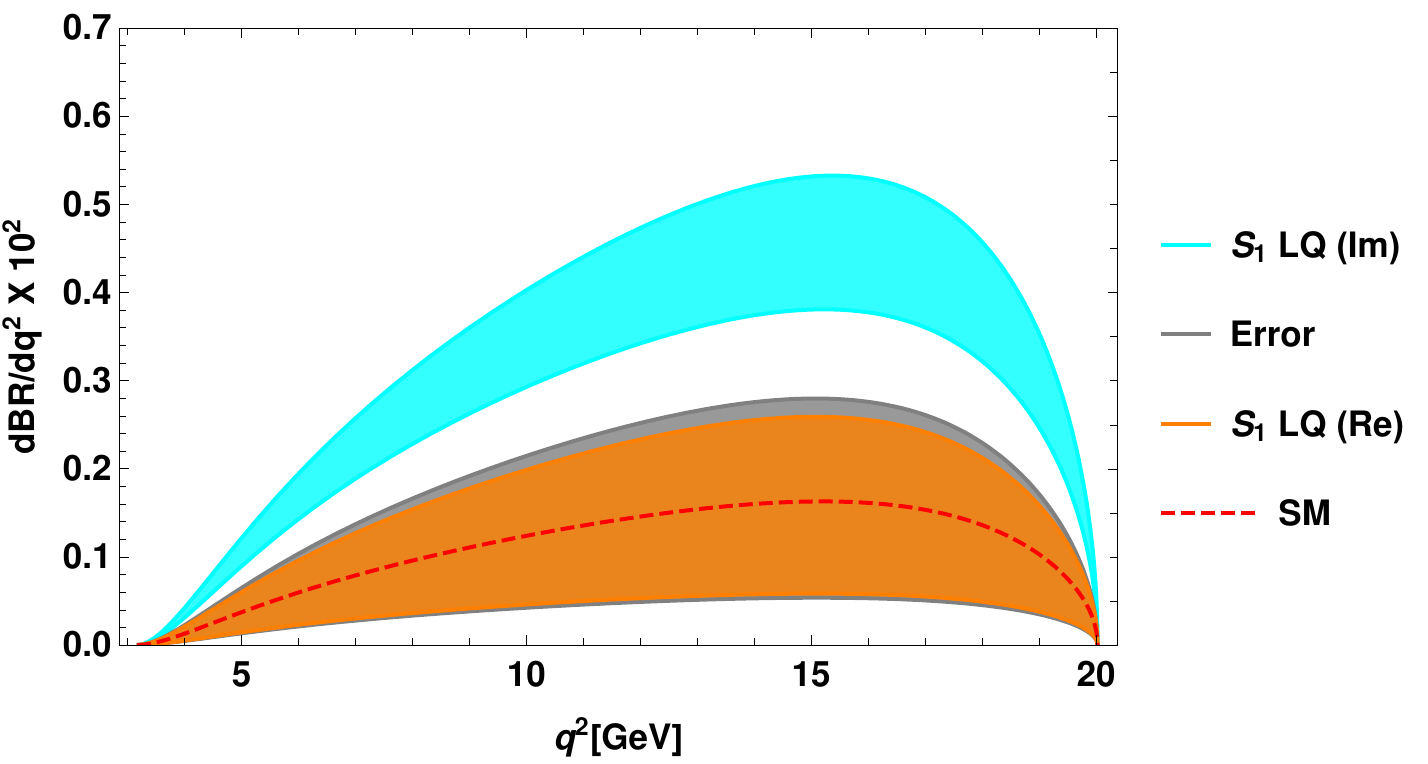}\\
\includegraphics[scale=0.5]{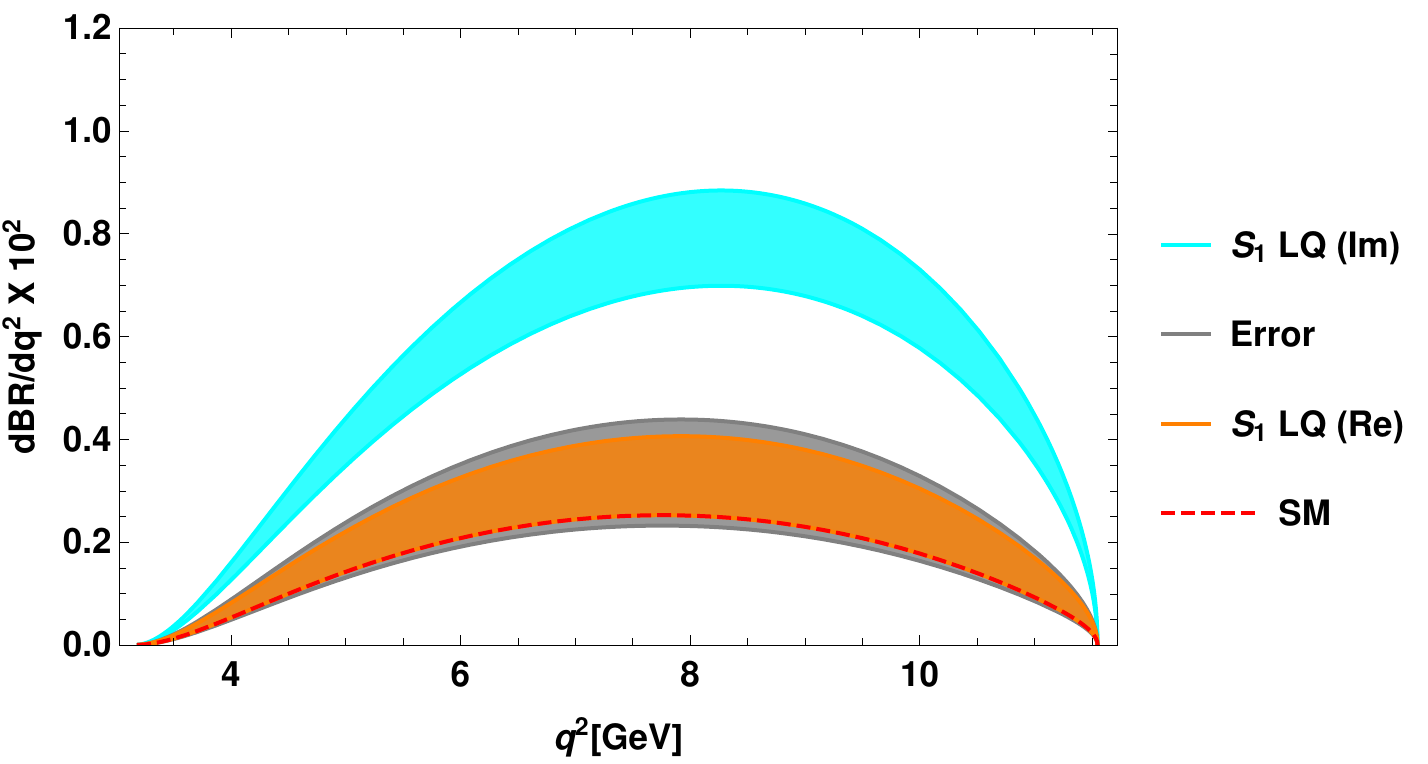}
\quad
\includegraphics[scale=0.5]{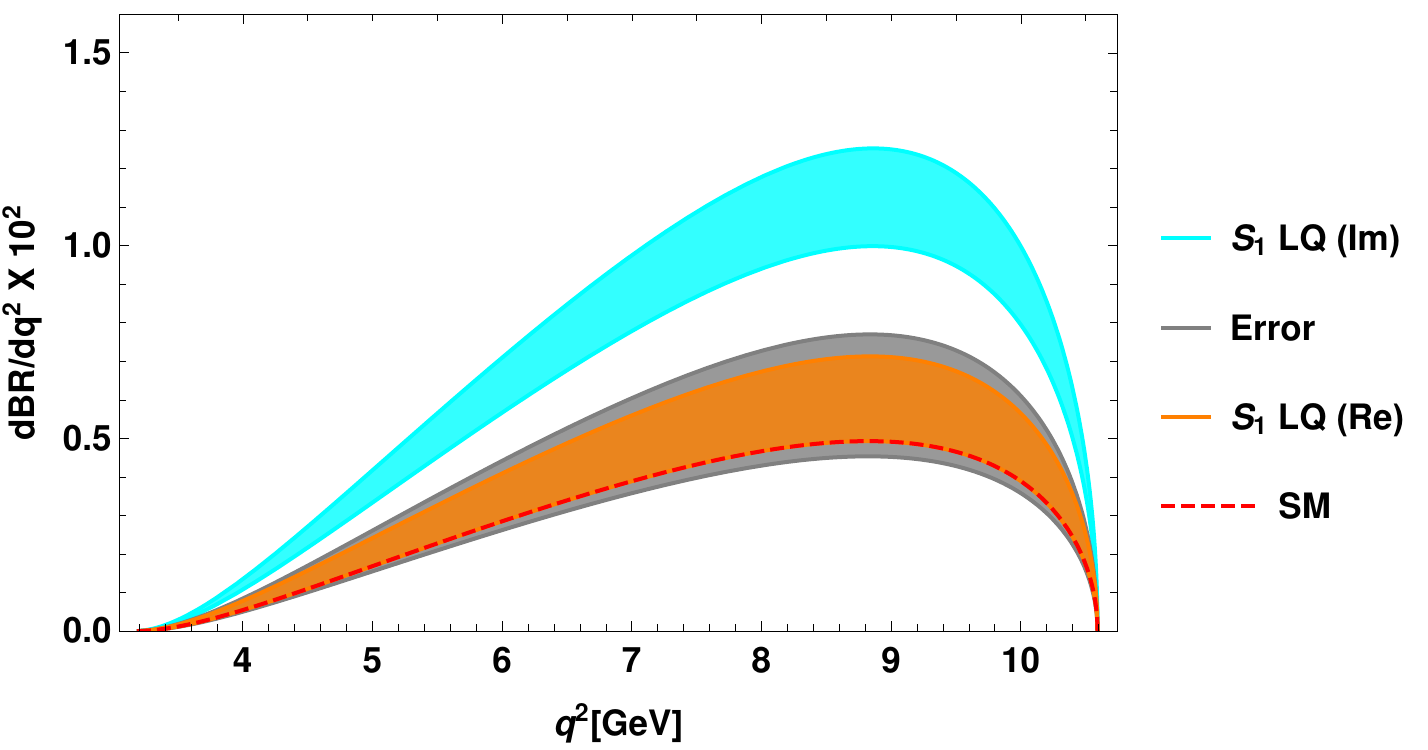}\\
\caption{The variation of branching ratios of $\bar B_s \to K^+l^- \bar \nu_\tau $ (top-left panel), $\bar B_s \to K^{* +} l^- \bar \nu_\tau $ (top-right panel), $\bar B_s \to D_s^+l^- \bar \nu_\tau$ (bottom-left panel) and $\bar B_s \to D_s^{*+}l^- \bar \nu_\tau$ (bottom-right panel) processes with respect to $q^2$ in the  $S_1$ scalar leptoquark model. Here cyan bands are due to the contribution from $S_1$ leptoquark with coupling as complex. The orange  bands stand for the allowed regions of real leptoquark coulings. The red dashed lines represent the standard model contributions with their corresponding theoretical uncertainties are shown in gray bands. } \label{S1-Br}
\end{figure}

This subsection is dedicated to the analysis of $B_s \to (K^{(*)}, D_s^{(*)}) \tau \bar \nu_\tau$  physical observables by using the singlet $S_1(\bar 3,1,1/3)$ scalar LQ. This LQ will contribute additional $V_L, ~S_R$ and $T_L$ coefficients to the SM. The constraint on the the  couplings (real and complex) and masses associated with this leptoquark are obtained by using the Br($B_{u,c} \to \tau \nu_\tau$), Br($B \to \pi \tau \tau$), $R_\pi^l$, $R_{D^{(*)}}$ and $R_{J/\psi}$ observables, as discussed in section III. Using the allowed space of real and complex couplings and the leptoquark mass  from Table  \ref{Tab:real} and \ref{Tab:complex}\,, we show the variation of the branching ratios of $B_s \to K \tau \bar \nu_\tau$ (top-left panel), $B_s \to K^* \tau \bar \nu_\tau$ (top-right panel), $B_s \to D_s \tau \bar \nu_\tau$ (bottom-left panel) and $B_s \to D_s^* \tau \bar \nu_\tau$ (bottom-right panel) processes with respect to $q^2$  in Fig. \ref{S1-Br}\,.  Here the cyan bands represent the contributions coming due to the $S_1$ SLQ exchange, where coupling is  complex.  The orange  bands are due to the SLQ contributions with real couplings. 
The red dashed lines present the central values of standard model and their  corresponding theoretical uncertainties, arising due to the uncertainties associated with hardonic form factors and CKM matrix elements, are shown in gray color. We found that the branching ratios of all these processes deviate significantly from their corresponding SM predictions due to the additional contribution from  the complex couling of $S_1$ scalar leptoquark.  Though the  region of real coupling case provide deviation from the SM results, these are comparatively less than the  case of complex coupling. The numerical values of the braching ratios of these decay modes for the SM and for all the cases of $S_1$ leptoquark couplings  are presented  in the Table \ref{Tab:S1}\,.

\begin{figure}[htb] 
\centering
\includegraphics[scale=0.5]{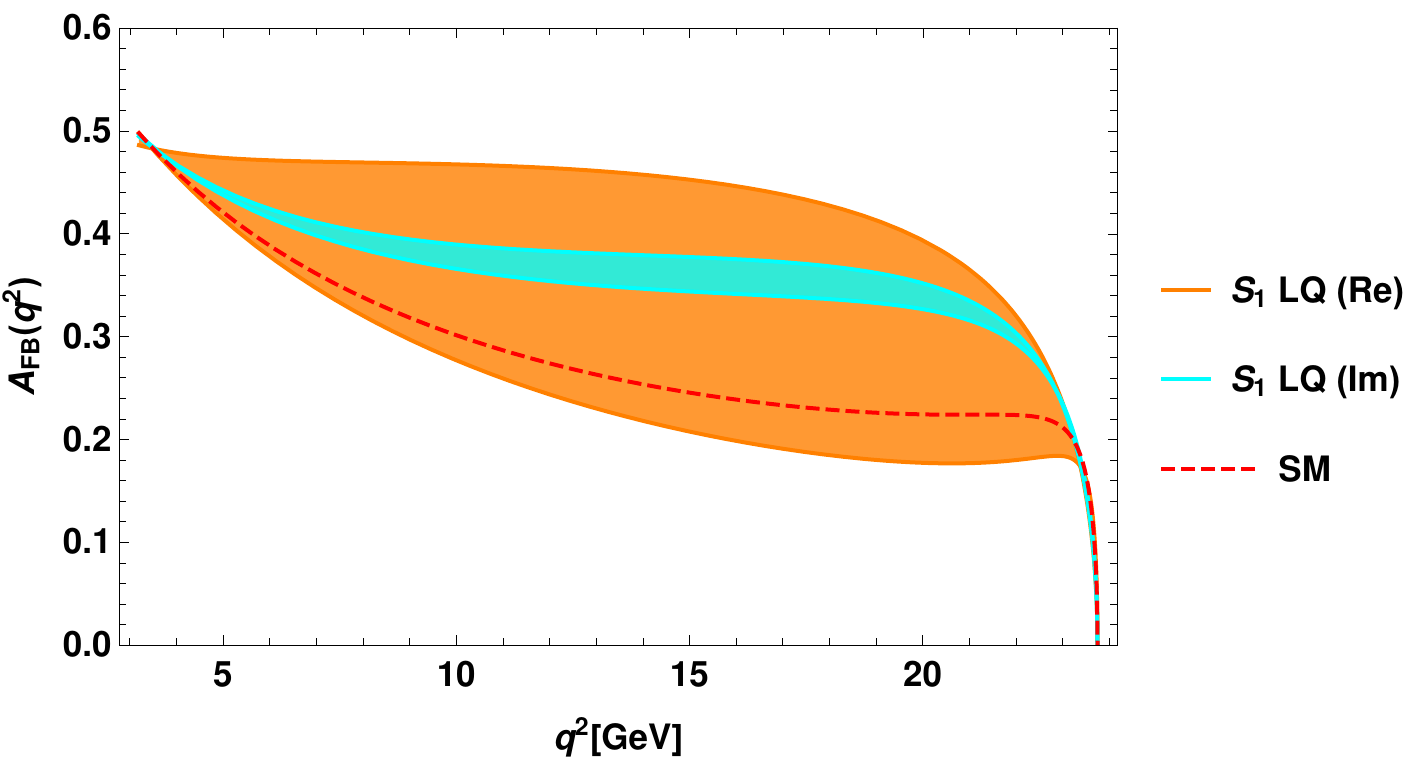}
\quad
\includegraphics[scale=0.5]{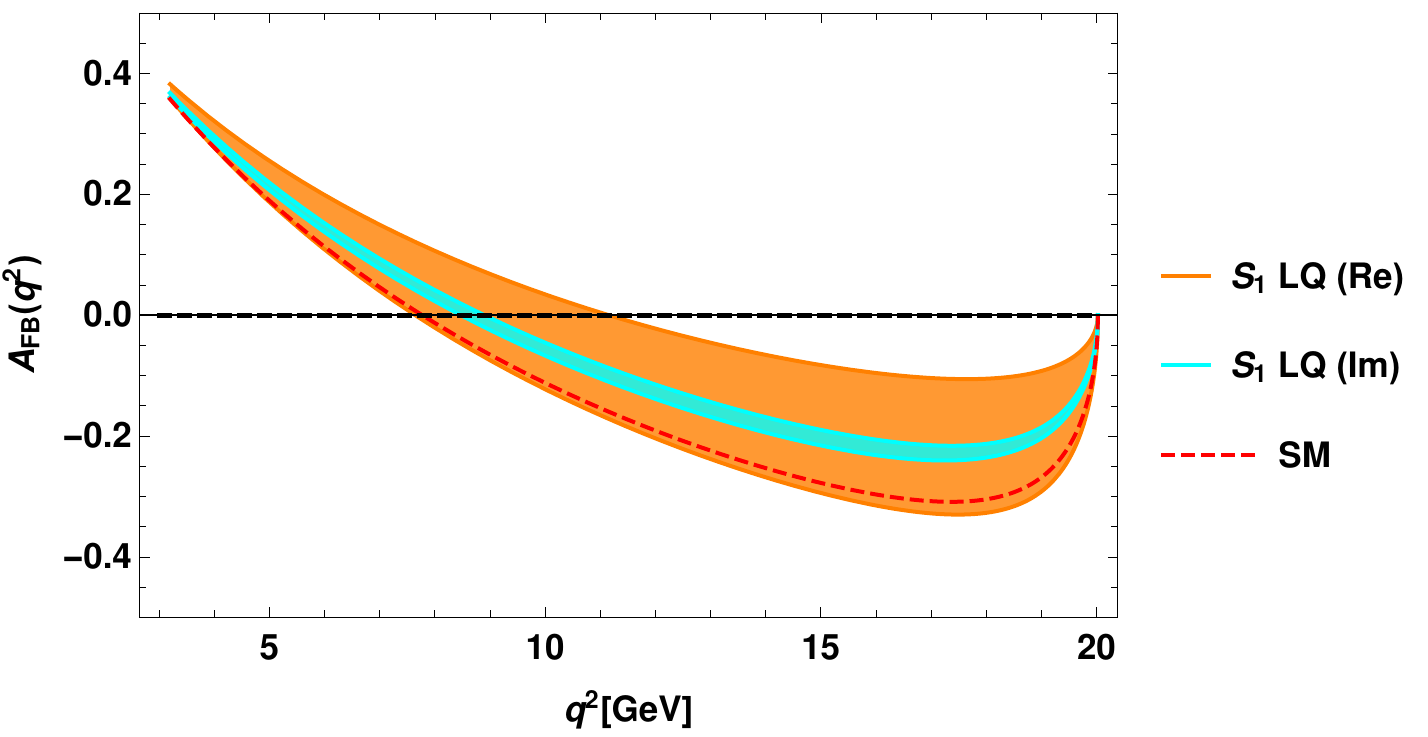}\\
\includegraphics[scale=0.5]{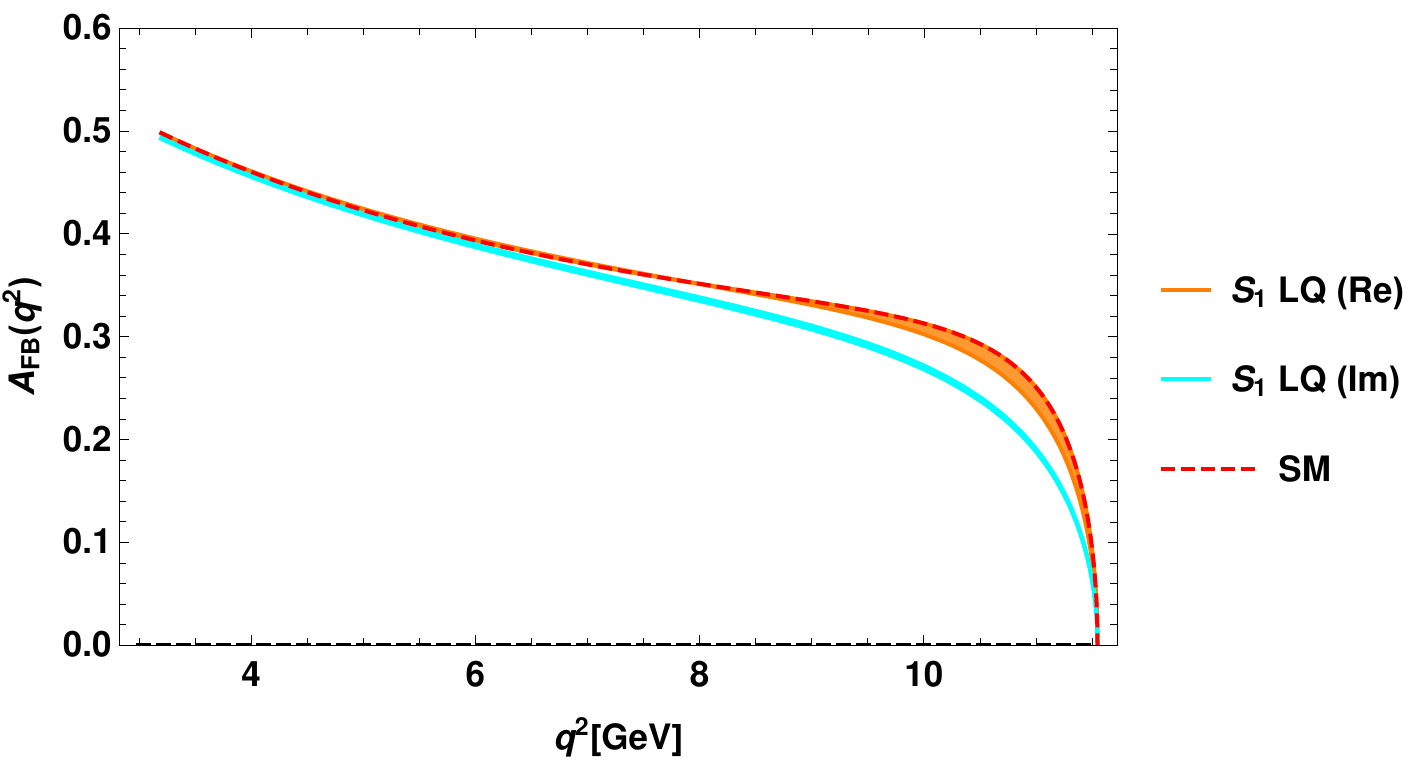}
\quad
\includegraphics[scale=0.5]{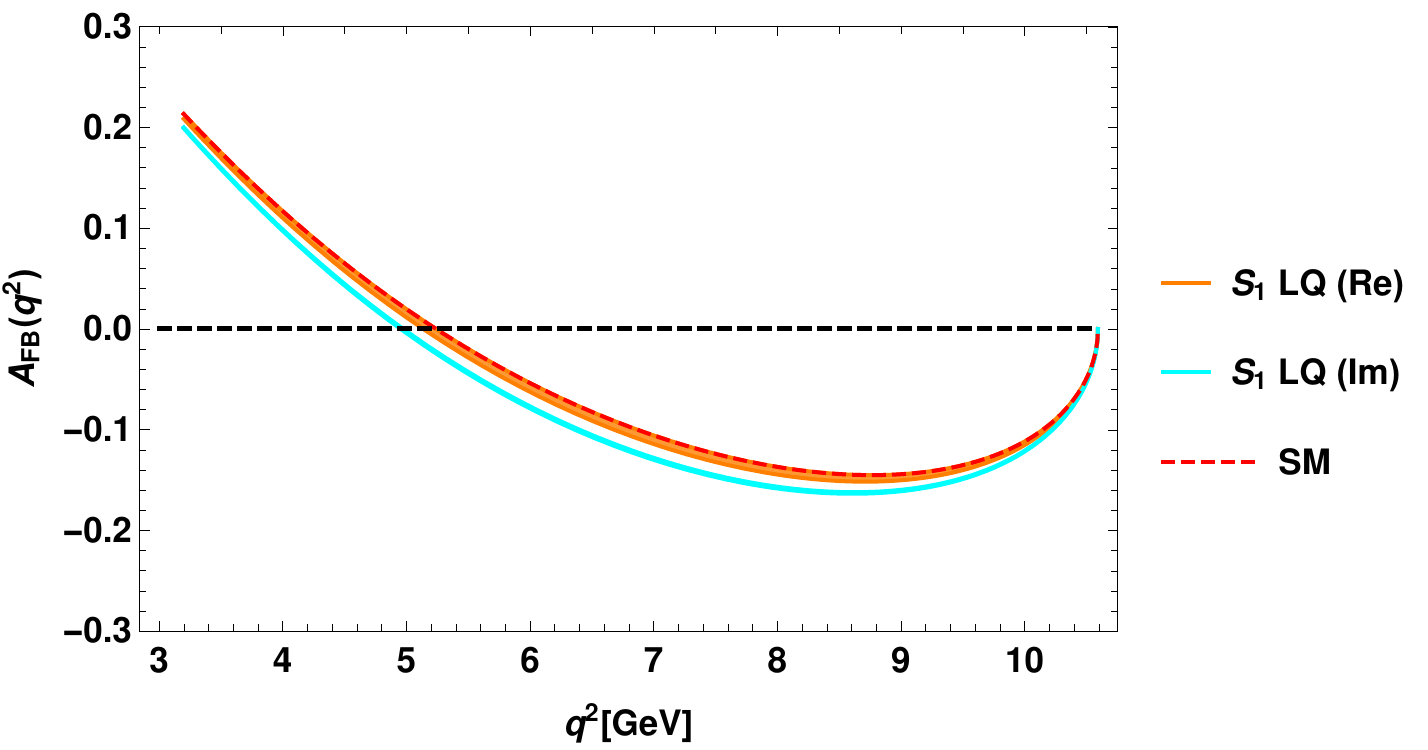}\\
\caption{The variation of forward-backward asymmetry of  $\bar B_s \to K^+\tau^- \bar \nu_\tau$ (top-left panel), $\bar B_s \to K^{* +} \tau^- \bar \nu_\tau $ (top-right panel), $\bar B_s \to D_s^+\tau^- \bar \nu_\tau $ (bottom-left panel) and $\bar B_s \to D_s^{*+}\tau^- \bar \nu_\tau $ (bottom-right panel) processes with respect to $q^2$ in the  $S_1$ scalar leptoquark model.} \label{S1-FB}
\end{figure}

Beyond the branching ratios, another interesting observable is the zero crossing of forward-backward asymmetry. The $q^2$ variation of the forward-backward asymmetries of $B_s \to K \tau \bar \nu_\tau$ (top-left panel), $B_s \to K^* \tau \bar \nu_\tau$ (top-right panel), $B_s \to D_s \tau \bar \nu_\tau$ (bottom-left panel) and $B_s \to D_s^* \tau \bar \nu_\tau$ (bottom-right panel) decay modes in the $S_1$ scalar leptoquark model are presented in Fig. \ref{S1-FB}\,. The  case of real leptoquark coupling provides significant deviation in  the forward-backward asymmetries of $B_s \to K^{(*)} \tau \bar \nu_\tau$ processes from their SM values, where as the case of complex leptoquark coupling provides comparatively less deviation.  The  constrained  couplings(real and complex) have almost negligible impact on the forward-backward asymmetries linked to the $B_s \to D_s^{(*)} \tau \bar \nu_\tau$ channels.   The numerical values of the forward-backward asymmetry  are given in Table \ref{Tab:S1}\,. 

\begin{figure}[h]
\centering
\includegraphics[scale=0.5]{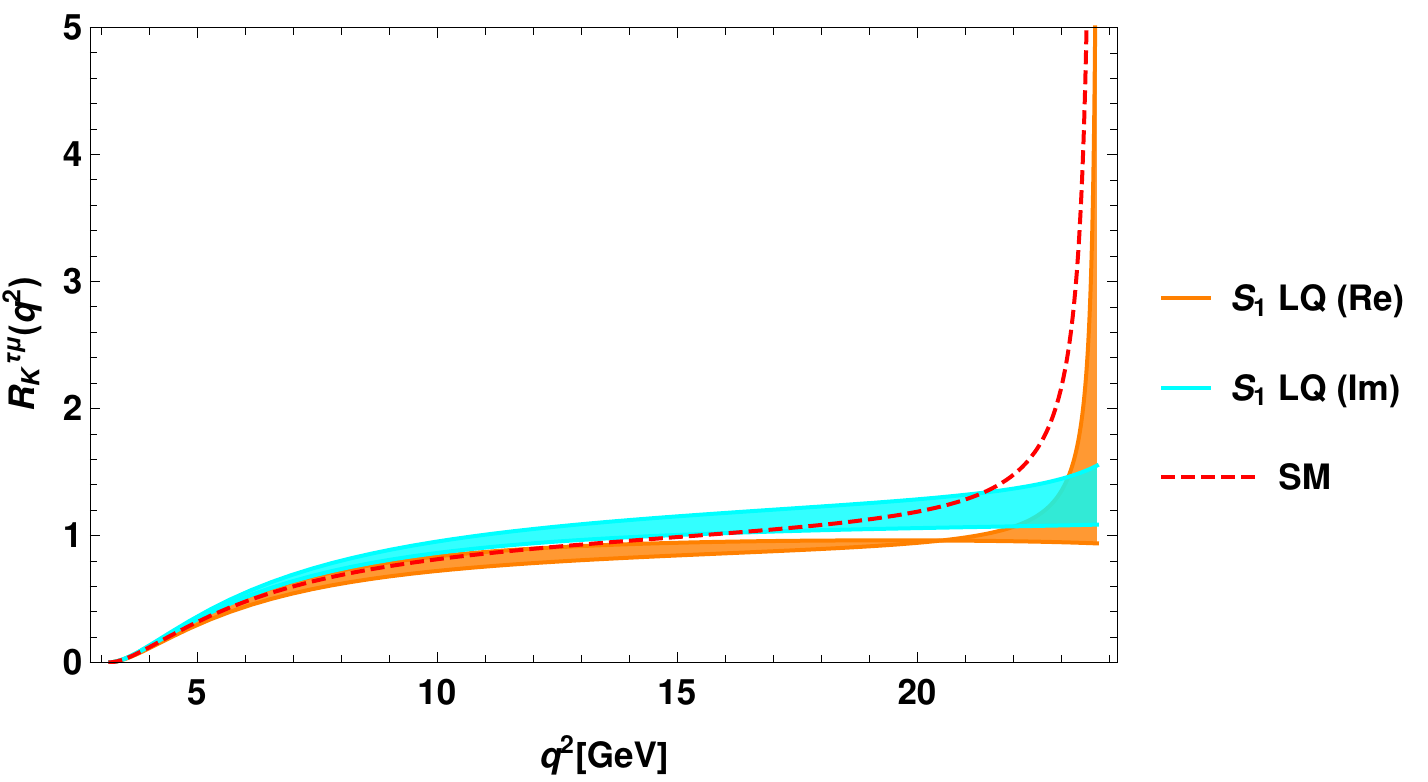}
\quad
\includegraphics[scale=0.5]{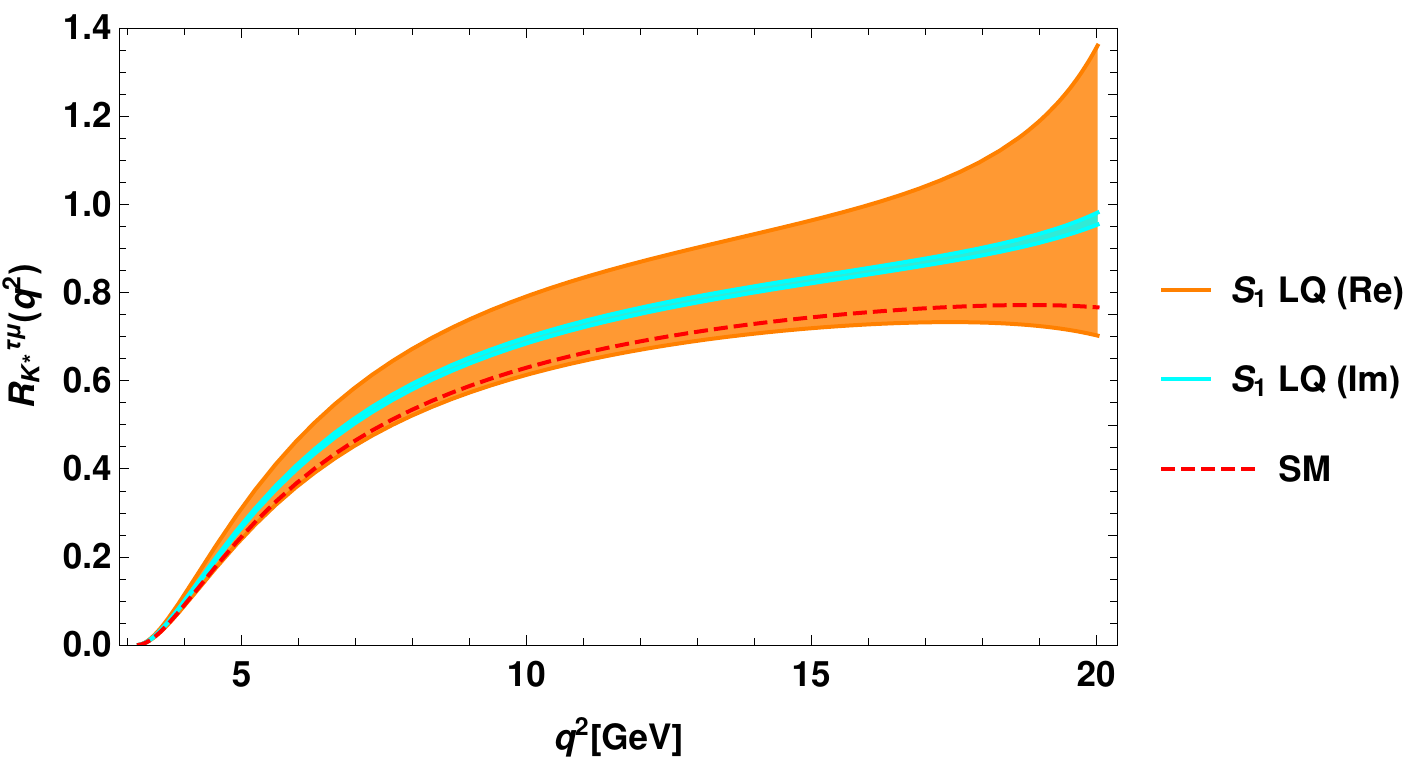}\\
\includegraphics[scale=0.5]{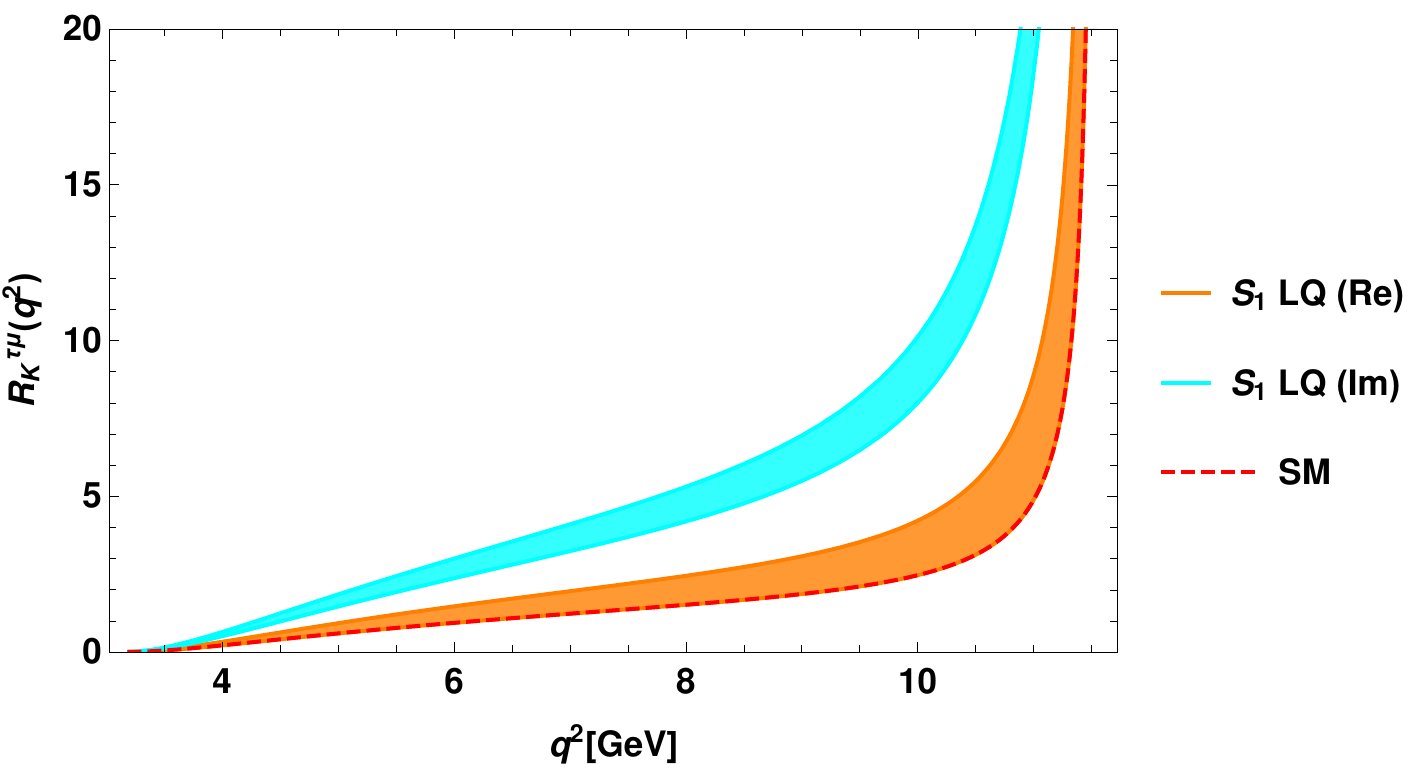}
\quad
\includegraphics[scale=0.5]{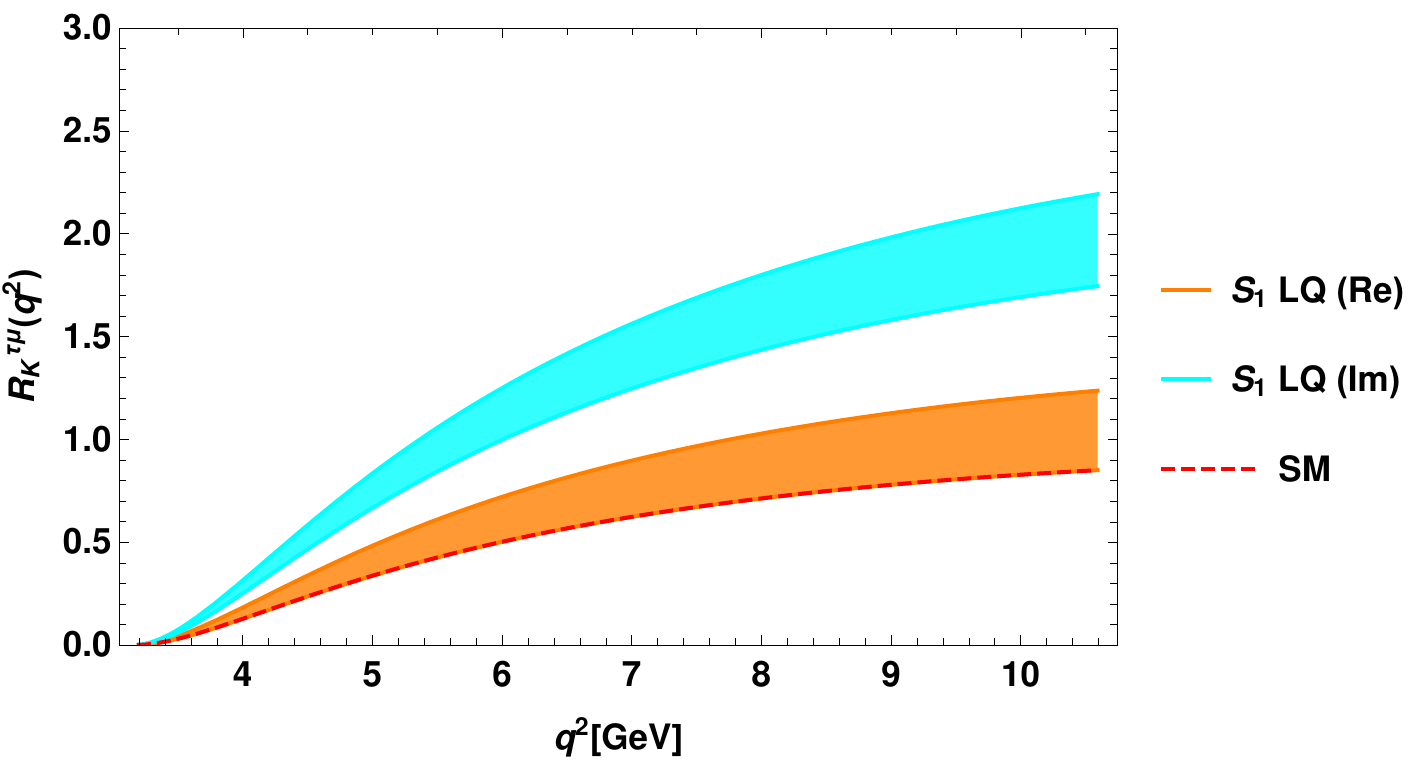}\\
\caption{The variation of lepton non-universality prameters of  $\bar B_s \to K^+\tau^- \bar \nu_\tau $ (top-left panel), $\bar B_s \to K^{* +} \tau^- \bar \nu_\tau $ (top-right panel), $\bar B_s \to D_s^+\tau^- \bar \nu_\tau $ (bottom-left panel) and $\bar B_s \to D_s^{*+}\tau^- \bar \nu_\tau $ (bottom-right panel) processes with respect to $q^2$ in the  $S_1$ scalar leptoquark model.} \label{S1-LNU}
\end{figure}

Fig. \ref{S1-LNU}\, depicts the variation of the  lepton non-universality parameters $R_K^{\tau \mu}$ (top-left panel), $R_{K^*}^{\tau \mu}$ (top-right panel), $R_{D_s}^{\tau \mu}$ (bottom-left panel) and $R_{D_s^*}^{\tau \mu}$ (bottom-right panel) with respect to $q^2$.  It is found that,  the allowed real coupling region has more effect on  $R_{K^{(*)}}^{\tau \mu}$  LNU parameters and the complex leptoquark couplings affect the $R_{D_s^{(*)}}^{\tau \mu}$ parameters. In Table \ref{Tab:S1}, the numerical values of the  lepton non-universality parameters are shown.

\begin{figure}[h]
\centering
\includegraphics[scale=0.5]{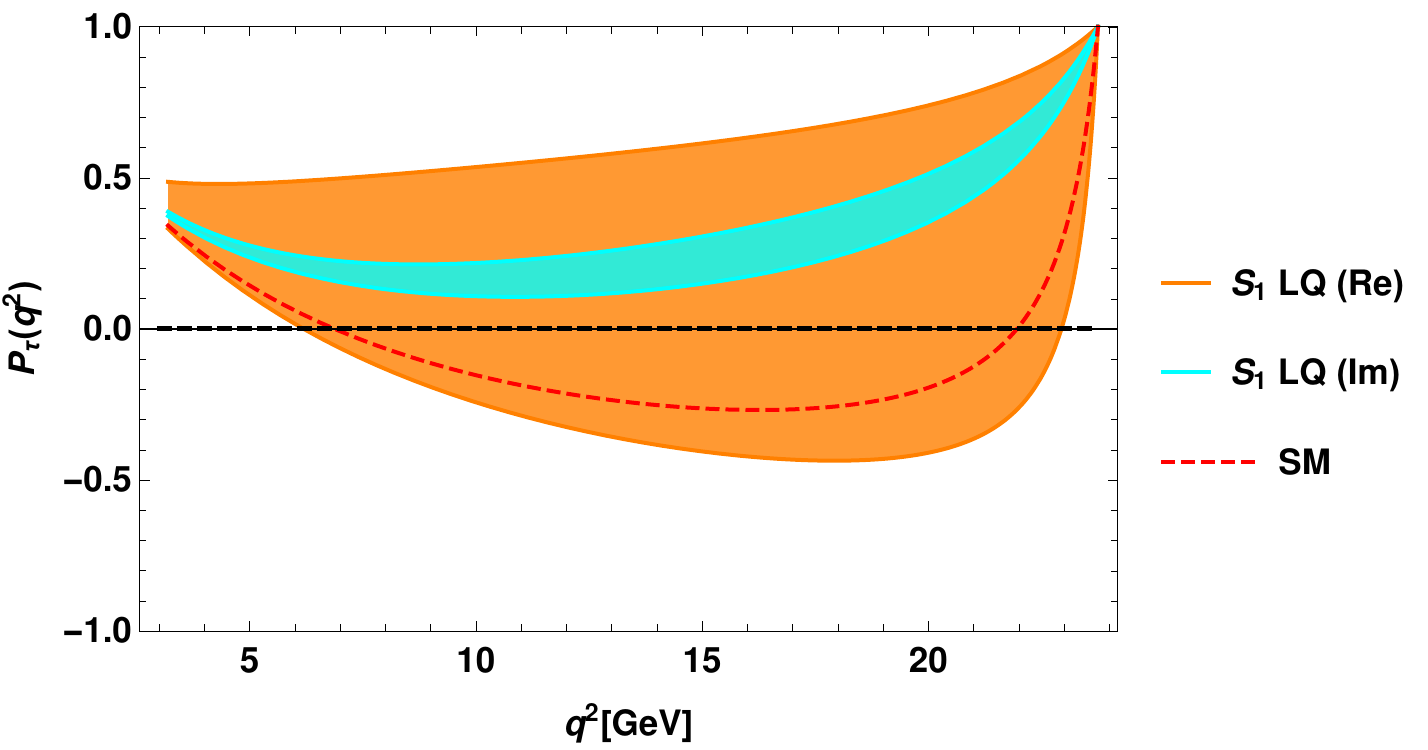}
\quad
\includegraphics[scale=0.5]{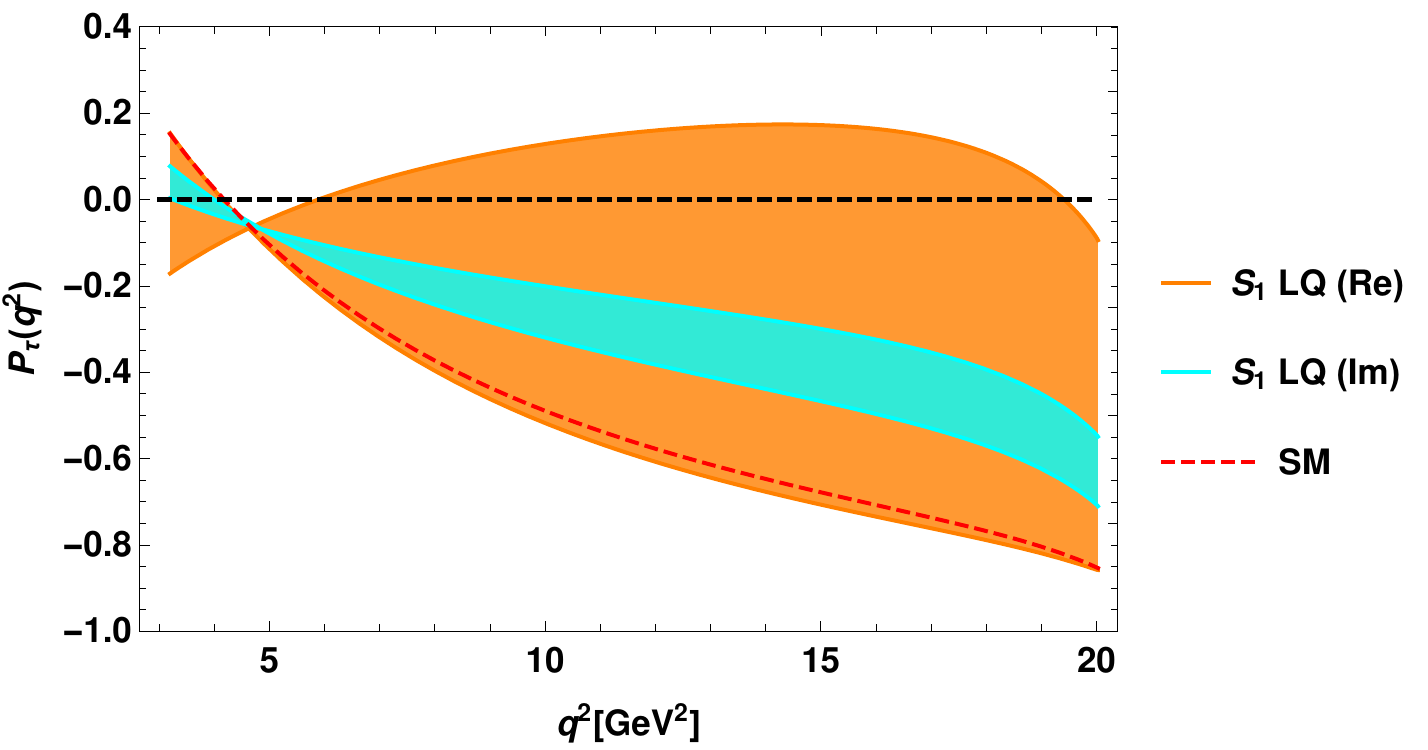}\\
\includegraphics[scale=0.5]{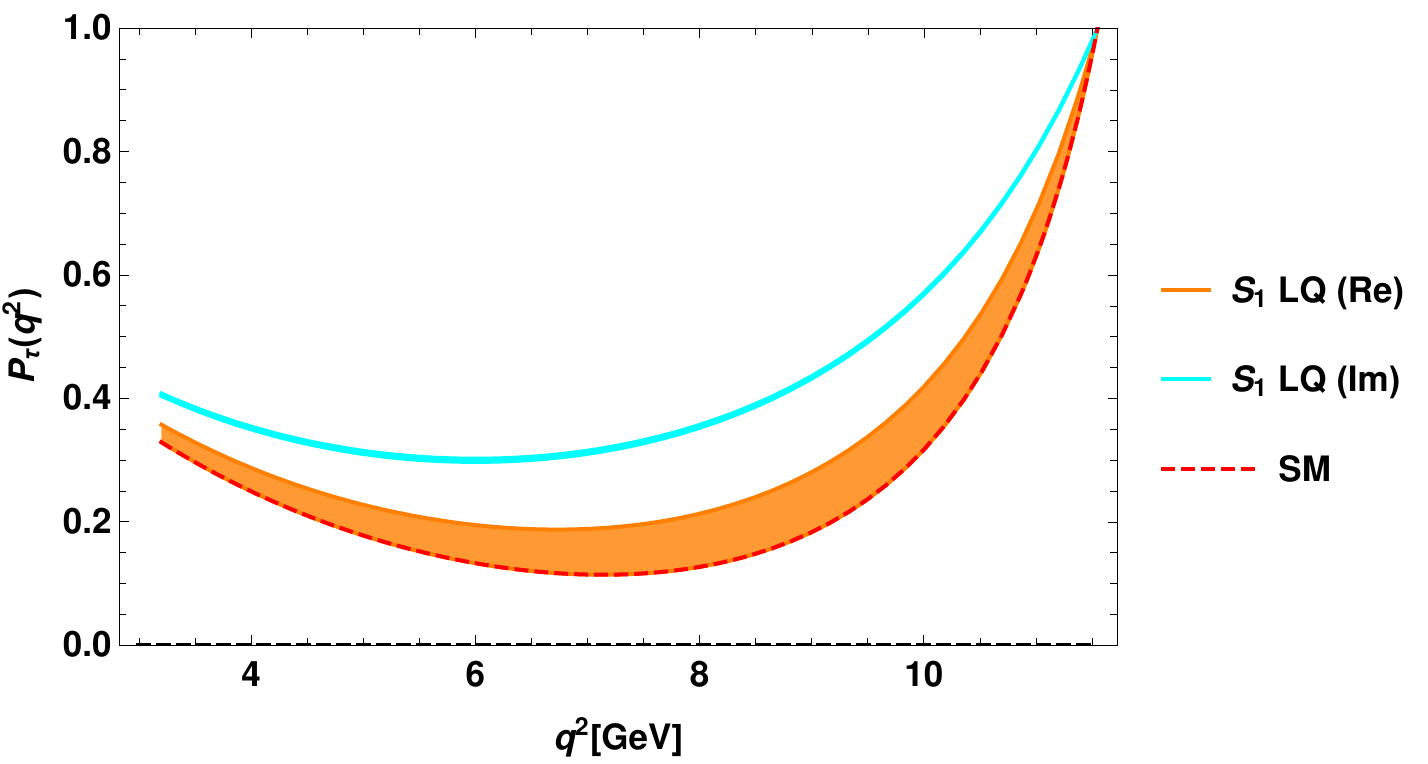}
\quad
\includegraphics[scale=0.5]{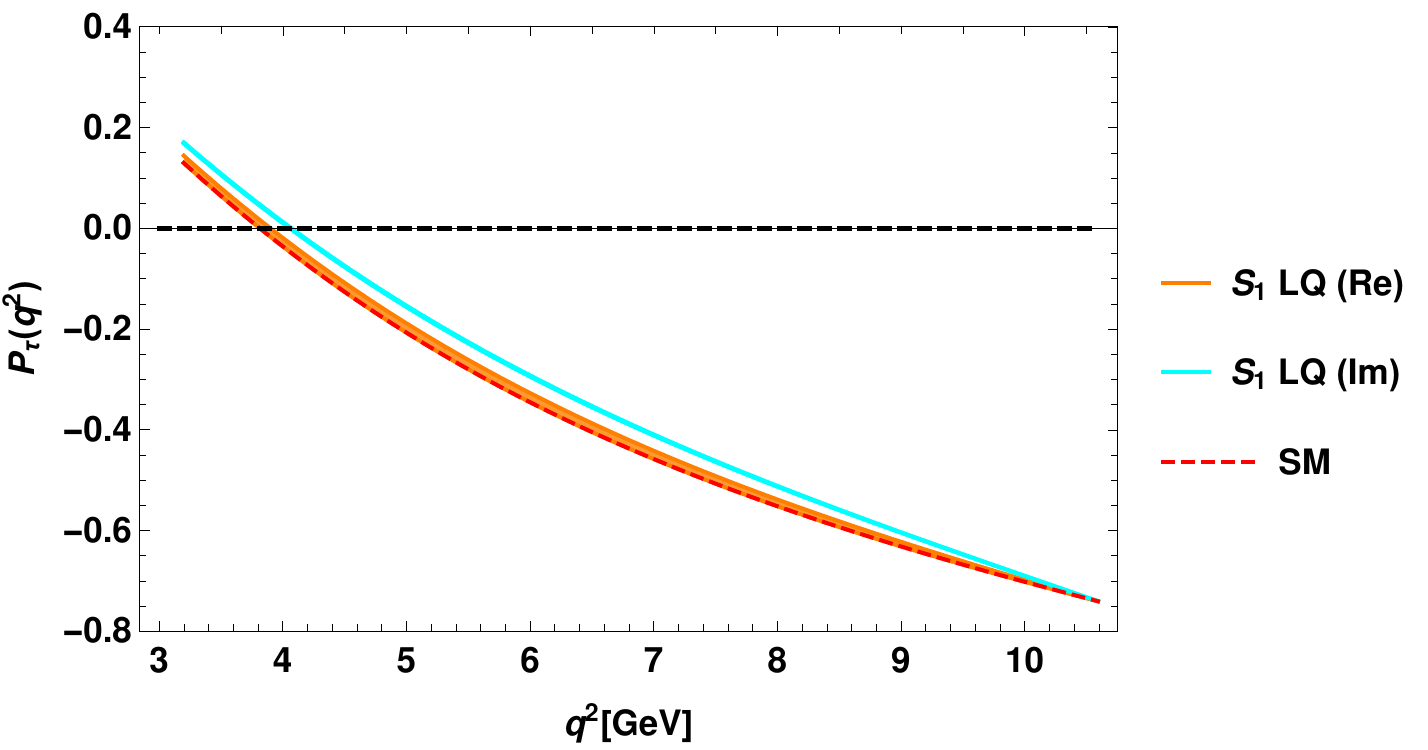}\\
\caption{The variation of lepton polarization asymmetry parameters of  $\bar B_s \to K^+\tau^- \bar \nu_\tau $ (top-left panel), $\bar B_s \to K^{* +} \tau^- \bar \nu_\tau $ (top-right panel), $\bar B_s \to D_s^+\tau^- \bar \nu_\tau $ (bottom-left panel) and $\bar B_s \to D_s^{*+}\tau^- \bar \nu_\tau $ (bottom-right panel) processes with respect to $q^2$ in the  $S_1$ scalar leptoquark model.} \label{S1-LP}
\end{figure}

\begin{figure}[h]
\centering
\includegraphics[scale=0.5]{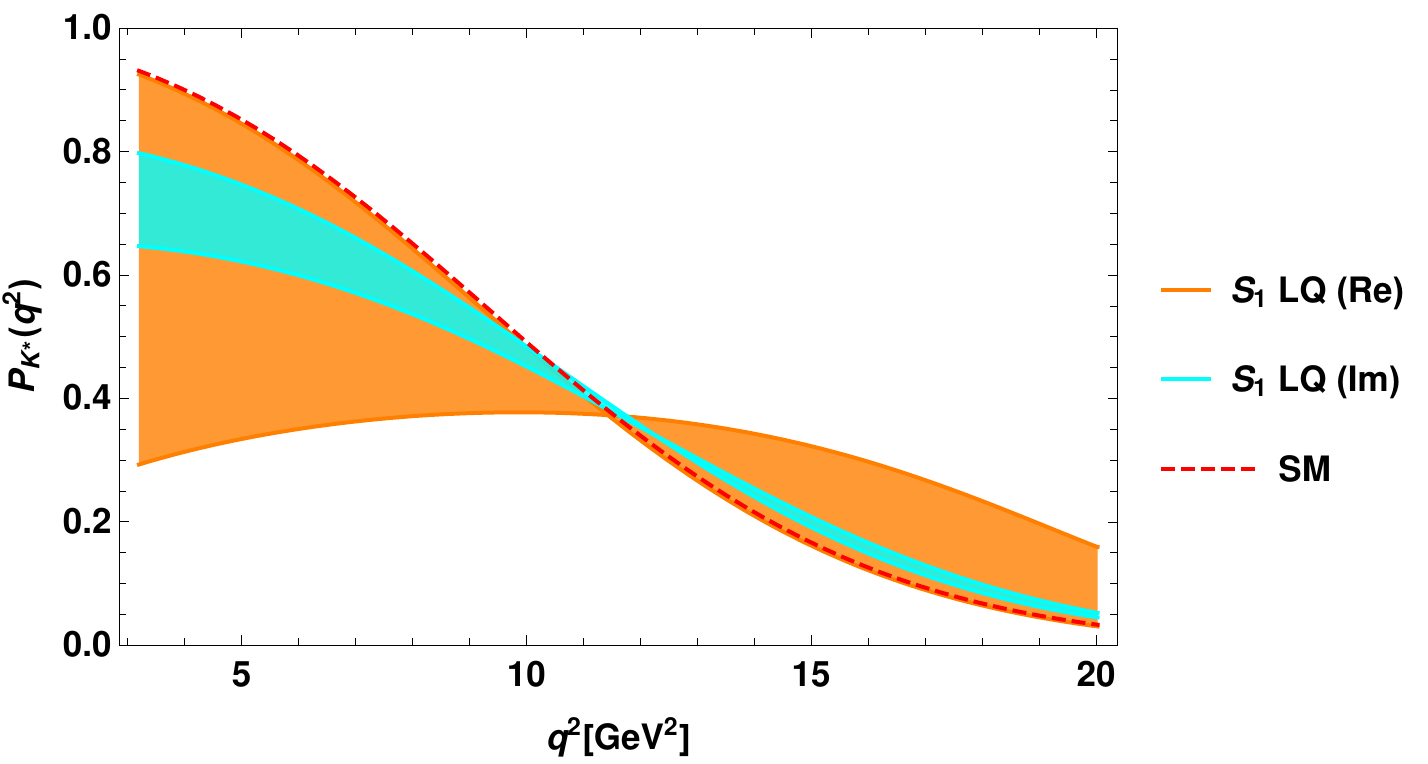}
\quad
\includegraphics[scale=0.5]{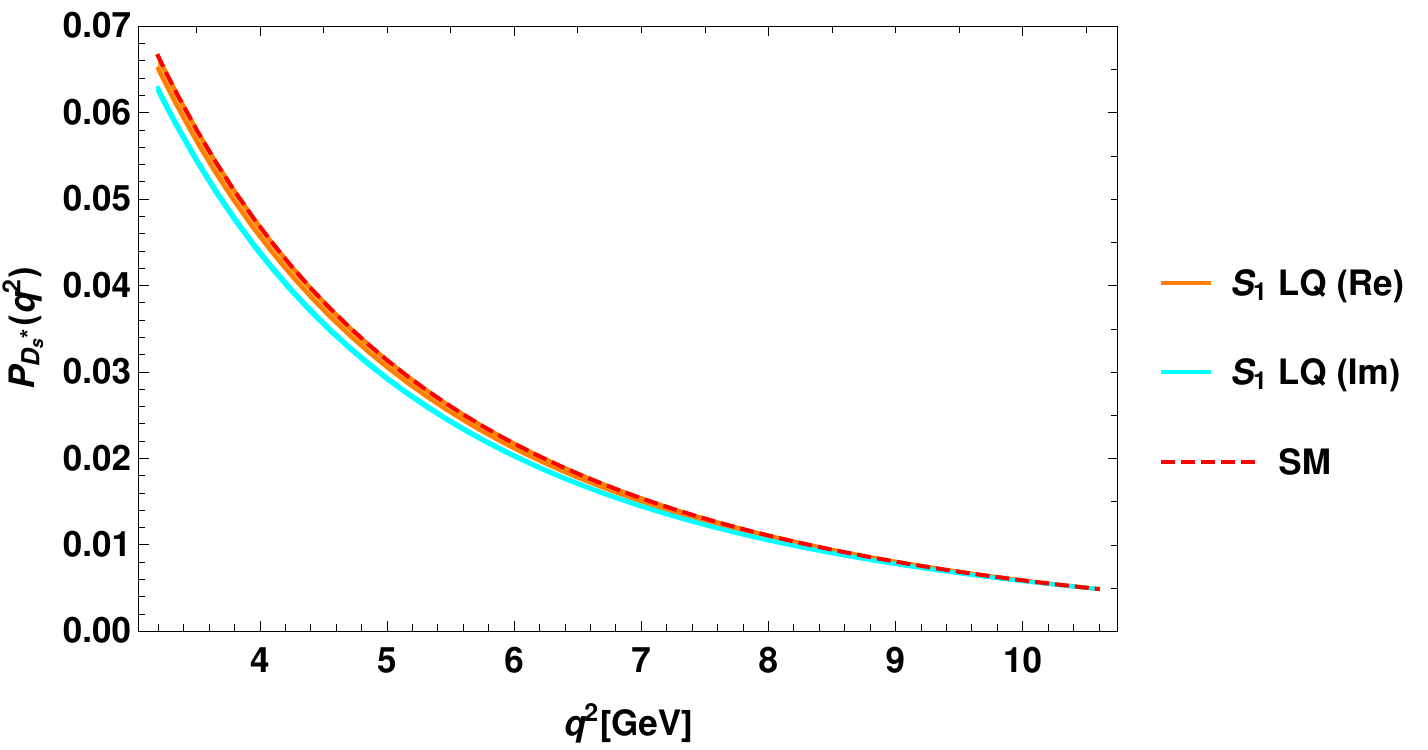}
\caption{The variation of hardon polarization asymmetry parameters of  $\bar B_s \to K^{* +} \tau^- \bar \nu_\tau $ (left panel) and $\bar B_s \to D_s^{*+}\tau^- \bar \nu_\tau $ (right panel) processes with respect to $q^2$ in the  $S_1$ scalar leptoquark model.} \label{S1-HP}
\end{figure}

The plots in the Fig. \ref{S1-LP}\,, show the effect of new parameter space on the $\tau$-polarization asymmetries of $B_s \to K \tau \bar \nu_\tau$ (top-left panel), $B_s \to K^* \tau \bar \nu_\tau$ (top-right panel), $B_s \to D_s \tau \bar \nu_\tau$ (bottom-left panel) and $B_s \to D_s^* \tau \bar \nu_\tau$ (bottom-right panel) processes. We observe profound  deviation in the  polarization asymmety of 
$B_s \to K^{(*)} \tau \bar \nu_\tau$ decay modes due to the $S_1$ SLQ contribution with the coupling as real. The contribution from the complex LQ couplings also deviate the $\tau$-polarization asymmetry parameters from their SM predictions. The complex coulings region affect the $\tau$-polarization asymmetry  $B_s \to D_s \tau \bar \nu_l$ significantly. There is no much impact of new physics on the $B_s \to D_s^* \tau \bar \nu_l$ decay proess. 
The $K^*$ $(D_s^*)$ polarization asymmetry plot for the $B_s \to K^* \tau \bar \nu_\tau$ $(B_s \to D_s^* \tau \bar \nu_\tau)$ mode is presented in the left panel (right panel) of Fig. \ref{S1-HP}\,. The predicted  numerical values of all these observables are given in \ref{Tab:S1}\,.

\begin{table}[htb]
\centering
\caption{The predicted values of the branching ratios and other physical observables of $B_s \to (K^{(*)}, D_s^{(*)}) \tau \bar \nu_\tau$  proesses in the SM and in the $S_1$ scalar leptoquark model. Here RC represents the real coupling  region and CC stands for complex coupling. } \label{Tab:S1}
\begin{tabular}{|c|c|c|c|c|c|}
\hline
~&~Observables~&~Values for SM~&~Values for RC~~&~Values for CC\\
\hline
\hline
 $B_s$~&~Br~&~$(6.65\pm 0.532 ) \times 10^{-5}$~&~$(5.35-9.36) \times 10^{-5}$~~&$(2.43-3.056) \times 10^{-4}$\\
$\downarrow$~&~$\langle A_{FB}^\tau\rangle$~&~$0.278\pm 0.022$~&~$0.25-0.42$~&~$0.348-0.369$\\ 
$K$~&~$\langle P_\tau \rangle$~&~$-0.1567\pm 0.013$~&~$-0.283\to 0.657$~&$0.22-0.368$\\ 
~&~$\langle R_K^{\tau \mu} \rangle$~&~$0.6365$~&~$0.512-0.896$~&$2.33-2.93$\\

\hline
 $B_s$~&~Br~&~$(1.85\pm 0.148 ) \times 10^{-4}$~&~$(0.68-2.95) \times 10^{-4}$~&$(4.38-6.09) \times 10^{-4}$\\
$\downarrow$~&~$\langle A_{FB}^\tau\rangle$~&~$-0.182\pm 0.015$~&~$-0.196\to -0.015$~&$-0.131\to -0.111$\\ 
$K^*$~&~$\langle P_\tau \rangle$~&~$-0.585\pm 0.045$~&~$-0.61\to 0.122$~&$-0.405\to-0.27$\\ 
~&~$\langle P_{K^*} \rangle$~&~$0.191\pm 0.015$~&~$0.185-0.313$~&$0.213-0.225$\\ 
~&~$\langle R_{K^*}^{\tau \mu} \rangle$~&~$0.54$~&~$0.198-0.858$~&$1.275-1.772$\\

\hline
 $B_s$~&~Br~&~$(1.39\pm 0.111 ) \times 10^{-2}$~&~$(1.39-2.26) \times 10^{-2}$~&$(3.93-4.67) \times 10^{-2}$\\
$\downarrow$~&~$\langle A_{FB}^\tau\rangle$~&~$0.358\pm 0.029$~&~$0.353-0.358$~&$0.329-0.333$\\ 
$D_s$~&~$\langle P_\tau \rangle$~&~$0.2\pm 0.016$~&~$0.2-0.284$~&$0.418-0.422$\\ 
~&~$\langle R_{D_s}^{\tau \mu} \rangle$~&~$0.6415$~&~$0.6415-1.04$~&$1.808-2.2867$\\

\hline
$B_s$~&~Br~&~$(2.23\pm 0.178 ) \times 10^{-2}$~&~$(2.23-3.215) \times 10^{-2}$~&$(4.5-5.62) \times 10^{-2}$\\
$\downarrow$~&~$\langle A_{FB}^\tau\rangle$~&~$-0.0996\pm 0.008$~&~$-0.107\to -0.0996$~~&$-0.12\to -0.118$\\ 
$D_s^*$~&~$\langle P_\tau \rangle$~&~$-0.514\to 0.0411$~&~$-0.514\to -0.5$~&$-0.4766\to -0.4746$\\ 
~&~$\langle P_{D_s^*} \rangle$~&~$0.0113\to 0.0009$~&~$0.011-0.0113$~&$0.0107-0.0108$\\ 
~&~$\langle R_{D_s^*}^{\tau \mu} \rangle$~&~$0.463$~&~$0.463-0.6673$~&$0.9313-1.167$\\

\hline
\end{tabular}
\end{table}

\subsection{$S_3$ scalar leptoquark}
\begin{figure}[h]
\centering
\includegraphics[scale=0.5]{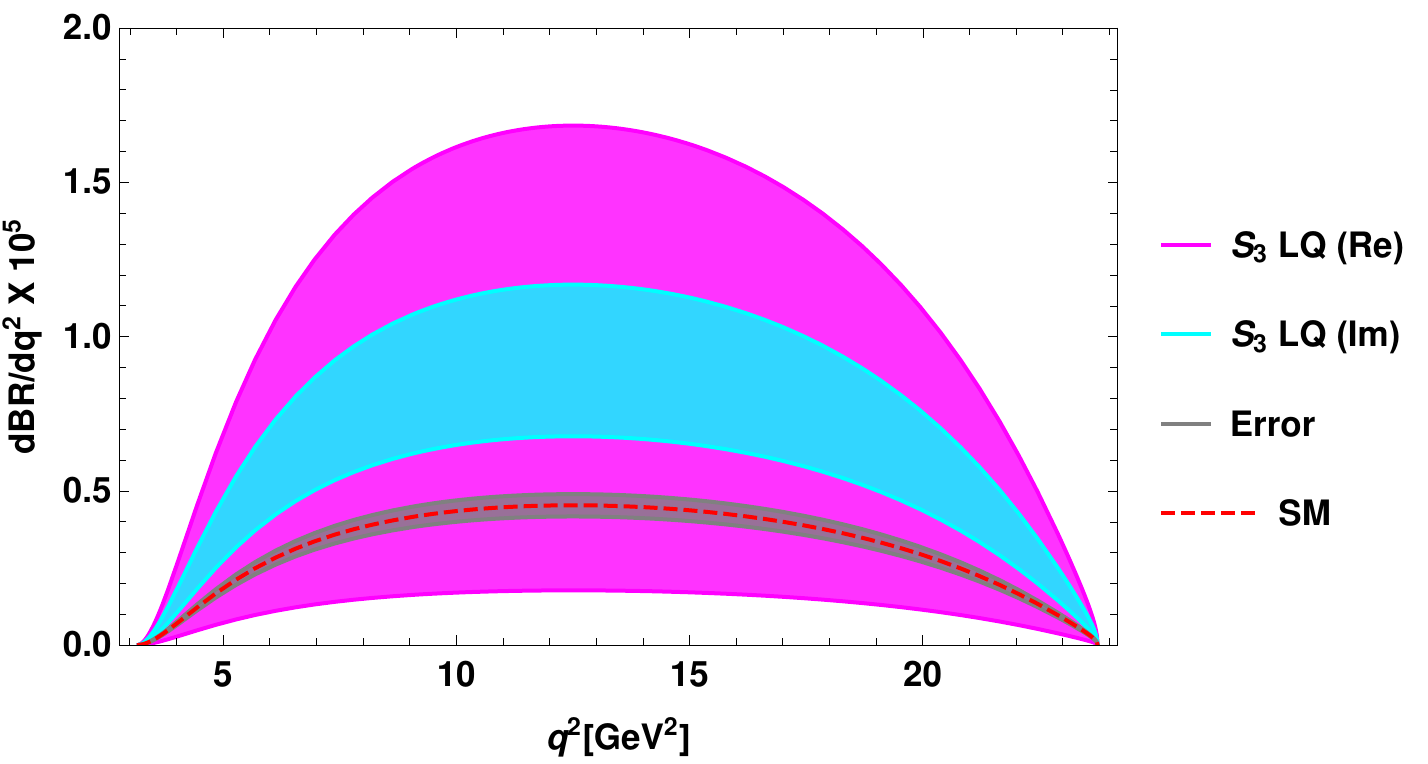}
\quad
\includegraphics[scale=0.5]{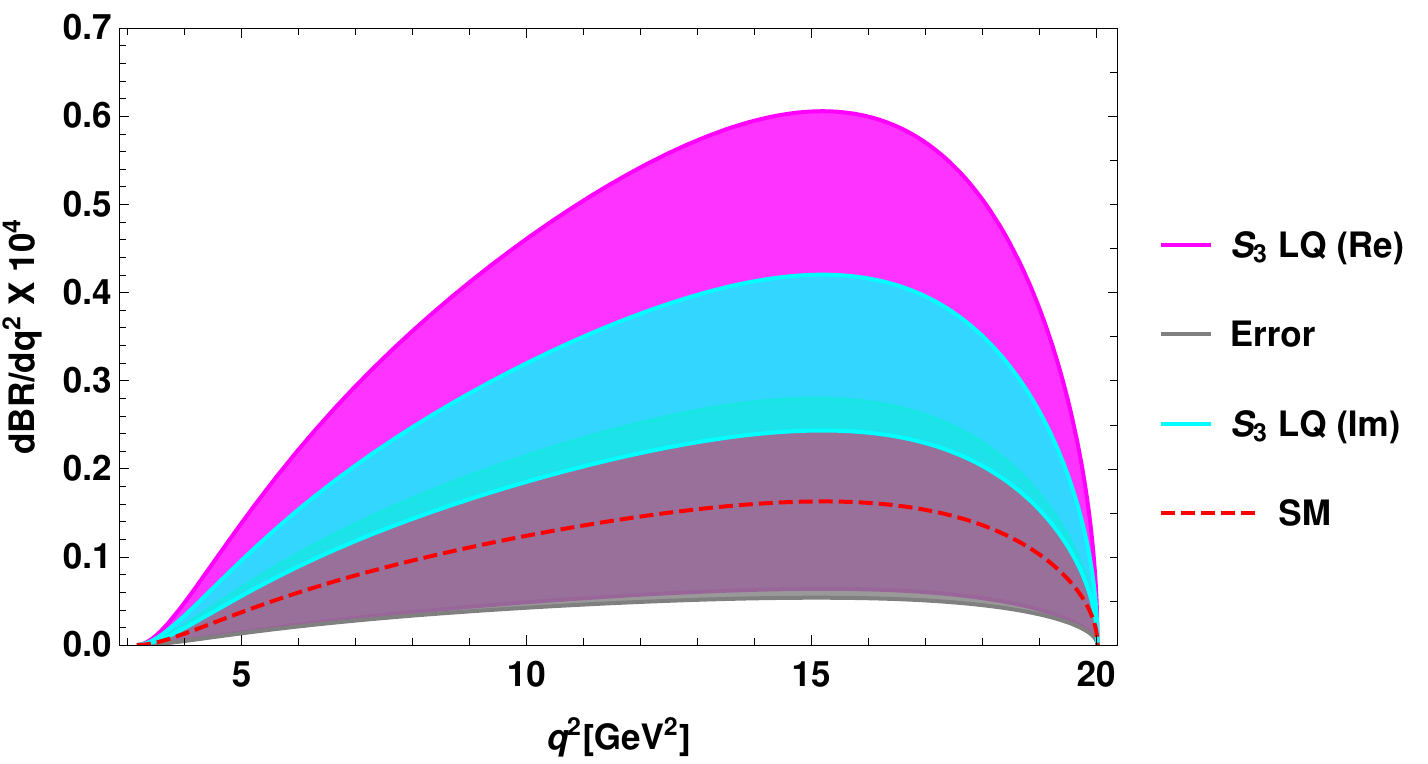}\\
\includegraphics[scale=0.5]{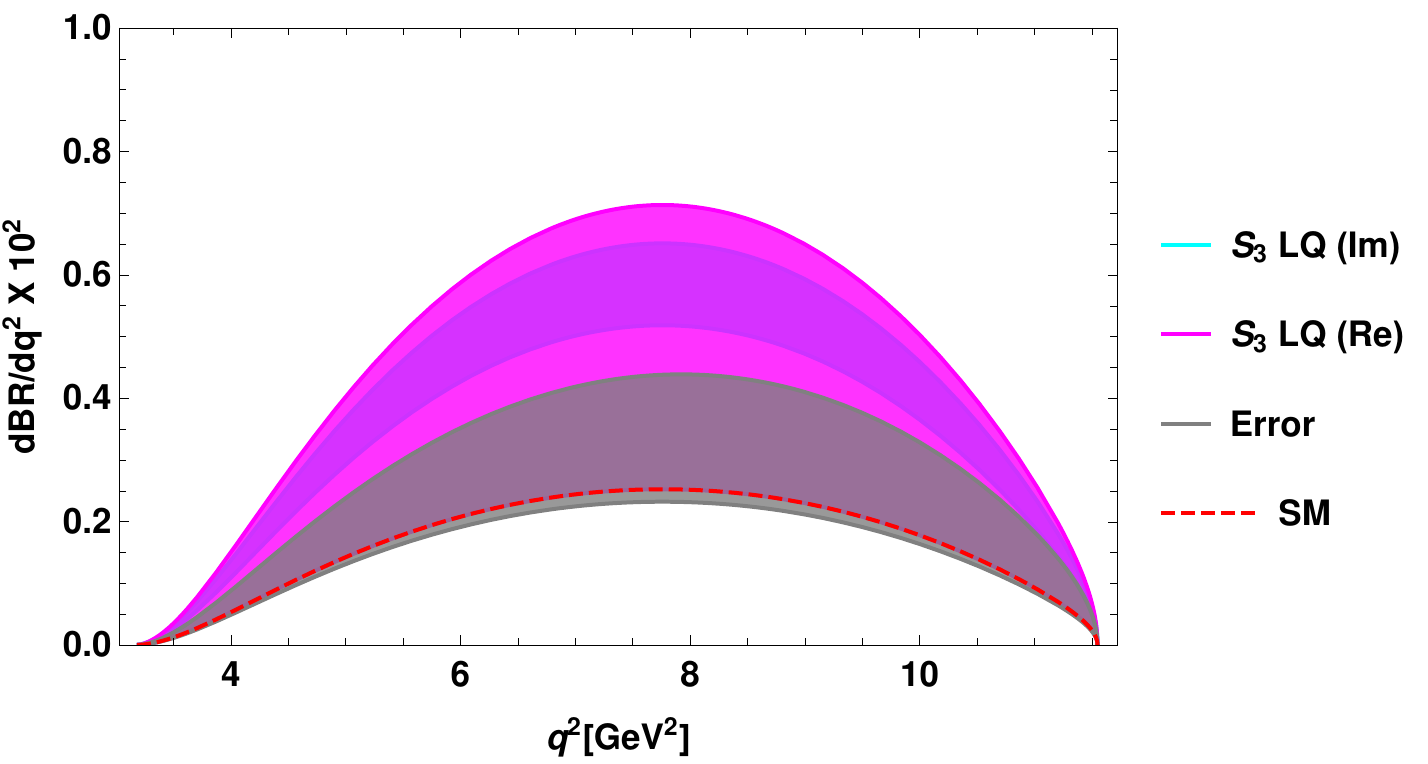}
\quad
\includegraphics[scale=0.5]{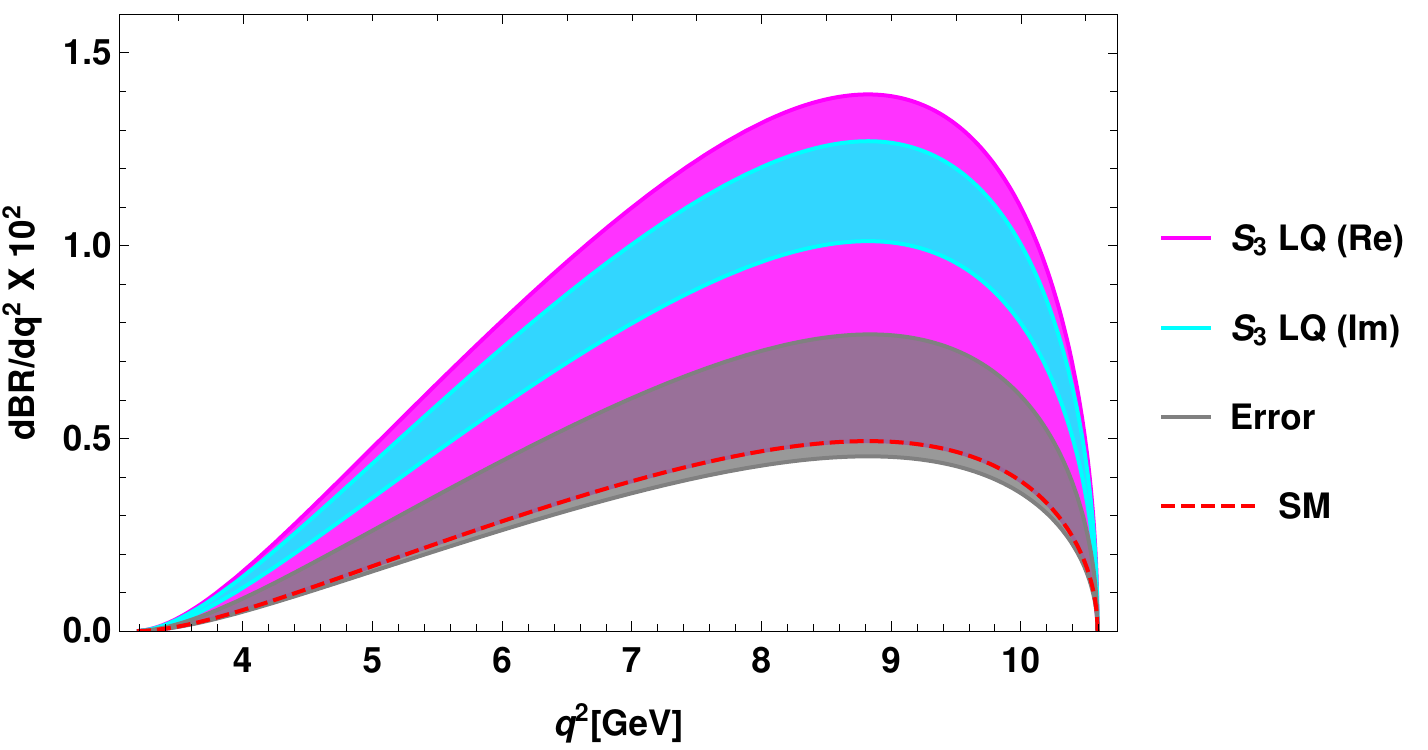}\\
\caption{The variation of branching ratios of $\bar B_s \to K^+\tau^- \bar \nu_\tau $ (top-left panel), $\bar B_s \to K^{* +} \tau^- \bar \nu_\tau $ (top-right panel), $\bar B_s \to D_s^+\tau^- \bar \nu_\tau $ (bottom-left panel) and $\bar B_s \to D_s^{*+}\tau^- \bar \nu_\tau $ (bottom-right panel) processes with respect to $q^2$ in the  $S_3$ scalar leptoquark model. Here cyan bands represent the case of complex $S_3$ leptoquark coupling and the magenta  bands stand for real coupling. } \label{S3-Br}
\end{figure}

The triplet $S_3(\bar 3, 1,1/3)$ scalar leptoquark  contributes only additional $V_L$ Wilson coefficient to the SM. The allowed new parameter space obtained from the available experimental data on relevant observables,   for both real and complex coupling cases are already provided in section III. Now using the constrained parameters, the branching ratios of $B_s \to K \tau \bar \nu_\tau$ (top-left panel), $B_s \to K^* \tau \bar \nu_\tau$ (top-right panel), $B_s \to D_s \tau \bar \nu_\tau$ (bottom-left panel) and $B_s \to D_s^* \tau \bar \nu_\tau$ (bottom-right panel) decay modes with $q^2$ in the $S_3$ scalar letoquark model  are shown in Fig. \ref{S3-Br}\,. Here the cyan color bands are arising due to the constrained complex coupligs  and the magneta  bands represent the new physics contributions to the branching ratios predicted from the  allowed region of $S_3$ leptoquark with real couplings. We found that, the branching ratios of all these decay modes deviate significantly from SM for the case of both complex and  real couplings. The predicted branching ratios for all the cases of new couplings are given in Table \ref{Tab:S3}\,. 

\begin{figure}[h]
\centering
\includegraphics[scale=0.5]{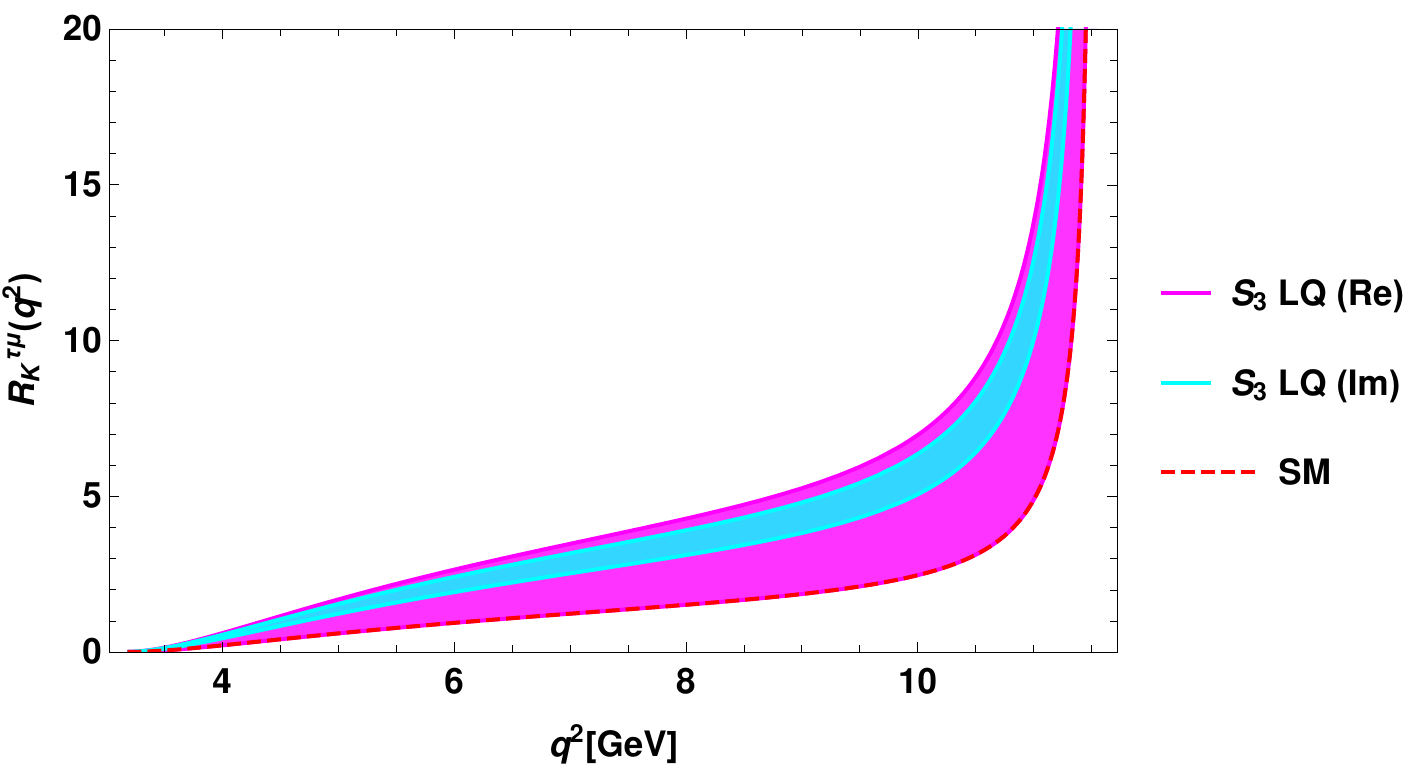}
\quad
\includegraphics[scale=0.5]{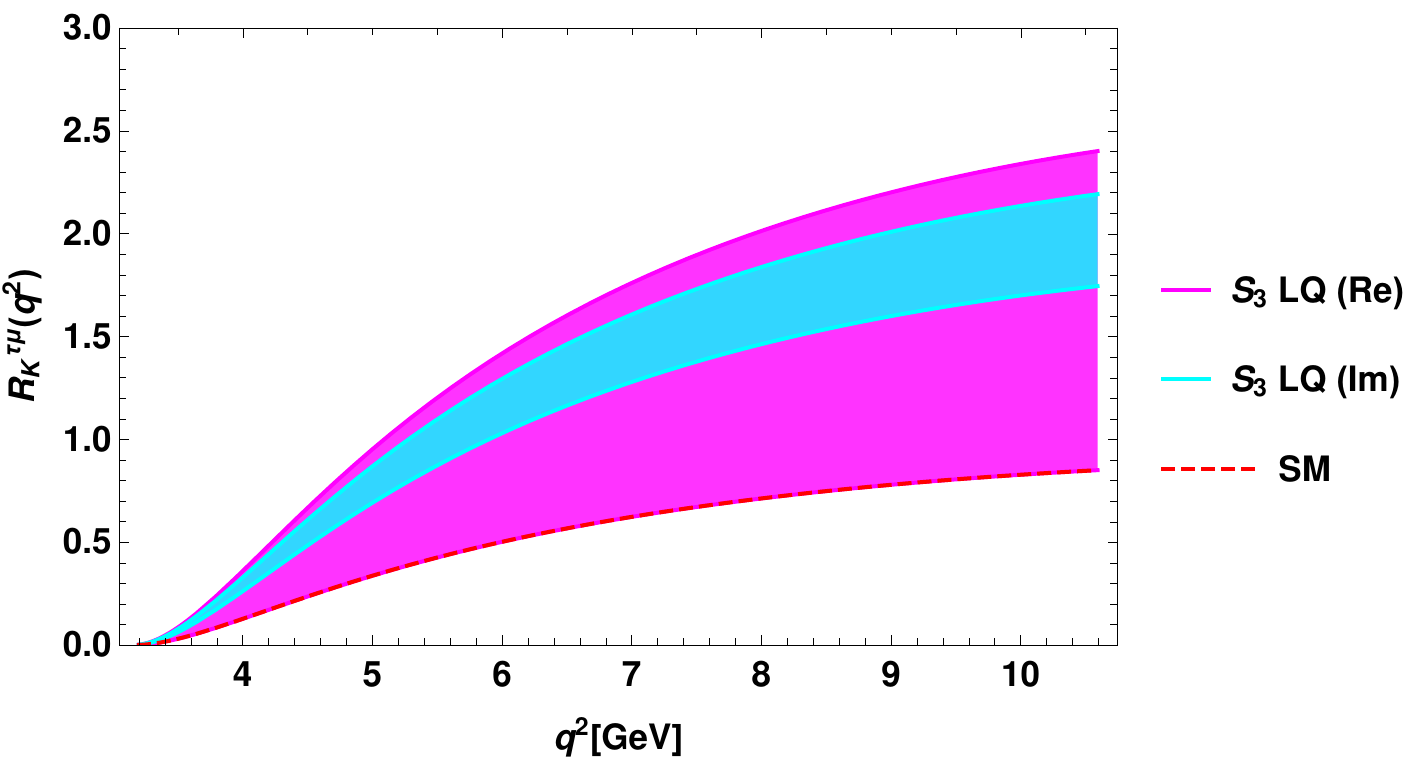}\\
\caption{The variation of lepton nonuniversality prameters of  $\bar B_s \to D_s^+\tau^- \bar \nu_\tau $ (left panel) and $\bar B_s \to D_s^{*+}\tau^- \bar \nu_\tau $ (right panel) processes with respect to $q^2$ in the  $S_3$ scalar leptoquark model.} \label{S3-LNU}
\end{figure}

In Fig. \ref{S3-LNU}\,, the lepton non-universality parameters for  $B_s \to D_s \tau \bar \nu_\tau$ and $B_s \to D_s^{*} \tau \bar \nu_\tau$ are shown in the left and right panels, respectively. For this  observables, the complex leptoquark coupling are found to be more effective in comparison to the case of real coupling. However, the impact of  $S_3$ leptoquark on the $R_{K^{(*)}}^{\tau \mu}$  parameter of $B_s \to K^{(*)} \tau \bar \nu_\tau$ processes are found to be negligible. The numerical values of the LNU  parameters of all these decay modes are presented in Table \ref{Tab:S3}\,. We don't find any deviation in the forward-backward asymmetry, lepton and hadron polarization asymmetry parameters of semileptonic $B_s$ decay processes due to the additional NP contributions from $S_3$ scalar LQ.   


\begin{table}[htb]
\centering
\caption{The predicted values of the branching ratios and other physical observables of $B_s \to (K^{(*)}, D_s^{(*)} \tau \bar \nu_\tau)$  proesses in the SM and in the $S_3$ scalar leptoquark model. Here RC represents the real coupling  and CC stands for complex coupling. } \label{Tab:S3}
\begin{tabular}{|c|c|c|c|}
\hline
~Observables~~&~Values for RC~&~Values for CC\\
\hline
\hline
~Br($B_s \to K \tau\bar \nu_l$)~&~$(0.26-2.47) \times 10^{-4}$~~&$(0.993-1.71) \times 10^{-4}$\\

\hline
~Br($B_s \to K^* \tau\bar \nu_l$)&~$(0.73-6.9) \times 10^{-4}$~&$(2.77-4.78) \times 10^{-4}$\\

\hline
~Br($B_s \to D_s \tau\bar \nu_l$)~&~$(2.26-3.94) \times 10^{-2}$~&$(2.86-3.59) \times 10^{-2}$\\
~$\langle R_{D_s}^{\tau \mu} \rangle$~~&~$0.6415-1.812$~&$1.317-1.654$\\

\hline
~Br($B_s \to D_s^* \tau\bar \nu_l$)~&~$(4.82-6.3) \times 10^{-2}$~~&$(4.6-5.74) \times 10^{-2}$\\ 
~$\langle R_{D_s^*}^{\tau \mu} \rangle$~~&~$0.463-1.307$~~&$0.95-1.193$\\

\hline
\end{tabular}
\end{table}

\subsection{$R_2$ scalar leptoquark}
\begin{figure}[h]
\centering
\includegraphics[scale=0.5]{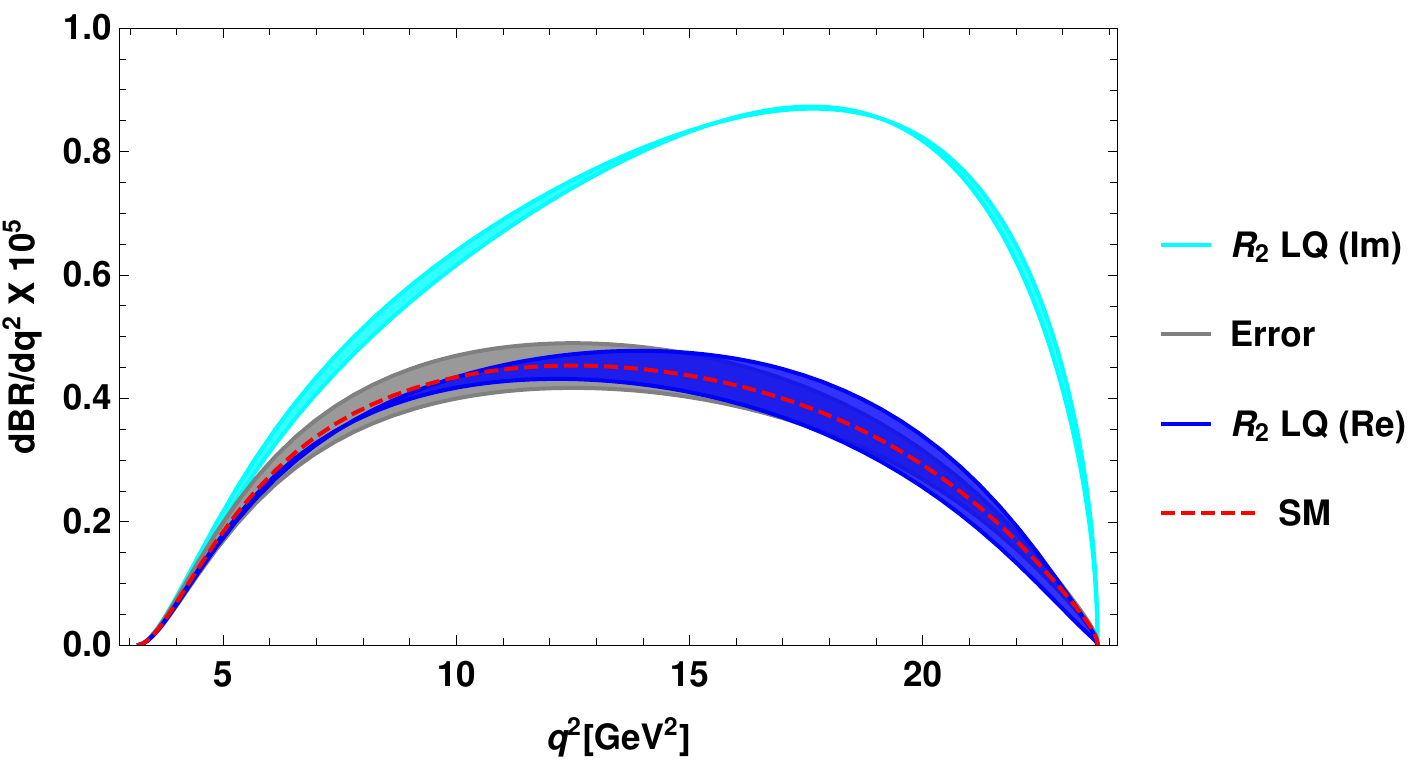}
\quad
\includegraphics[scale=0.5]{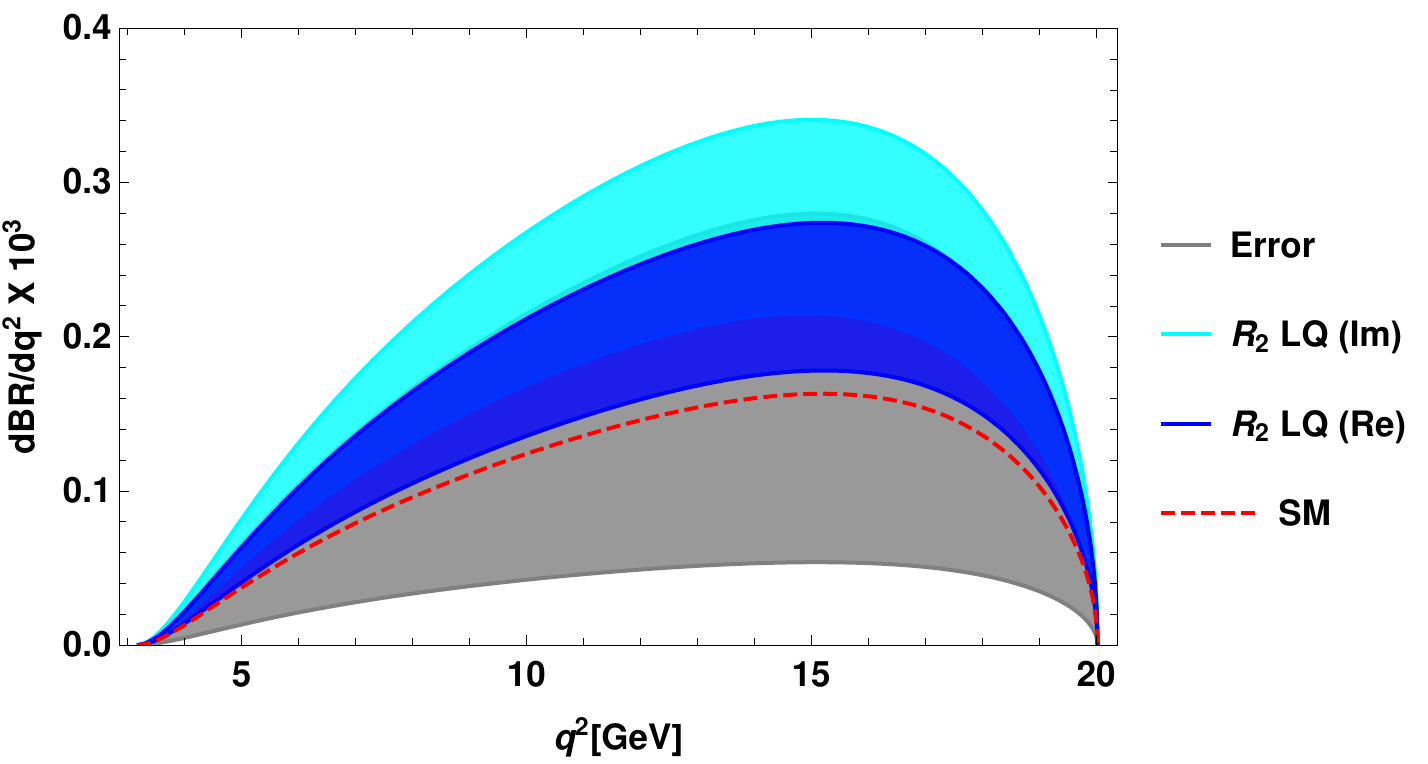}\\
\includegraphics[scale=0.5]{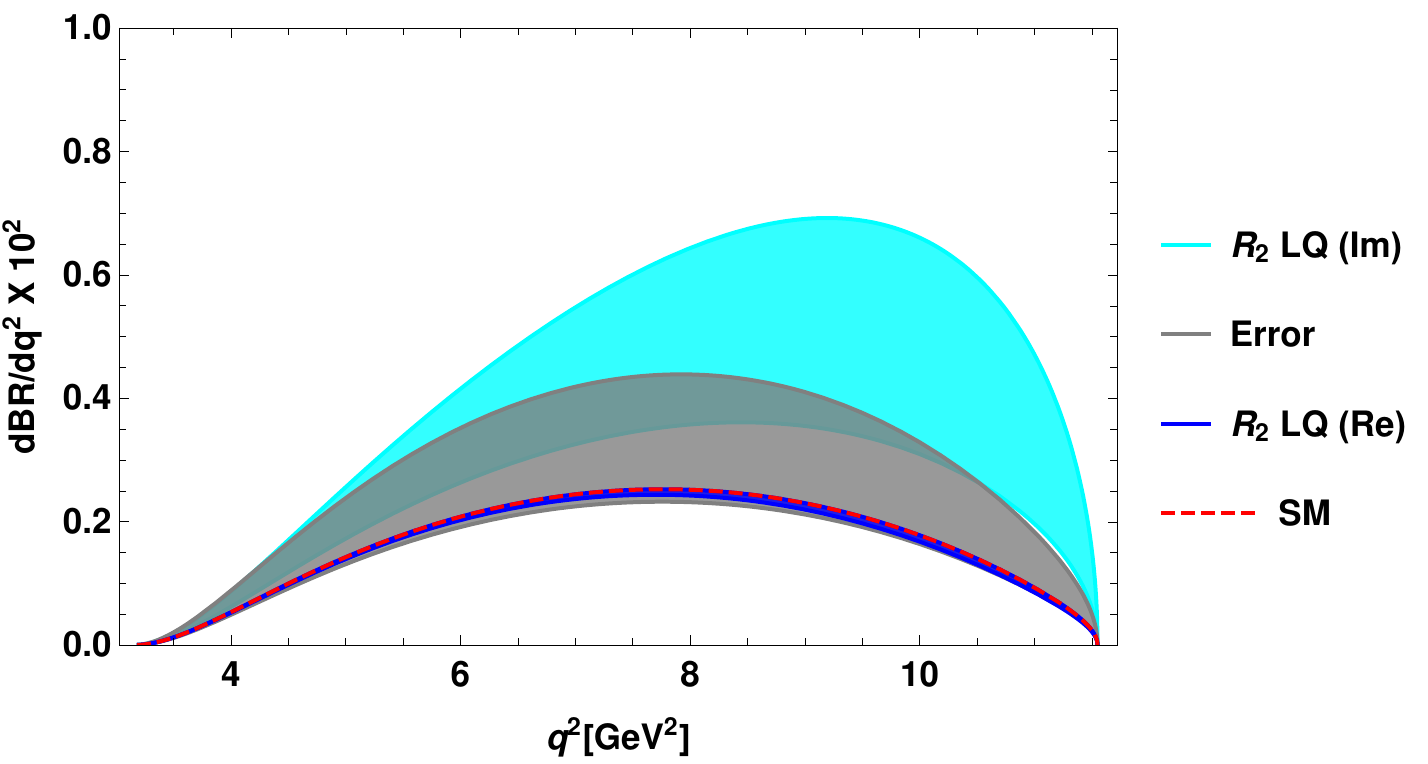}
\quad
\includegraphics[scale=0.5]{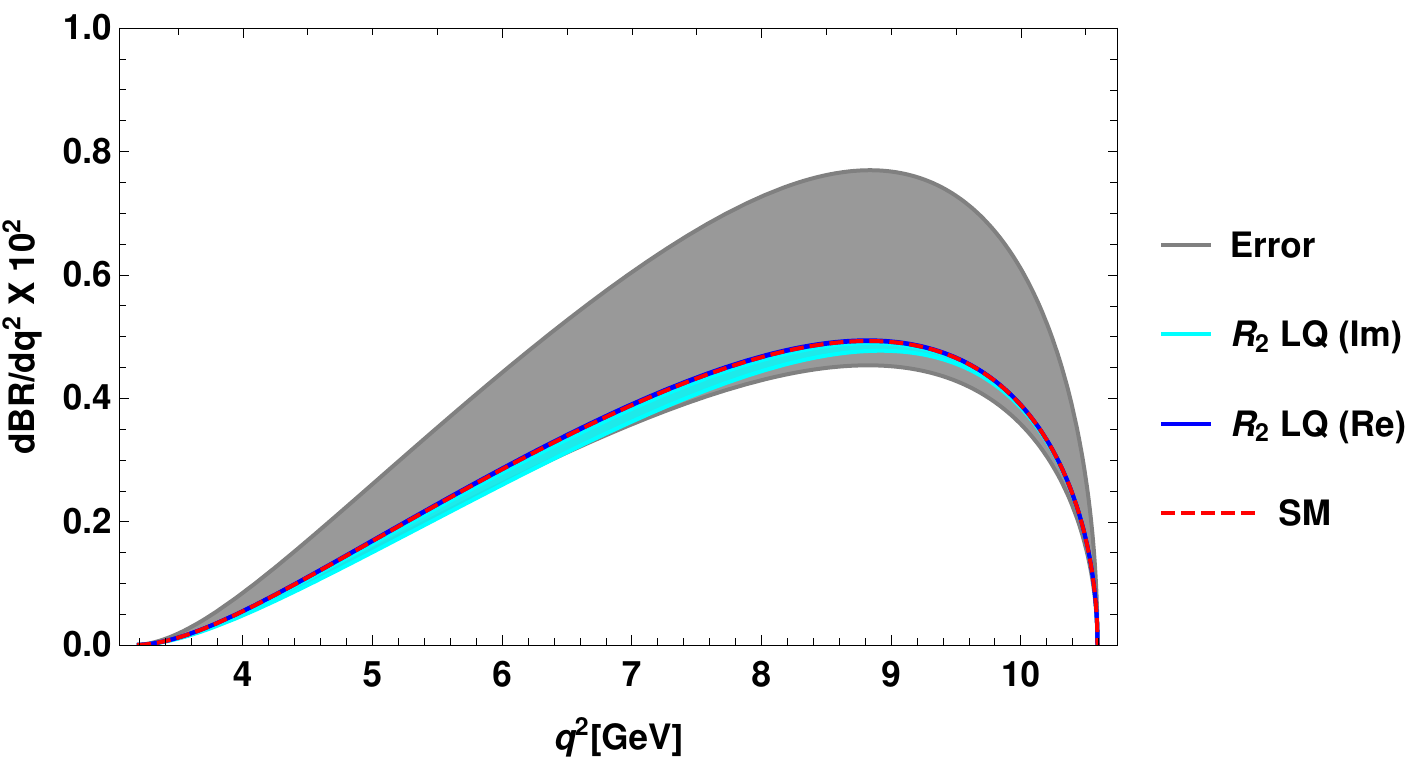}\\
\caption{The variation of branching ratios of $\bar B_s \to K^+\tau^- \bar \nu_\tau $ (top-left panel), $\bar B_s \to K^{* +} \tau^- \bar \nu_\tau $ (top-right panel), $\bar B_s \to D_s^+\tau^- \bar \nu_\tau $ (bottom-left panel) and $\bar B_s \to D_s^{*+}\tau^- \bar \nu_\tau$ (bottom-right panel) processes with respect to $q^2$ in the  $R_2$ scalar leptoquark model. Here cyan bands represent the case of complex $S_3$ leptoquark coupling and the blue bands stand for  real coupling.} \label{R2-Br}
\end{figure}

After discussing the impact of $S_{1,3}$ scalar leptoquarks on the  flavor observables of $B_s$ decay modes mediated by the  $b \to (u,c) \tau \bar \nu_\tau$ transitions, we now proceed to check the same physical observables in the $R_2(3,2,7/6)$ SLQ model. $R_2$  LQ is doublet  under $SU(2)$ and contributes only scalar and tensor type couplings to the SM. The variation of  branching ratios of $B_s \to K \tau \bar \nu_\tau$ (top-left panel), $B_s \to K^* \tau \bar \nu_\tau$ (top-right panel), $B_s \to D_s \tau \bar \nu_\tau$ (bottom-left panel) and $B_s \to D_s^* \tau \bar \nu_\tau$ (bottom-right panel) processes with $q^2$, obtained by using the allowed parameter space of $R_2$ leptoquark (\ref{Tab:real}, \ref{Tab:complex})   are given in Fig. \ref{R2-Br}\,. Here the blue bands are due to the constrained real couplings and the cyan bands are for the complex $R_2$ leptoquark couplings. We observe that,  the branching ratios of $B_s \to (K, D_s)$ modes show profound deviatation  from their corresponding  SM results due to the additional complex coupling contributions. The real coupling has more effect on Br$(B_s \to K^* \tau \bar \nu_\tau)$  in comarison to the case of complex parmeters. Where as there is no deviation in the branching ratio of $B_s \to D_s^* \tau \bar \nu_\tau$ due to the presence of $R_2$ leptoquark. The predicted values of branching ratios  for both real and complex couplings  cases are presented in Table \ref{Tab:R2}\,. 

\begin{figure}[h]
\centering
\includegraphics[scale=0.5]{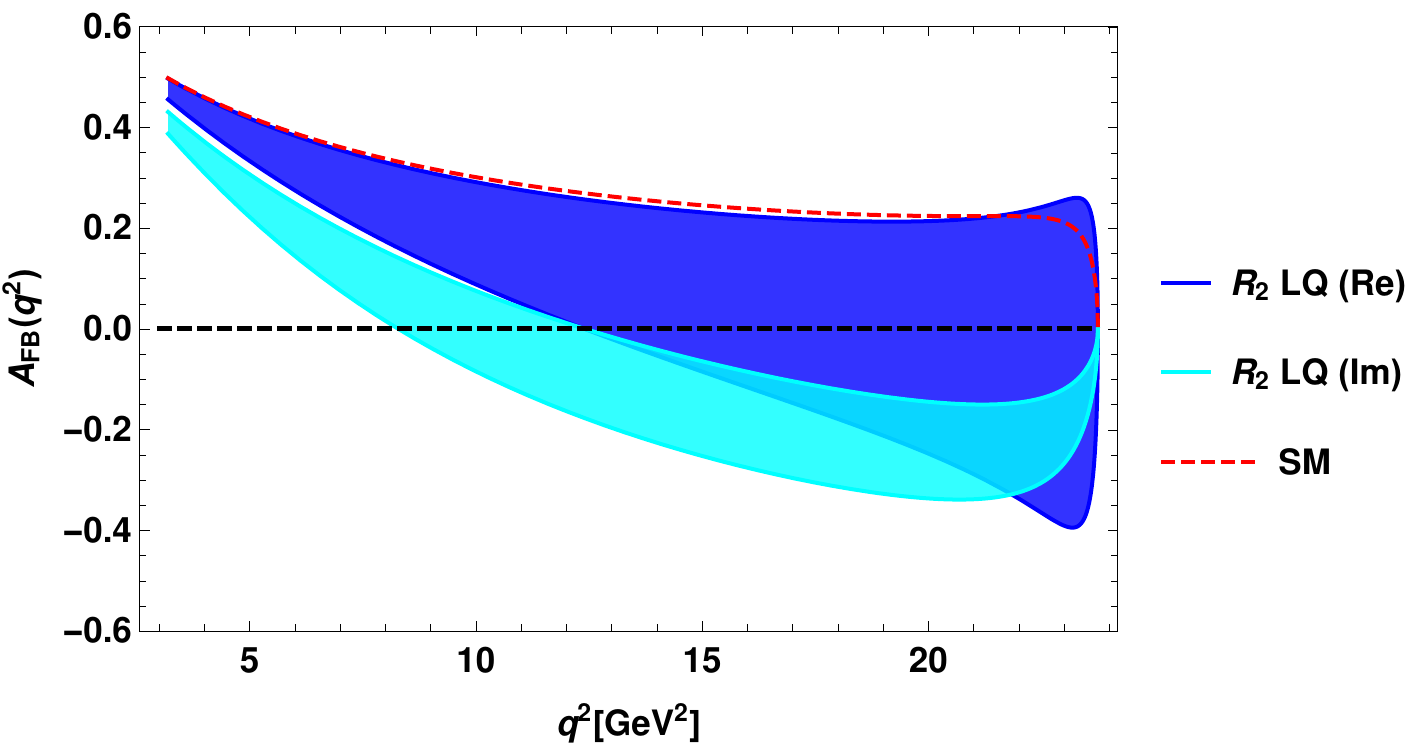}
\quad
\includegraphics[scale=0.5]{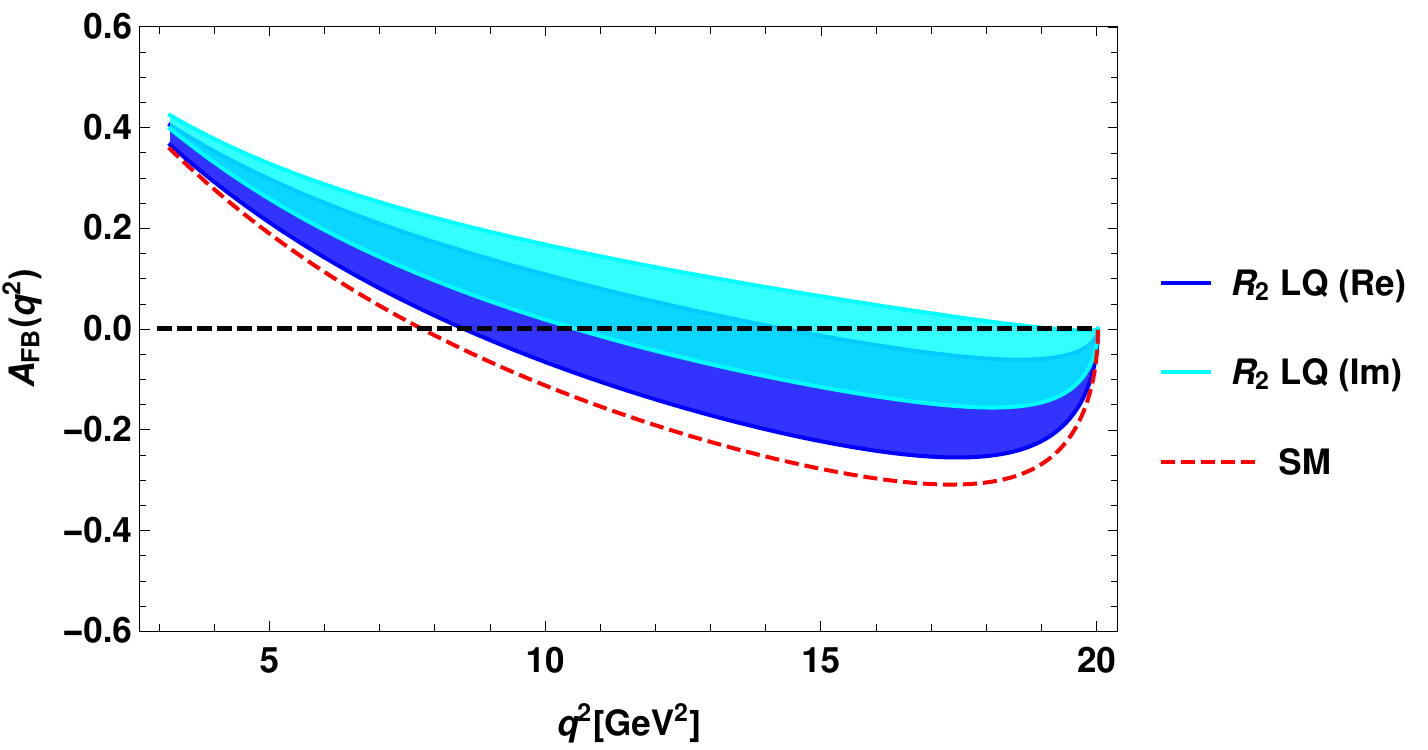}\\
\includegraphics[scale=0.5]{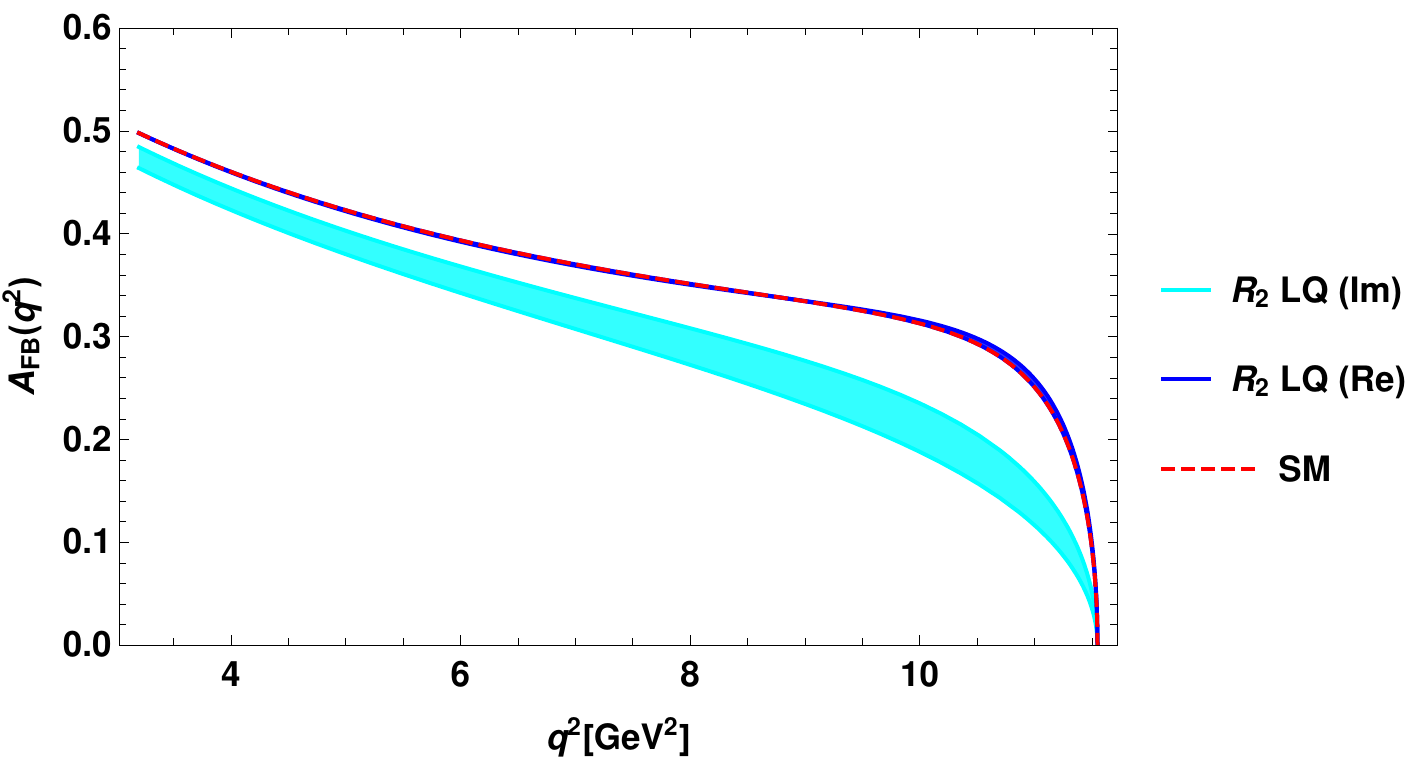}
\quad
\includegraphics[scale=0.5]{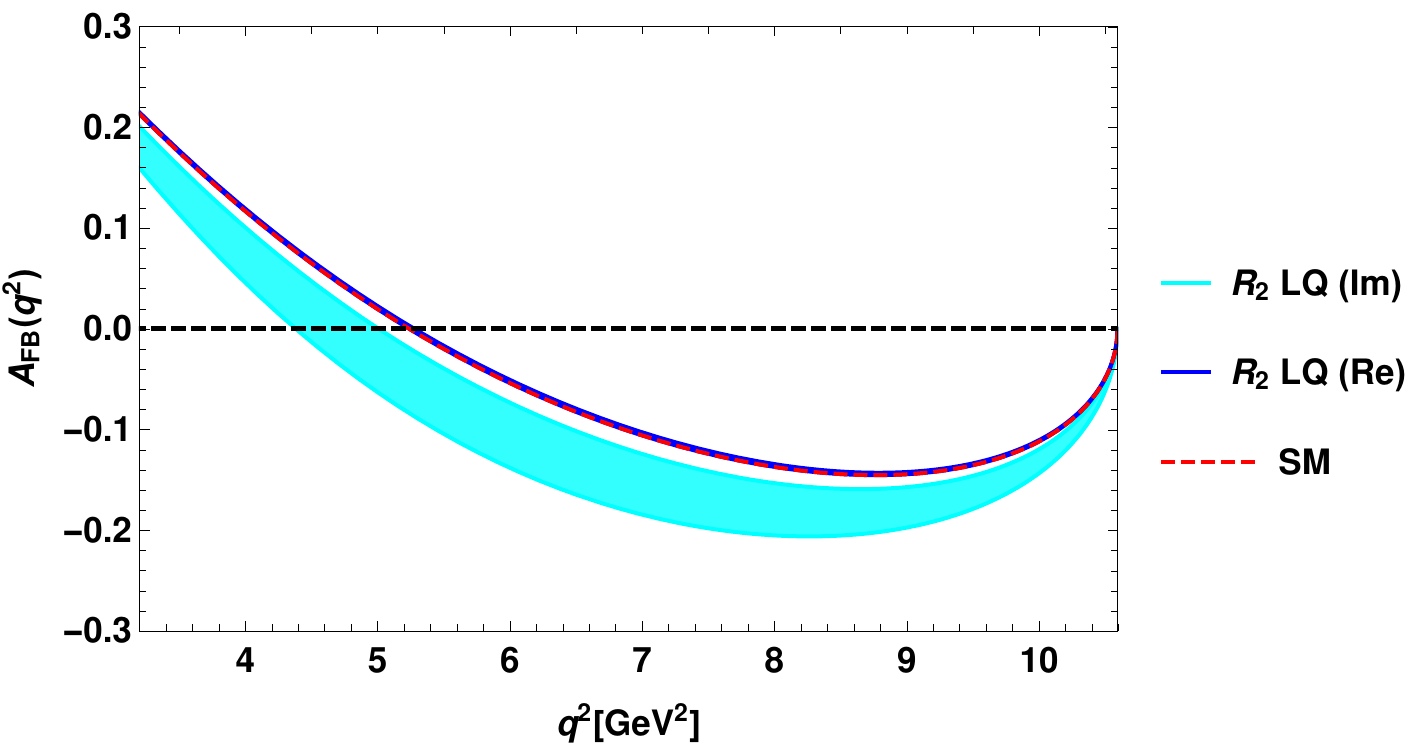}\\
\caption{The variation of forward-backward asymmetry of  $\bar B_s \to K^+\tau^- \bar \nu_\tau $ (top-left panel), $\bar B_s \to K^{* +} \tau^- \bar \nu_\tau$ (top-right panel), $\bar B_s \to D_s^+\tau^- \bar \nu_\tau$ (bottom-left panel) and $\bar B_s \to D_s^{*+}\tau^- \bar \nu_\tau $ (bottom-right panel) processes with respect to $q^2$ in the  $R_2$ scalar leptoquark model.} \label{R2-FB}
\end{figure}

Fig. \ref{R2-FB} describes the variation of the forward-backward asymmetry of $B_s \to K \tau \bar \nu_\tau$ (top-left panel), $B_s \to K^* \tau \bar \nu_\tau$ (top-right panel), $B_s \to D_s \tau \bar \nu_\tau$ (bottom-left panel) and $B_s \to D_s^* \tau \bar \nu_\tau$ (bottom-right panel) with respect to $q^2$. The $A_{FB}$ of $B_s \to K^{(*)}$  deviate significantly due to additional $R_2$ leptoquark whereas very  minor effect of new parameters are observed for  $B_s \to D_s^{(*)}$ modes. The numerical  values of  forward-backward asymmetries for both real and complex couplings  cases are given in Table \ref{Tab:R2}\,.

\begin{figure}[h]
\centering
\includegraphics[scale=0.5]{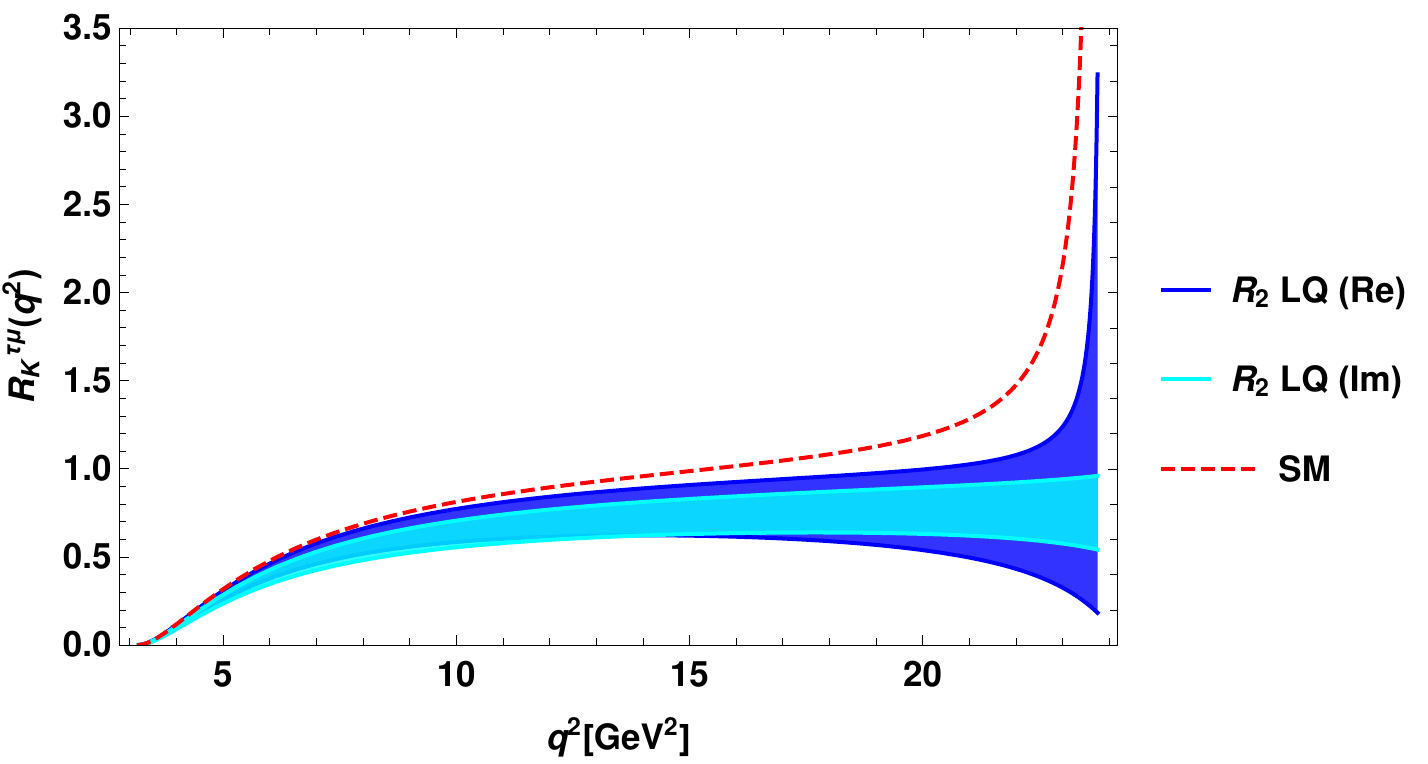}
\quad
\includegraphics[scale=0.5]{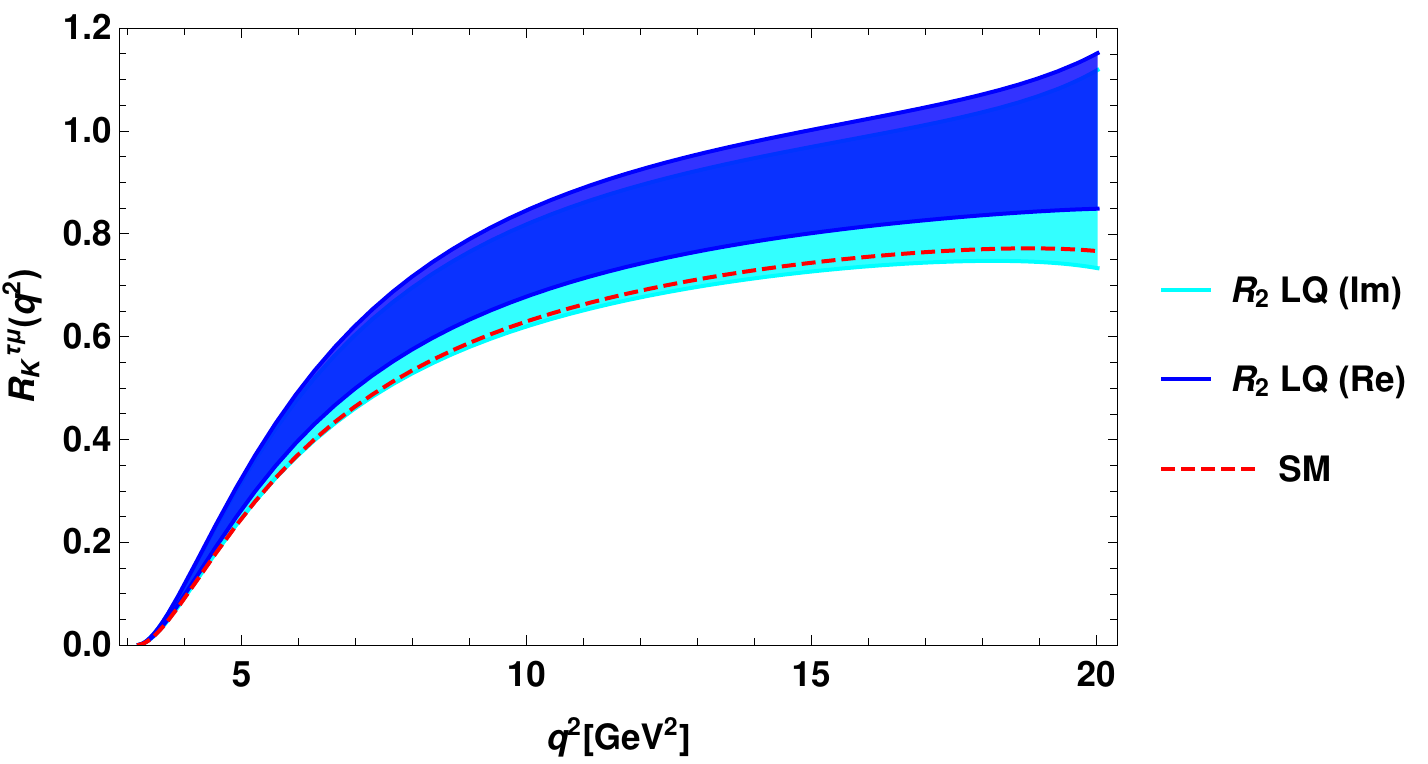}\\
\includegraphics[scale=0.5]{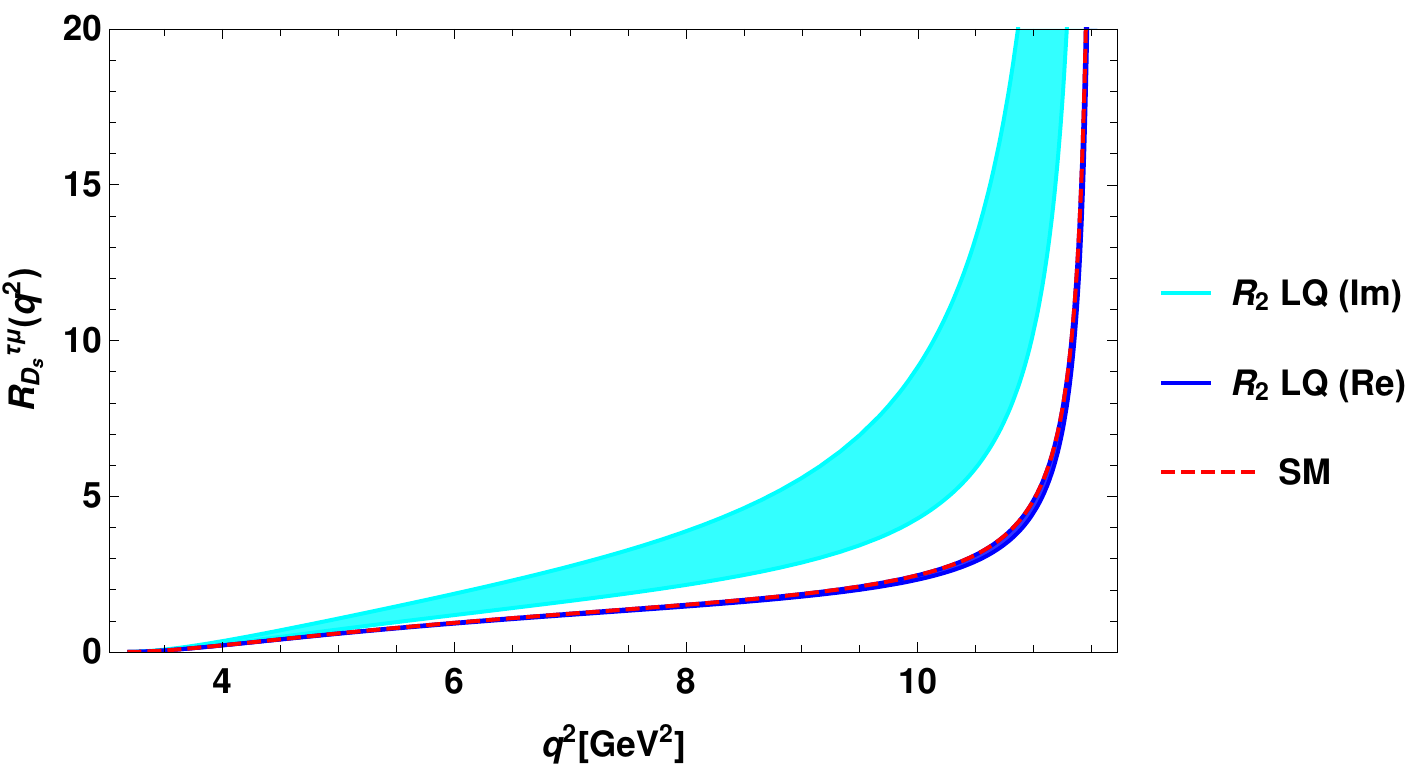}
\quad
\includegraphics[scale=0.5]{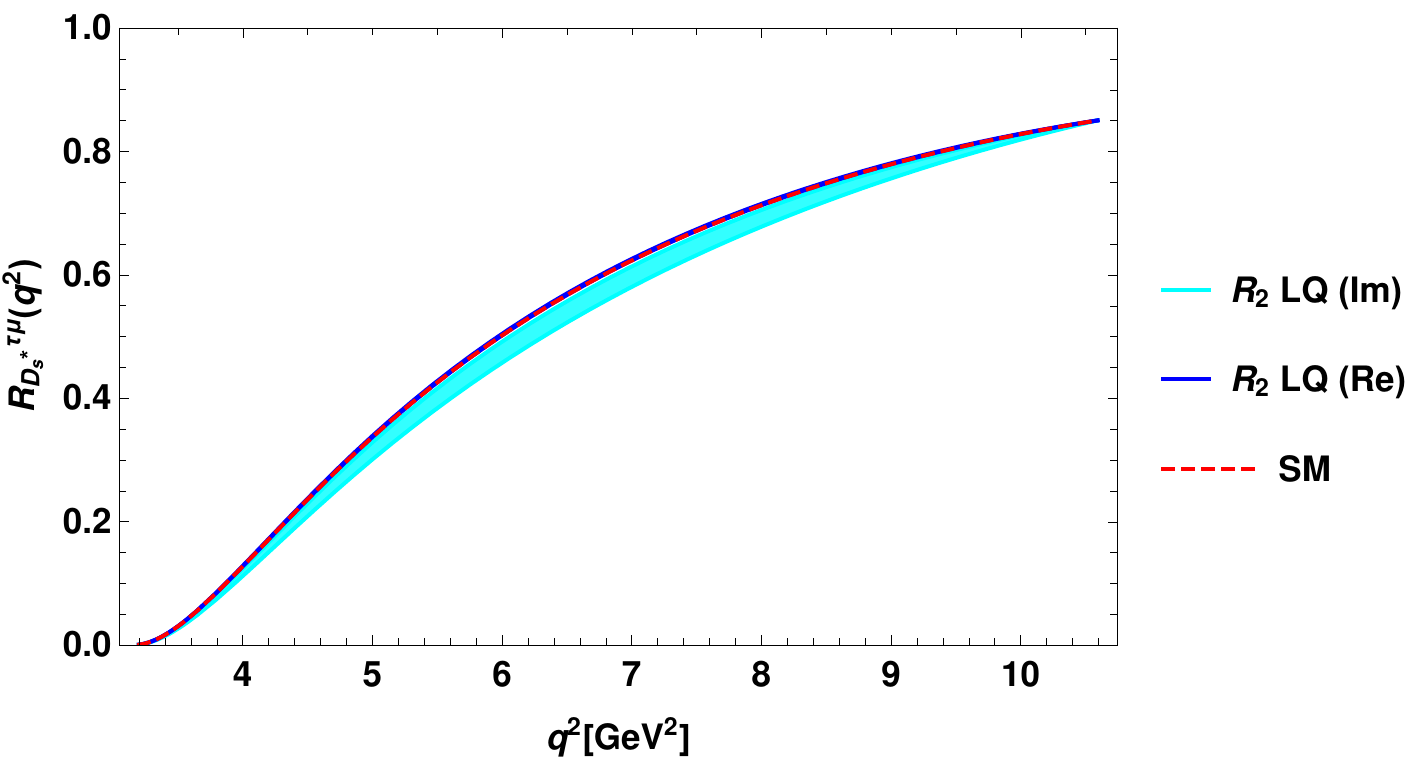}\\
\caption{The variation of lepton nonuniversality prameters of  $\bar B_s \to K^+\tau^- \bar \nu_\tau$ (top-left panel), $\bar B_s \to K^{* +} \tau^- \bar \nu_\tau$ (top-right panel), $\bar B_s \to D_s^+\tau^- \bar \nu_\tau$ (bottom-left panel) and $\bar B_s \to D_s^{*+}\tau^- \bar \nu_\tau $ (bottom-right panel) processes with respect to $q^2$ in the  $R_2$ scalar leptoquark model.} \label{R2-LNU}
\end{figure}

In Fig. \ref{R2-LNU}\,, we depict the variation of  $R_K^{\tau \mu}$ (top-left panel), $R_{K^*}^{\tau \mu}$ (top-right panel), $R_{D_s}^{\tau \mu}$ (bottom-left panel) and $R_{D_s^*}^{\tau \mu}$ (bottom-right panel) LNU parameters. The impact of  $R_2$ leptoquark on the $R_{D_s^{(*)}}^{\tau \mu}$ LNU parameters are found to be negligible. Table \ref{Tab:R2}\, contains the numerical values of all the LNU parameters.

\begin{figure}[h]
\centering
\includegraphics[scale=0.5]{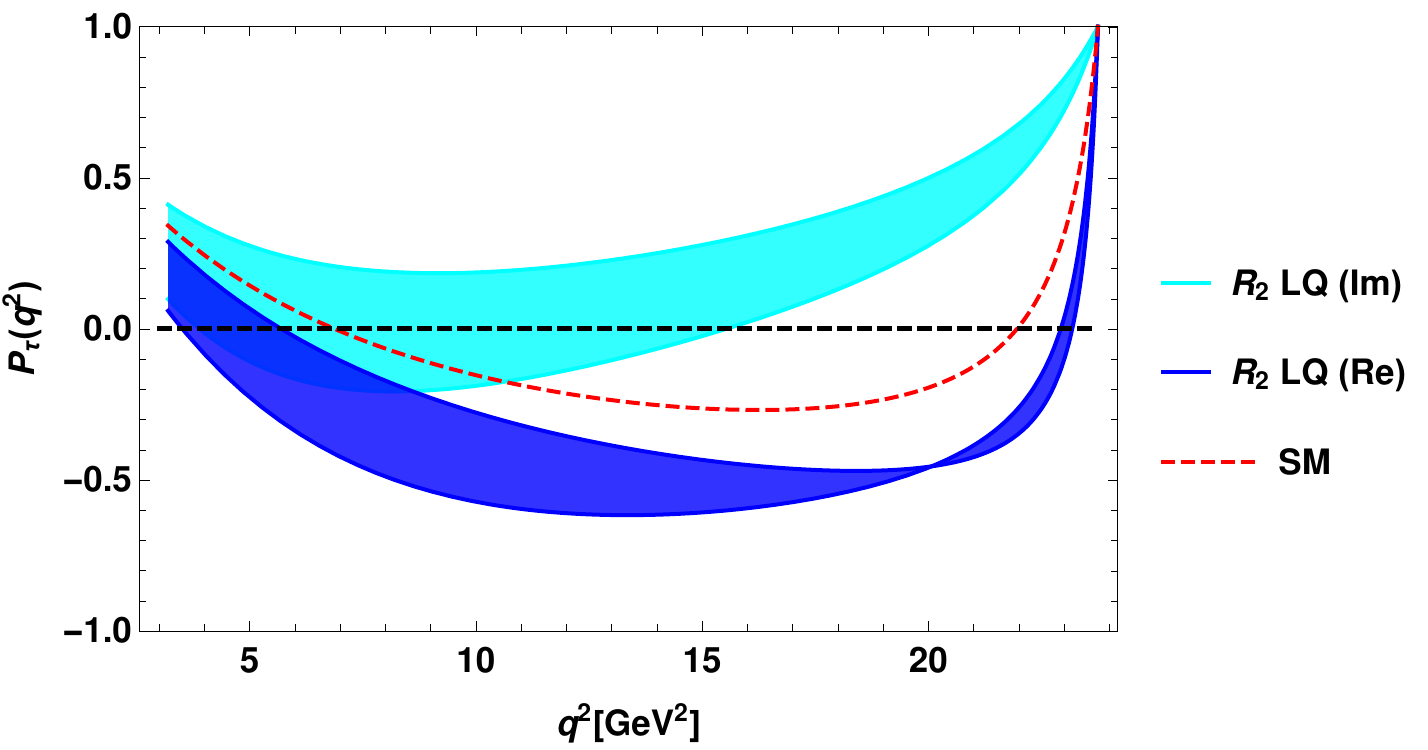}
\quad
\includegraphics[scale=0.5]{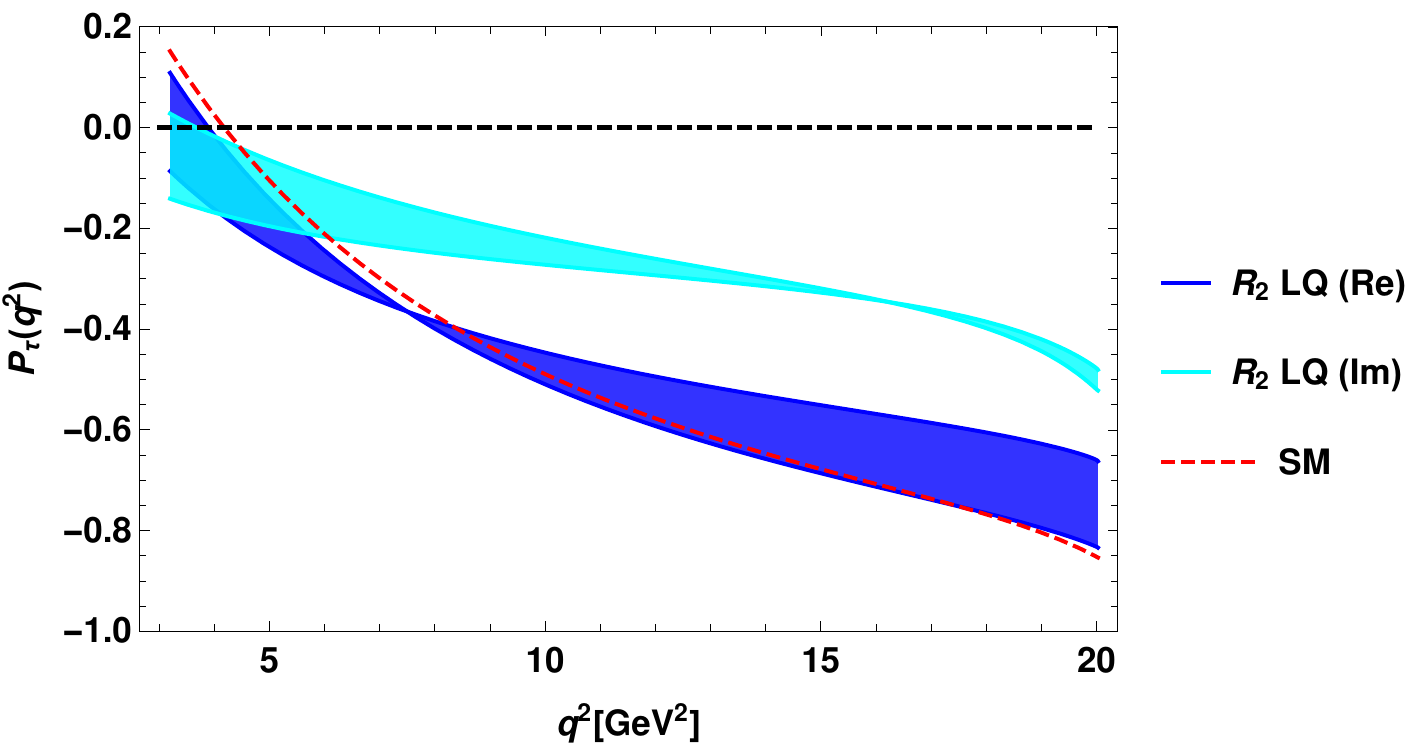}\\
\includegraphics[scale=0.5]{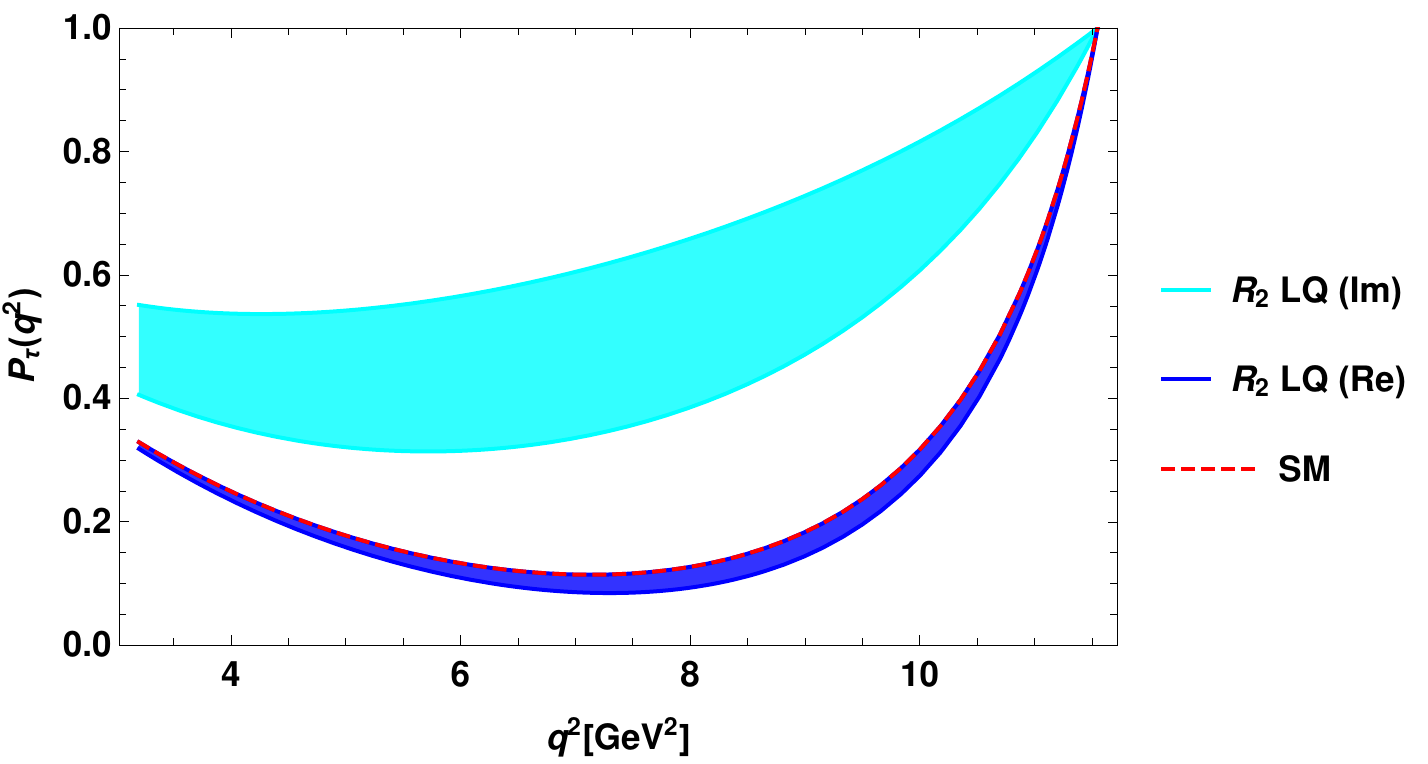}
\quad
\includegraphics[scale=0.5]{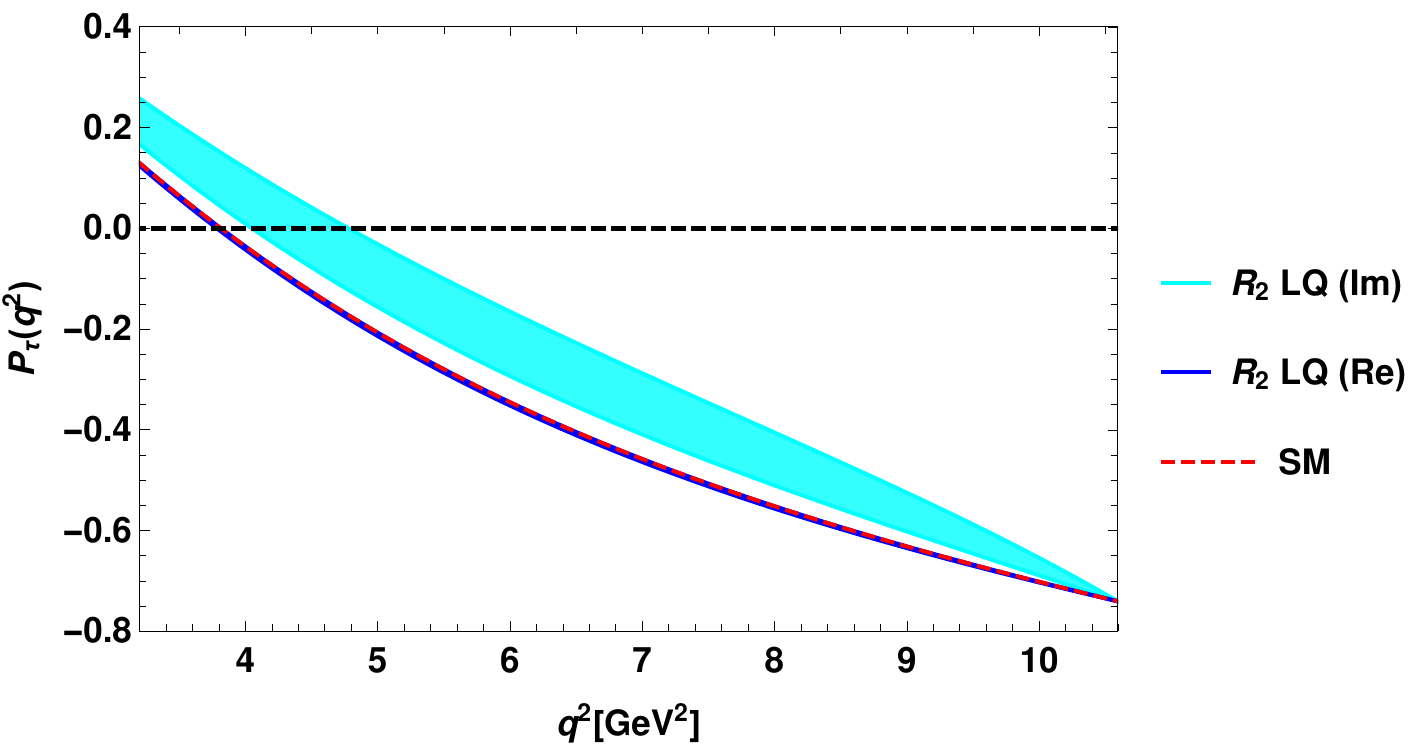}\\
\caption{The variation of lepton polarization asymmetry parameters of  $\bar B_s \to K^+\tau^- \bar \nu_\tau $ (top-left panel), $\bar B_s \to K^{* +} \tau^- \bar \nu_\tau $ (top-right panel), $\bar B_s \to D_s^+\tau^- \bar \nu_\tau$ (bottom-left panel) and $\bar B_s \to D_s^{*+}\tau^- \bar \nu_\tau $ (bottom-right panel) processes with respect to $q^2$ in the  $R_2$ scalar leptoquark model.} \label{R2-LP}
\end{figure}
\begin{figure}[h]
\centering
\includegraphics[scale=0.5]{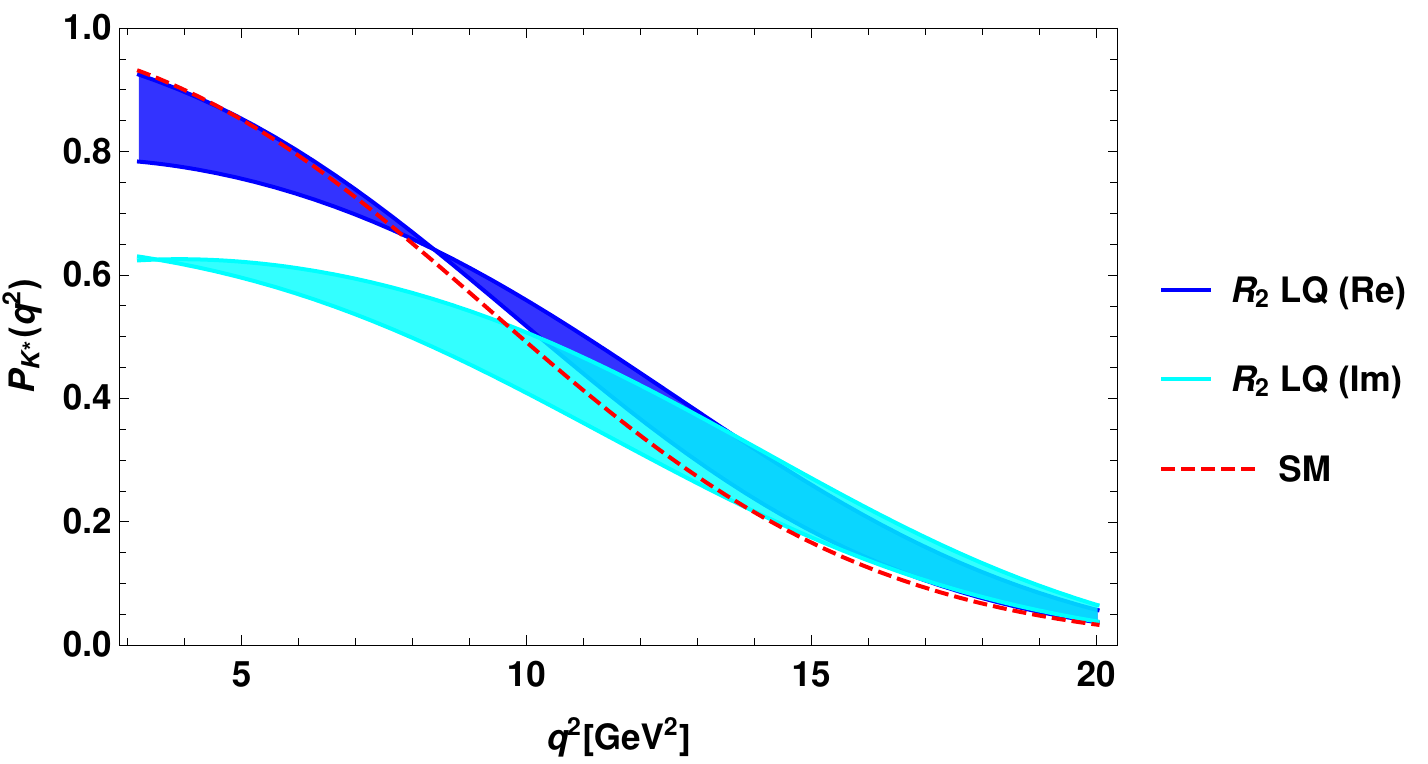}
\quad
\includegraphics[scale=0.5]{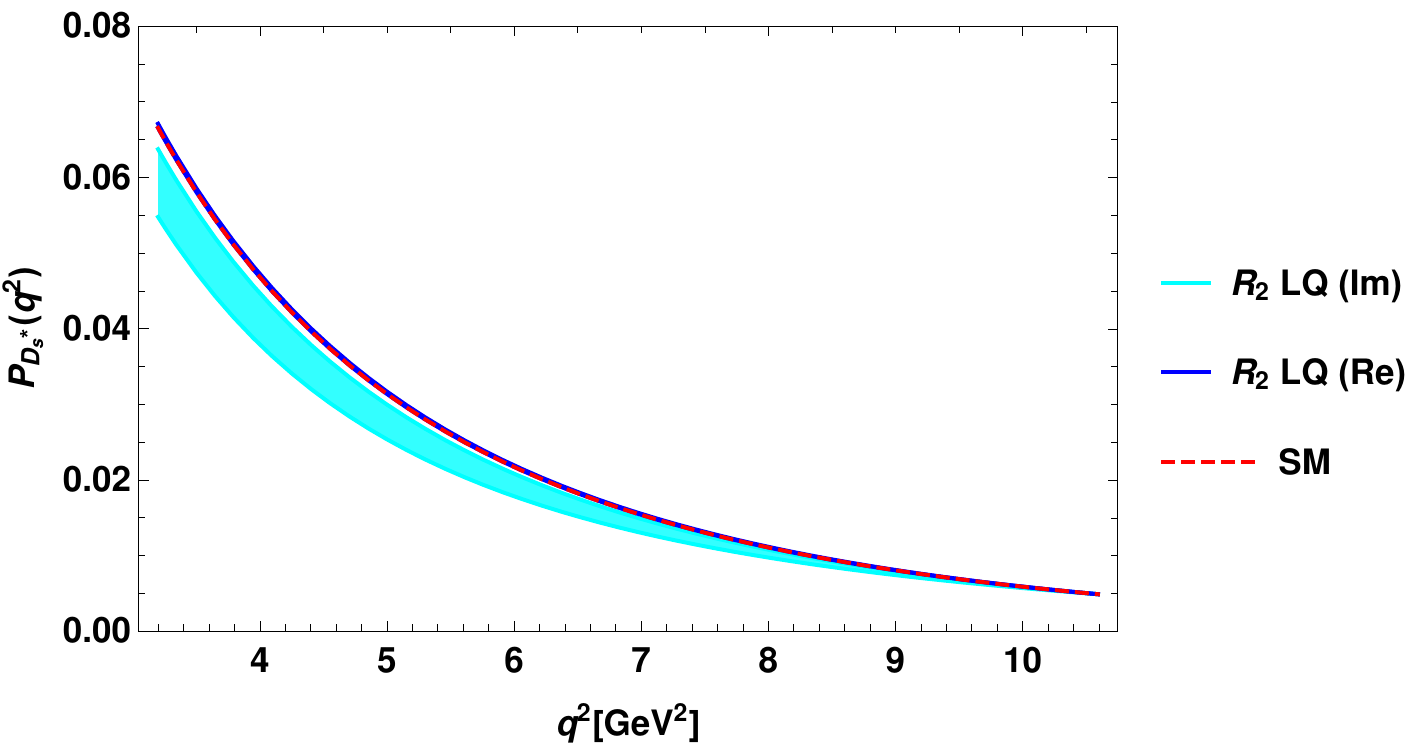}
\caption{The variation of hardon polarization asymmetry parameters of  $\bar B_s \to K^{* +} \tau^- \bar \nu_\tau $ (left panel) and $\bar B_s \to D_s^{*+}\tau^- \bar \nu_\tau $ (right panel) processes with respect to $q^2$ in the  $R_2$ scalar leptoquark model.} \label{R2-HP}
\end{figure}

The $\tau$  polarization asymmetries of all these decay modes are given in Fig. \ref{R2-LP}\,. The left and right panel of Fig. \ref{R2-HP} presents the $K^*$ and $D_s^*$ polarization asymmetry parameters respectively. In Table \ref{Tab:R2}\,, the predicted numerical values of lepton and hardon polarization asymmetries are listed. 

\begin{table}[htb]
\centering
\caption{The predicted values of the branching ratios and other physical observables of $B_s \to (K^{(*)}, D_s^{(*)} \tau \bar \nu_\tau)$  proesses in the SM and in the $R_2$ scalar leptoquark model. Here RC represents the real coupling  region and CC stands for complex coupling.} \label{Tab:R2}
\begin{tabular}{|c|c|c|c|}
\hline
~&~Observables~~&~Values for RC~~&~Values for CC\\
\hline
\hline
 $B_s$~&~Br~~&~$(6.2-7.0) \times 10^{-5}$~&~$(1.25-1.278) \times 10^{-4}$~\\
$\downarrow$~&~$\langle A_{FB}\rangle$~&~$-0.029\to 0.27$~&~$-0.204\to -0.029$\\ 
$K$~&~$\langle P_\tau \rangle$~~&~$-0.516\to-0.322$~&~$0.054-0.347$\\ 
~&~$\langle R_K^{\tau \mu} \rangle$~&~$0.593-0.671$~&~$1.198-1.224$\\

\hline
 $B_s$~&~Br~~&~$(2.0-3.15) \times 10^{-4}$~&~$(2.47-3.93) \times 10^{-4}$~\\
$\downarrow$~&~$\langle A_{FB}^\tau\rangle$~&~$-0.134\to 0.047$~&~$-0.041\to 0.1143$\\ 
$K^*$~&~$\langle P_\tau \rangle$~&~~$-0.6\to -0.5$~&~$-0.31\to -0.278$\\ 
~&~$\langle P_{K^*} \rangle$~&~$0.21-0.281$~&$0.19-0.285$ \\ 
~&~$\langle R_{K^*}^{\tau \mu} \rangle$~&~$0.591-0.917$~&~$0.719-1.146$\\

\hline
 $B_s$~&~Br~~&~$(1.34-1.39) \times 10^{-2}$~&~$(2.0-3.74) \times 10^{-2}$~\\
$\downarrow$~&~$\langle A_{FB}^\tau\rangle$~&~~$0.358$~&~$0.255-0.3$\\ 
$D_s$~&~$\langle P_\tau \rangle$~~&~$0.1674-0.2$~&~$0.45-0.7$\\ 
~&~$\langle R_{D_s}^{\tau \mu} \rangle$~~&~$0.16-0.6415$~&~$0.934-1.72$\\

\hline
 $B_s$~&~Br~~&~$(2.23-2.24) \times 10^{-2}$~&~$(2.1-2.2) \times 10^{-2}$~\\
$\downarrow$~&~$\langle A_{FB}^\tau\rangle$~~&~$-0.0996\to -0.097$~&~$-0.164\to -0.115$\\ 
$D_s^*$~&~$\langle P_\tau \rangle$~~&~$-0.519\to -0.514$~&~$-0.475\to -0.374$\\ 
~&~$\langle P_{D_s^*} \rangle$~~&~$0.0113-0.0114$&$0.0099-0.011$\\ 
~&~$\langle R_{D_s^*}^{\tau \mu} \rangle$~&~$0.43-0.464$~&~$0.438-0.457$\\

\hline
\end{tabular}
\end{table}

\section{Summary and Conclusion}

To summarize the article, we have studied the rare semileptonic decay processes of $B_s$ meson i.e,  $B_s \to (K^{(*)}, D_s^{(*)}) \tau \bar \nu_\tau$ mediated by the $b \to (u,c) \tau \bar \nu_l$ quark level transitions in the context of scalar leptoquark model. Our main motivation was to see how the singlet $S_1(\bar 3,1,1/3)$, doublet $R_2(\bar 3,2,7/6)$ and the triplet $S_3(\bar 3,3,1/3)$ leptoquarks affect the braching ratios and other physical observables like forward-backward asymmetry, lepton non-universality parameter,  lepton and hardon polarization asymmetry associated with these decay modes. The $S_1$ leptoquark contributes additional vector, scalar and tensor couplings to the standard model, where as the $S_3$ leptoquark contributes only $V_L$ coefficient and the $R_2$ leptoquark provides scalar and tensor couplings contributions. We have considered  that the new physics contributes only  to the tau lepton and the contribution to the first and second generation leptons are assumed to be SM  like.   We then considered two valid cases of leptoquark couplings, real and complex. We have used the experimental limit on the branching ratios of $B_u \to\tau \nu_\tau$ and $B \to\pi \tau \tau$ and the $R_\pi^l$ parameters to constrain the new leptoquark couplings related to the $b \to u \tau\bar \nu_\tau$ transition. For the case of $b \to c \tau \bar \nu_\tau$ processes, we have computed the allowed parameter space by using the branching ratio of $B_c \to \tau \nu_\tau$ obtained by using the life time of $B_c$ meson and the $R_{D^{(*)}}, ~R_{J/\psi}$ parameters. 
Using the allowed parameter space, we have estimated the branchng ratios and other observables of   $B_s \to (K^{(*)}, D_s^{(*)}) \tau \bar \nu_\tau$  modes.

We have observed that the branching ratios of all these decay processes deviate significantly from their corresponding SM results due to the new contributions from the complex couplings of $S_1$ leptoquark. However the  region of the constrained real couplings show negligible effect on the branching ratios of these decay modes. The impact of $S_1$ leptoquark on the forward-backward asymmetry of $B_s \to K^{(*)} \tau \bar \nu_\tau$  are found to be more sizable  in  comparison to the $B_s \to D_s^{(*)} \tau \bar \nu_\tau$ decay processes, where as opposite results are observed in the case of lepton non-universality parameters. The constrained couplings of  $S_1$ leptoquark affect the $R_{D_s^{(*)}}^{\tau  \mu}$ parameters more  comparatively than the   $R_{K^{(*)}}^{\tau  \mu}$ observables.  The lepton and hardon polarization asymmetries of all the decay modes have also shifted profoundly except the $B_s \to D_s^* \tau \bar \nu_\tau$ process.

The branching ratios of all the discussed semileptonic decay modes of $B_s$ meson have shown significant deviation in the $S_3$ leptoquark model. Both the real coupling and the complex coupling have sizable  impact on the banching ratios. Though the $R_{K^{(*)}}$ LNU parameters have shown negligible deviation, the $R_{D_s^{(*)}}^{\tau  \mu}$ parameters have  shifted  from their SM results.  The new couplings of $S_3$ leptoquark have  not shown any effect on the  forward-backward asymmetry, lepton and hardon polarization asymmetry.

The $R_2$ leptoquark have provided significant impact on the  branching ratios and other observables of $B_s \to K^{(*)} \tau \bar \nu_\tau$. However it has observed that, this leptoquark is very less sensitive to the observables of $B_s \to D_s^{(*)} \tau \bar \nu_\tau$ processes.

 To conclude, we have observed that the  physical observables of rare semileptonic decay modes of $B_s$ meson are too sensitive to the new physics contribution to the SM arising due to the scalar leptoquark exchange. As like $R_{D^{(*)}},~R_{J/\psi}$ parameters associated with the $B \to D^{(*)} \tau \bar \nu_\tau$ processes, the $B$-factories as well as the LHCb should check the  violation of lepton universality  in their corresponding  $B_s$ decay modes i.e, $B_s \to  D_s^{(*)}\tau \bar \nu_\tau$, the observation of
which would provide the indirect hints for possible existence of leptoquarks.

\bibliography{BL}

\end{document}